\documentclass[12pt]{article}
\usepackage{amssymb,amsmath}
\usepackage{comment}
\usepackage[dvipdfmx]{graphicx}
\usepackage{bm}
\usepackage{color}
\usepackage{here}
\usepackage{enumerate}
\usepackage{here}
\usepackage{subfigure} 
\newcommand{\Zb}{\mathbb{Z}}
\newcommand{\Rb}{\mathbb{R}}
\newcommand{\Cb}{\mathbb{C}}

\DeclareMathOperator*{\Tr}{{\rm Tr}}

\newcommand{\psib}{\bar{\psi}}

\newcommand{\im}{\mathrm{Im}}

\newcommand{\sigmab}{\bar{\sigma}}

\newcommand{\sigmad}{\sigma^{\dagger}}

\newcommand{\vtheta}{\vartheta}

\newcommand{\Ih}{\hat{I}}

\renewcommand{\Re}{\operatorname{Re}}
\renewcommand{\Im}{\operatorname{Im}}

\newcommand{\parallelsum}{\mathbin{\|}}

\numberwithin{equation}{section}

\begin{document}

%%%%%%%%%%%%%%%%%%%%%%%%%%%%%%%%%%%%%%%%%%%%
\thispagestyle{empty}
\begin{flushright}
DCPT-20/11
\\
%%%%%%%%%%%%%%%%%%%%%%%%%%%%%%%%%%%%%%%%%%%%%%%%%%%%%%%%%%%%%%%%%
%input \\
%%%%%%%%%%%%%%%%%%%%%%%%%%%%%%%%%%%%%%%%%%%%%%%%%%%%%%%%%%%%%%%%%

\end{flushright}
\vskip2cm
\begin{center}
{\Large \bf Singular BPS boundary conditions \\ 
\vskip0.5cm
in $\mathcal{N} = (2,2)$ supersymmetric gauge theories
}

\vskip1.5cm
Tadashi Okazaki\footnote{tadashi.okazaki@durham.ac.uk}

\bigskip
and
\\
\bigskip
Douglas J. Smith\footnote{douglas.smith@durham.ac.uk}

\bigskip
{\it Department of Mathematical Sciences, Durham University,\\
Lower Mountjoy, Stockton Road, Durham DH1 3LE, UK}

\end{center}

%%%%%%%%%%%%%%%%%%%%%%%%%%%%%%%%%%%%%%%%%%%%
\vskip1cm
\begin{abstract}
We derive general BPS boundary conditions in two-dimensional $\mathcal{N}=(2,2)$ supersymmetric gauge theories. We analyze the solutions of these boundary
conditions, and in particular those
that allow the bulk fields to have poles at the boundary. 
We also present the brane configurations for the half- and quarter-BPS boundary conditions 
of the $\mathcal{N}=(2,2)$ supersymmetric gauge theories in terms of branes in Type IIA string theory. 
We find that 
both A-type and B-type brane configurations are lifted to M-theory 
as a system of M2-branes ending on an M5-brane wrapped on a product of a holomorphic curve in $\mathbb{C}^2$ 
with a special Lagrangian 3-cycle in $\mathbb{C}^3$. 
\end{abstract}
%%%%%%%%%%%%%%%%%%%%%%%%%%%%%%%%%%%%%%%%%%

\newpage
\setcounter{tocdepth}{3}
\tableofcontents
%%%%%%%%%%%%%%%%%%%%%%%%%%

%%%%%%%%%%%%%%%%%%%%%%%%%%%%%%%%%%%%%%%%%%%%%%%%%%%%%%%%%%%%%%%%%
%%%%%%%%%%%%%%%%%%%%%%%%%%%%%%%%%%%%%%%%%%%%%%%%%%%%%%%%%%%%%%%%%
\section{Introduction and summary}
\label{sec_intro}
%%%%%%%%%%%%%%%%%%%%%%%%%%%%%%%%%%%%%%%%%%%%%%%%%%%%%%%%%%%%%%%%%
%%%%%%%%%%%%%%%%%%%%%%%%%%%%%%%%%%%%%%%%%%%%%%%%%%%%%%%%%%%%%%%%%
The BPS boundary conditions in supersymmetric field theories have been studied in different dimensions and with various amounts of supersymmetry, 
e.g. 4d $\mathcal{N}=4$ super Yang-Mills (SYM) theories \cite{Gaiotto:2008sa,Gaiotto:2008ak, Hashimoto:2014vpa, Hashimoto:2014nwa,Gaiotto:2016hvd}, 
4d $\mathcal{N}=2$ supersymmetric gauge theories \cite{Dimofte:2012pd}, 
3d $\mathcal{N}=2$ supersymmetric field theories \cite{Gadde:2013wq, Okazaki:2013kaa,Aprile:2016gvn,Yoshida:2014ssa,Dimofte:2017tpi,Brunner:2019qyf,Costello:2020ndc},  
3d $\mathcal{N}=4$ supersymmetric gauge theories \cite{Bullimore:2016nji, Chung:2016pgt, Costello:2018fnz, Hanany:2018hlz,Okazaki:2019bok}, 
3d $\mathcal{N}\ge 4$ Chern-Simons matter theories  \cite{Berman:2009xd, Hosomichi:2014rqa, Okazaki:2015fiq} 
and 5d $\mathcal{N}=1$ supersymmetric gauge theories \cite{Gaiotto:2015una}. 
An alternative, classically equivalent, approach is to couple to a boundary action in such a way that
supersymmetry is preserved without boundary conditions, allowing off-shell
supersymmetry \cite{Belyaev:2008xk}. This approach has been applied in the context of
ABJM and supersymmetric Chern-Simons theories \cite{Berman:2009kj, Faizal:2011cd, Faizal:2016skd}.

In particular supersymmetric boundary conditions in 2d $\mathcal{N}=(2,2)$ supersymmetric field theories 
which describe D-branes \cite{Ooguri:1996ck,Govindarajan:1999js,Hori:2000ck,Govindarajan:2000ef,Hori:2000ic} 
have attracted much attention in both physics and mathematics. 
For example, 
they can provide a physical setup to address the homological mirror symmetry conjecture \cite{Kontsevich:1994dn} 
and gauge theoretic definition of knot homology \cite{Gaiotto:2015aoa, Gaiotto:2015zna}. 
The 2d $\mathcal{N}=(2,2)$ supersymmetric field theory admits two types of half-BPS boundary conditions, 
that is the A-type and B-type boundary conditions \cite{Witten:1991zz}
which preserve 1d $\mathcal{N}=2_A$ and $\mathcal{N}=2_B$ supersymmetries respectively. 
The D-brane of type B in a Calabi-Yau manifold $X$ is argued to be 
equivalent to the bounded derived category of coherent sheaves on $X$ 
\cite{Douglas:2000gi, Aspinwall:2001pu, Sharpe:1999qz, Lerche:2001vj, Hellerman:2001ct, Katz:2002gh} 
while that of type A is closely related to the Fukaya category \cite{Witten:1992fb, Kapustin:2001ij}.

The detailed analysis of B-type boundary conditions for Abelian gauge theories were presented in \cite{Herbst:2008jq} 
and the basic B-type boundary conditions associated to the Neumann boundary conditions on the gauge field 
for non-Abelian gauge theories were studied in \cite{Honda:2013uca,Hori:2013ika}. 
On the other hand, to our knowledge, A-type boundary conditions for gauge theories have been much less studied in the literature. 

In this paper we examine more general A-type and B-type boundary conditions as well as quarter-BPS boundary conditions 
in $\mathcal{N}=(2,2)$ supersymmetric non-Abelian gauge theories 
for which the gauge field may not satisfy the Neumann boundary condition. 
Although the problem of describing all half-BPS boundary conditions is enormously involved 
and our understanding of the boundary conditions for the non-Abelian gauge theories is far from complete, 
we find new types of BPS boundary conditions which admit singular solutions. 
Singular boundary conditions were found in higher dimensional BPS boundary conditions; 
e.g. 
the half-BPS boundary conditions \cite{Witten:2011zz, Mazzeo:2013zga} in 5d SYM theory, 
the BPS boundary conditions \cite{Gaiotto:2008sa} and the quarter-BPS boundary conditions \cite{Hashimoto:2014vpa,Hashimoto:2014nwa} in 4d $\mathcal{N}=4$ SYM theory 
and the $\mathcal{N}=(0,4)$ boundary conditions \cite{Chung:2016pgt} in 3d $\mathcal{N}=4$ gauge theories. 
Such higher dimensional and highly supersymmetric cases are described 
by Nahm's equation \cite{Nahm:1979yw} 
as the Nahm pole boundary condition. 
We argue that the B-type boundary conditions in $\mathcal{N}=(2,2)$ gauge theories, 
which are distinguished from the Nahm pole boundary condition, 
allow the bulk fields to have singularities even for the Abelian gauge theory 
when the gauge fields are not subject to the ordinary Neumann boundary condition. 
In particular, when the gauge field is subject to the Dirichlet boundary condition 
and the chiral multiplet scalar field satisfies the Neumann-type boundary condition, 
such singular solutions naturally arise without any boundary degrees of freedom. 
We further discuss the quarter BPS boundary conditions which have many more solutions, but much richer than the half-BPS case. 
They also admit singular solutions and mixed Dirichlet-Neumann boundary conditions.

We also construct the BPS boundary conditions in $\mathcal{N}=(2,2)$ supersymmetric gauge theories by using brane configurations in Type IIA string theory 
by introducing additional branes to the Hanany-Hori brane setup \cite{Hanany:1997vm}. 
For each of A-type and B-type boundary conditions, 
one can additionally introduce two kinds of NS5-branes and two kinds of D4-branes. 
The Neumann and Dirichlet boundary conditions for the vector multiplet are realized 
by the NS5-branes and D4-branes respectively 
and the two possible choices for each brane corresponds to those of boundary conditions for the chiral multiplet. 
We find singular solutions to some of the boundary conditions, as could be
anticipated from our brane configurations which include systems of D2 and D4
branes oriented such that they are T-dual to the D1-D3 system realising the
Nahm equation \cite{Diaconescu:1996rk, Tsimpis:1998zh}.
Furthermore, the presence of both kinds of branes can preserve 1d $\mathcal{N}=1$ supersymmetry, 
which realizes the quarter-BPS boundary conditions. 
We find that the boundary conditions on fermions crucially determine 
the boundary branes as well as the boundary conditions on the bosonic fields. 
Both A-type and B-type boundary conditions can be lifted to M-theory 
as a system of a single M5-brane wrapped on a product of a holomorphic curve in $\mathbb{C}^2$ 
with a special Lagrangian 3-cycle in $\mathbb{C}^3$ as well as M2-branes. 

The organization of the paper is straightforward. 
In section \ref{sec_halfbc} we compute the half-BPS boundary conditions 
for 2d $\mathcal{N}=(2,2)$ supersymmetric gauge theories and argue that they admit singularities. 
In section \ref{sec_quarter} we analyze the quarter-BPS boundary conditions. 
In section \ref{sec_IIA} we construct brane configurations realizing these BPS boundary conditions in Type IIA string theory 
which generalizes the Hanany-Hori brane setup \cite{Hanany:1997vm}. We also
describe the M-theory lift of these Type IIA brane configurations, showing that
there is a rich set of duality relations, unifying all the configurations we
consider.
In appendix \ref{sec_4dN1} we give our conventions and notations 
of superspace and supermultiplets. 
 
%%%%%%%%%%%%%%%%%%%%%%%%%%%%%%%%%%
%%%%%%%%%%%%%%%%%%%%%%%%%%%%%%%%%%
\section{Half-BPS boundary conditions}
\label{sec_halfbc}
%%%%%%%%%%%%%%%%%%%%%%%%%%%%%%%%%%
%%%%%%%%%%%%%%%%%%%%%%%%%%%%%%%%%%
We consider 2d $\mathcal{N}=(2,2)$ supersymmetric gauge theories on a half-space $\mathbb{R}\times \mathbb{R}_+$ 
of the Minkowski space with coordinated $(x^0, x^1)$ whose boundary is at $x^1=0$. 
We analyze the supersymmetric boundary conditions imposed on the bulk fields 
without any boundary degrees of freedom. It would also be interesting to explore
boundary conditions in the presence of coupling to boundary fields but we leave
that for future work. Similarly it would be interesting to derive our results
within the framework of supersymmetry without boundary conditions by including
boundary interactions, developed by Belyaev and van Niewenhuizen \cite{Belyaev:2008xk}.

The 2d $\mathcal{N}=(2,2)$ supersymmetric gauge theories can be constructed 
in terms of three supermultiplets, the chiral, twisted chiral and vector
multiplets. The twisted chiral multiplets arise in 2d, but the other multiplets
come from direct dimensional reduction of hypermultiplet and vector multiplets
in 4d $\mathcal{N} = 1$ theories. These 2d
theories have R-symmetry group $U(1)_A \times U(1)_V$ which arises from the
4d $U(1)$ R-symmetry and the dimensional reduction to 2d. This R-symmetry group
may be broken depending on the field content and superpotential.

The chiral multiplet consists of a complex scalar field $\phi$ and Dirac fermions $\psi_{\pm}$, $\overline{\psi}_{\pm}$. 
The scalar $\phi$ has no $U(1)_A$ charge while the fermions $\psi_{-}$, $\overline{\psi}_{+}$ carry charge $+1$ 
and $\overline{\psi}_{-}$, $\psi_{+}$ have charge $-1$. 
The $U(1)_{V}$ charges of the fields can be shifted by a constant. 

The vector multiplet has a two-dimensional gauge field $A_{\mu}$, 
a complex scalar field $\sigma$ and gauginos $\lambda_{\pm}$, $\overline{\lambda}_{\pm}$ as Dirac fermions. 
They transform as the adjoint representation under the gauge group $G$. 
The gauge field is neutral under the R-symmetry group. 
The scalar field $\sigma$ has no charge under the $U(1)_{V}$ but carries the $U(1)_{A}$ R-charge $+2$. 
The gauginos $\lambda_{\pm}$ and $\overline{\lambda}_{\pm}$ carry the $U(1)_{V}$ R-charge $-1$ and $+1$. 
The gauginos $\overline{\lambda}_{+}$ and $\lambda_{-}$ have the $U(1)_{A}$ R-charge $+1$ 
while the other gauginos have the $U(1)_{A}$ R-charge $-1$. 

%{\bf [The notations for lower-indices on fermions $+$, $-$ should be fixed. Have used notation $\epsilon^1 = \epsilon_2 = \epsilon_{+}$ and $\epsilon^2 = -\epsilon_1 = -\epsilon_{-}$.]}
We use notation for 2d spinors $\psi_{\alpha}$ where $\alpha$ can take values
$\alpha = 1 = -$ or $\alpha = 2 = +$, and $\overline{\psi}_{\alpha}$ denotes the
complex conjugate of $\psi_{\alpha}$, or Hermitian conjugate for vector or
matrix valued spinors. Our conventions for spinors are summarized in
appendix~\ref{sec_spinorconv}.

In order to have normalizable supersymmetric ground states, 
the gauge theory must have large enough numbers of matter multiplets. 
For $G=U(N_c)$ gauge theory with 
$N_f$ chiral multiplets transforming in the fundamental representation 
there are $\left( \begin{smallmatrix} N_f\\ N_c \end{smallmatrix} \right)$ supersymmetric ground states
and supersymmetry is broken for $N_c > N_f$.\footnote{If we also have
$N_a$ chiral multiplets transforming in the antifundamental representation,
these statements hold with $N_f$ replaced by $\max (N_f, N_a)$.}
For $G=SU(N_c)$ gauge theory with 
$N_f$ fundamental chiral multiplets, 
there is no supersymmetric ground state for $1\le N_f\le N_c$ \cite{Hori:2006dk}. 
\footnote{See \cite{Hori:2011pd} for orthogonal and symplectic gauge groups. }

The Lagrangian densities of 2d $\mathcal{N}=(2,2)$ gauge theory are given by
\begin{eqnarray}
\mathcal{L}_{\mathrm{gauge}} & = & \Tr \left[ \frac{1}{2} \left( W^{\alpha}W_{\alpha}\vert_{\theta\theta} + h.c. \right) \right]_{4d \rightarrow 2d} \nonumber \\
 & = & \Tr \bigg[ -\frac{1}{2} v_{mn}v^{mn} - 2i \bar{\lambda}\bar{\sigma}^mD_m\lambda + DD + i \partial_m (\bar{\lambda}\bar{\sigma}^m\lambda) \bigg]_{4d \rightarrow 2d} \nonumber \\
 & = & \Tr \bigg[ -\frac{1}{2} F_{\mu\nu}F^{\mu\nu} - 2(D_{\mu}\sigma^{\dagger})(D^{\mu}\sigma) - g^2 [\sigma, \sigma^{\dagger}]^2 - 2i \bar{\lambda}\Gamma^{\mu}D_{\mu}\lambda \nonumber \\
 & & + 2\sqrt{2}g \left( \bar{\lambda}^1 [\sigmad, \lambda_1] - \bar{\lambda}^2 [\sigma, \lambda_2] \right) + DD + i \partial_{\mu} (\bar{\lambda}\Gamma^{\mu}\lambda) \bigg] \nonumber \\
\mathcal{L}_{\mathrm{chiral}} & = & \left[ \Phi^{\dagger} e^V \Phi \vert_{\theta\theta\bar{\theta}\bar{\theta}} \right]_{4d \rightarrow 2d} \\
 & = & \bigg[ g \phi^{\dagger}D\phi - (D_m\phi^{\dagger})(D^m \phi)
 - i \bar{\psi} \bar{\sigma}^m D_m \psi + F^{\dagger} F \nonumber \\
 & & + i \sqrt{2} g \left(\phi^{\dagger}\lambda\psi - \bar{\psi}\bar{\lambda}\phi \right) + \frac{i}{2}\partial_m(\bar{\psi}\bar{\sigma}^m\psi) \bigg]_{4d \rightarrow 2d} \nonumber \\
 & = & g \phi^{\dagger}D\phi - (D_{\mu}\phi^{\dagger})(D^{\mu} \phi)
 - g^2 \phi^{\dagger} \left( \sigma \sigma^{\dagger} + \sigma^{\dagger} \sigma \right) \phi
 - i \bar{\psi} \Gamma^{\mu} D_{\mu} \psi \nonumber \\
 & & + \sqrt{2}g \left( \bar{\psi}^1 \sigma \lambda_1 - \bar{\psi}^2 \sigma^{\dagger} \psi_2 \right)+ F^{\dagger} F + i \sqrt{2} g \left(\phi^{\dagger}\lambda\psi - \bar{\psi}\bar{\lambda}\phi \right) \nonumber \\ 
 & & + \frac{i}{2}\partial_{\mu}(\bar{\psi}\Gamma^{\mu}\psi) \\
\mathcal{L}_{\mathrm{FI}} & = & \Tr \left[ -\zeta D + \frac{\theta}{2\pi}F_{01} \right] \label{LFI}
\end{eqnarray}
where $\mathcal{L}_{\mathrm{gauge}}$ and $\mathcal{L}_{\mathrm{chiral}}$ are 
the kinetic terms of the vector multiplet and those of the chiral multiplet which arise from the dimensional
reduction of the 4d $\mathcal{N} = 1$ Lagrangians (in the WZ gauge), and
$\mathcal{L}_{\mathrm{FI}}$ are the 2d FI and theta angle terms (with $\theta$
here not to be confused with the Grassmann coordinates).
When the gauge group $G$ has an Abelian factor, the FI parameter $\zeta$ appears
and we can include the FI term in 2d. 
The final (total derivative) terms in the gauge and chiral Lagrangians are required for the Lagrangians to be
real in the presence of a boundary.
More generally we can include $N_f$ fundamental chiral superfields $\Phi^i$
(and also consider other representations) and include a superpotential
$\mathcal{W}(\Phi)$ and twisted superpotential
$\widetilde{\mathcal{W}}(\widetilde{\Phi})$ which are holomorphic functions of
chiral and twisted chiral superfields. These contribute to the action as
\begin{eqnarray}
\mathcal{L}_{\mathcal{W}} & = & \mathcal{W}\vert_{\theta\theta} + h.c. \\
\mathcal{L}_{\widetilde{\mathcal{W}}} & = & \widetilde{\mathcal{W}}\vert_{\theta^1 \bar{\theta}^2} + h.c.
\end{eqnarray}

%{\bf [To complete the Lagrangian densities]}

For the supercurrent we find the
following result for the gauge multiplet
\begin{align}
J^{\mu}_{-} & =  -i\left( F_{01} + ig[\sigma, \sigmad] + i \zeta \right) \epsilon^{\mu\nu}(\sigma_{\nu}\bar{\lambda})_{-} - \sqrt{2}i\left( D^{\mu} + \epsilon^{\mu\nu}D_{\nu} \right) \sigma \bar{\lambda}_{-}, \\
J^{\mu}_{+} & =  -i\left( F_{01} + ig[\sigma, \sigmad] - i \zeta \right) \epsilon^{\mu\nu}(\sigma_{\nu}\bar{\lambda})_{+} + \sqrt{2}i\left( D^{\mu} - \epsilon^{\mu\nu}D_{\nu} \right) \sigmad \bar{\lambda}_{+}, \\
\bar{J}^{\mu \, -} & =  -i\left( F_{01} - ig[\sigma, \sigmad] + i \zeta \right) \epsilon^{\mu\nu}(\sigmab_{\nu}\lambda)^{-} - \sqrt{2}i\left( D^{\mu} - \epsilon^{\mu\nu}D_{\nu} \right) \sigma \lambda^{-}, \\
\bar{J}^{\mu \, +} & = -i\left( F_{01} - ig[\sigma, \sigmad] - i \zeta \right) \epsilon^{\mu\nu}(\sigmab_{\nu}\lambda)^{+} + \sqrt{2}i\left( D^{\mu} + \epsilon^{\mu\nu}D_{\nu} \right) \sigmad \lambda^{+},
\end{align}
so explicitly
\begin{align}
J^{0}_{-} & =  -i\left( F_{01} + ig[\sigma, \sigmad] + i \zeta \right) \bar{\lambda}_{+} + \sqrt{2}i\left( D_{0} - D_{1} \right) \sigma \bar{\lambda}_{-}, \\
J^{0}_{+} & =  -i\left( F_{01} + ig[\sigma, \sigmad] - i \zeta \right) \bar{\lambda}_{-} - \sqrt{2}i\left( D_{0} + D_{1} \right) \sigmad \bar{\lambda}_{+}, \\
\bar{J}^0_{+} & =  i\left( F_{01} - ig[\sigma, \sigmad] + i \zeta \right) \lambda_{-} + \sqrt{2}i\left( D_{0} + D_{1} \right) \sigma \lambda_{+}, \\
\bar{J}^{0}_{-} & = i\left( F_{01} - ig[\sigma, \sigmad] - i \zeta \right) \lambda_{+} - \sqrt{2}i\left( D_{0} - D_{1} \right) \sigmad \lambda_{-}, \\
J^{1}_{-} & =  i\left( F_{01} + ig[\sigma, \sigmad] + i \zeta \right) \bar{\lambda}_{+} + \sqrt{2}i\left( D_{0} - D_{1} \right) \sigma \bar{\lambda}_{-}, \\
J^{1}_{+} & =  -i\left( F_{01} + ig[\sigma, \sigmad] - i \zeta \right) \bar{\lambda}_{-} + \sqrt{2}i\left( D_{0} + D_{1} \right) \sigmad \bar{\lambda}_{+}, \\
\bar{J}^1_{+} & =  i\left( F_{01} - ig[\sigma, \sigmad] + i \zeta \right) \lambda_{-} - \sqrt{2}i\left( D_{0} + D_{1} \right) \sigma \lambda_{+}, \\
\bar{J}^{1}_{-} & = -i\left( F_{01} - ig[\sigma, \sigmad] - i \zeta \right) \lambda_{+} - \sqrt{2}i\left( D_{0} - D_{1} \right) \sigmad \lambda_{-},
\end{align}
while for the fundamental chiral multiplet we have
\begin{align}
J^{0}_{-} & = \sqrt{2}((D_0 - D_1)\phi^{\dagger})\psi_{-} - 2ig\phi^{\dagger}\sigma \psi_{+} - g \phi^{\dagger}\bar{\lambda}_{+}\phi, \\
J^{0}_{+} & = \sqrt{2}((D_0 + D_1)\phi^{\dagger})\psi_{+} - 2ig\phi^{\dagger}\sigmad \psi_{-} + g \phi^{\dagger}\bar{\lambda}_{-}\phi, \\
J^{1}_{-} & =  \sqrt{2}((D_0 - D_1)\phi^{\dagger})\psi_{-} + 2ig\phi^{\dagger}\sigma \psi_{+} + g \phi^{\dagger}\bar{\lambda}_{+}\phi, \\
J^{1}_{+} & =  -\sqrt{2}((D_0 + D_1)\phi^{\dagger})\psi_{+} - 2ig\phi^{\dagger}\sigmad \psi_{-} + g \phi^{\dagger}\bar{\lambda}_{-}\phi, \\
\bar{J}^{0 \, -} & = \sqrt{2}\bar{\psi}^{-}(D_0 + D_1)\phi - 2ig\bar{\psi}^{+}\sigma \phi + g \phi^{\dagger}\lambda_{-}\phi , \\
\bar{J}^{0 \, +} & = \sqrt{2}\bar{\psi}^{+}(D_0 - D_1)\phi - 2ig\bar{\psi}^{-}\sigmad \phi + g \phi^{\dagger}\lambda_{+}\phi , \\
\bar{J}^{1 \, -} & =  -\sqrt{2}\bar{\psi}^{-}(D_0 + D_1)\phi - 2ig\bar{\psi}^{+}\sigma \phi - g \phi^{\dagger}\lambda^{+}\phi, \\
\bar{J}^{1 \, +} & =  \sqrt{2}\bar{\psi}^{+}(D_0 - D_1)\phi + 2ig\bar{\psi}^{-}\sigmad \phi - g \phi^{\dagger}\lambda^{-}\phi, \\
\bar{J}^{0}_{+} & = \sqrt{2}\bar{\psi}_{+}(D_0 + D_1)\phi + 2ig\bar{\psi}_{-}\sigma \phi + g \phi^{\dagger}\lambda_{-}\phi, \\
\bar{J}^{0}_{-} & = \sqrt{2}\bar{\psi}_{-}(D_0 - D_1)\phi + 2ig\bar{\psi}_{+}\sigmad \phi - g \phi^{\dagger}\lambda_{+}\phi, \\
\bar{J}^{1}_{+} & =  -\sqrt{2}\bar{\psi}_{+}(D_0 + D_1)\phi + 2ig\bar{\psi}_{-}\sigma \phi + g \phi^{\dagger}\lambda_{-}\phi, \\
\bar{J}^{1}_{-} & =  \sqrt{2}\bar{\psi}_{-}(D_0 - D_1)\phi - 2ig\bar{\psi}_{+}\sigmad \phi + g \phi^{\dagger}\lambda_{+}\phi \; .
\end{align}
We can easily include twisted mass $\widetilde{m}$ for the chiral multiplet by
simply shifting $\sigma \rightarrow \sigma - \widetilde{m}$ and
$\sigmad \rightarrow \sigmad - \bar{\widetilde{m}}$ in the
contributions from the chiral multiplet.

Finally, for the fundamental twisted chiral multiplet one finds
\begin{eqnarray}
J^{0}_{-} & = & \sqrt{2}((D_0 - D_1)\widetilde{\phi}^{\dagger})\widetilde{\psi}_{-} - 2ig\widetilde{\phi}^{\dagger}\sigma \bar{\widetilde{\psi}}_{+} - g \widetilde{\phi}^{\dagger}\bar{\lambda}_{+}\widetilde{\phi} \\
\bar{J}^{0}_{+} & = & \sqrt{2}((D_0 + D_1)\widetilde{\phi}^{\dagger})\bar{\widetilde{\psi}}_{+} - 2ig\widetilde{\phi}^{\dagger}\sigmad \widetilde{\psi}_{-} + g \widetilde{\phi}^{\dagger}\bar{\lambda}_{-}\widetilde{\phi} \\
J^{1}_{-} & = & \sqrt{2}((D_0 - D_1)\widetilde{\phi}^{\dagger})\widetilde{\psi}_{-} + 2ig\widetilde{\phi}^{\dagger}\sigma \bar{\widetilde{\psi}}_{+} + g \widetilde{\phi}^{\dagger}\bar{\lambda}_{+}\widetilde{\phi} \\
\bar{J}^{1}_{+} & = & -\sqrt{2}((D_0 + D_1)\widetilde{\phi}^{\dagger})\bar{\widetilde{\psi}}_{+} - 2ig\widetilde{\phi}^{\dagger}\sigmad \widetilde{\psi}_{-} + g \widetilde{\phi}^{\dagger}\bar{\lambda}_{-}\widetilde{\phi} \\
J^{0}_{+} & = & \sqrt{2}\widetilde{\psi}_{+}(D_0 + D_1)\widetilde{\phi} + 2ig\bar{\widetilde{\psi}}_{-}\sigma \widetilde{\phi} + g \widetilde{\phi}^{\dagger}\lambda_{-}\widetilde{\phi} \\
\bar{J}^{0}_{-} & = & \sqrt{2}\bar{\widetilde{\psi}}_{-}(D_0 - D_1)\widetilde{\phi} + 2ig\widetilde{\psi}_{+}\sigmad \widetilde{\phi} - g \widetilde{\phi}^{\dagger}\lambda_{+}\widetilde{\phi} \\
J^{1}_{+} & = & -\sqrt{2}\widetilde{\psi}_{+}(D_0 + D_1)\widetilde{\phi} + 2ig\bar{\widetilde{\psi}}_{-}\sigma \widetilde{\phi} + g \widetilde{\phi}^{\dagger}\lambda_{-}\widetilde{\phi} \\
\bar{J}^{1}_{-} & = & \sqrt{2}\bar{\widetilde{\psi}}_{-}(D_0 - D_1)\widetilde{\phi} - 2ig\widetilde{\psi}_{+}\sigmad \widetilde{\phi} + g \widetilde{\phi}^{\dagger}\lambda_{+}\widetilde{\phi}
\end{eqnarray}

We now have the condition to preserve supersymmetry at a boundary $x^1=0$. 
In the absence of boundary 
the conservation law $\partial_0 Q=0$ can be obtained by integrating the continuity equation $\partial_{\mu}J^{\mu}=0$ over the large volume.
However, in the presence of boundary 
it may be violated due to the net supercurrent through the surface. 
Thus we demand
%by demanding 
that the component of the supercurrent orthogonal to the boundary vanishes 
\begin{align}
0& 
%= \epsilon J^1 + \bar{\epsilon} \bar{J}^1 
= \epsilon^{\alpha} J^1_{\alpha} + \bar{\epsilon}_{\alpha} \bar{J}^{1 \, \alpha} \nonumber\\
&= \epsilon_{+} J^1_{-} - \epsilon_{-} J^1_{+} + \bar{\epsilon}_{-} \bar{J}^1_{+} - \bar{\epsilon}_{+} \bar{J}^1_{-}
\end{align}
at the boundary
where $\epsilon_{\pm}$ and $\overline{\epsilon}_{\pm}$ are the supersymmetry parameters. 

We can now build general $\mathcal{N} = (2,2)$ theories by combining these
multiplets, choosing the gauge groups and matter field representations. For
simplicity we focus mostly on gauge group $U(N_c)$ and $N_f$ fundamental
chiral multiplets.
For a vector multiplet and one fundamental chiral multiplet
this gives

%{\bf [To generalize to multiple chiral multiplets since SUSY is broken for $N_c> N_f$ in the following expression. ]}

\begin{align}
0 & =  i\; \Tr \bigg( \left( F_{01} + ig[\sigma, \sigmad] + i \zeta - ig\phi\phi^{\dagger} \right) \epsilon_{+}\bar\lambda_{+}
 + \left( F_{01} + ig[\sigma, \sigmad] - i \zeta + ig\phi\phi^{\dagger} \right) \epsilon_{-}\bar\lambda_{-} \nonumber \\
 & +  \sqrt{2} \left( \left( D_0 - D_1 \right) \sigma \right) \epsilon_{+}\bar\lambda_{-}
- \sqrt{2} \left( \left( D_0 + D_1 \right) \sigmad \right) \epsilon_{-}\bar\lambda_{+} \nonumber \\
& +  \left( F_{01} - ig[\sigma, \sigmad] - i \zeta + ig\phi\phi^{\dagger} \right) \bar{\epsilon}_{+}\lambda_{+}
 + \left( F_{01} - ig[\sigma, \sigmad] + i \zeta - ig\phi\phi^{\dagger} \right) \bar{\epsilon}_{-}\lambda_{-} \nonumber \\
 & +  \sqrt{2} \left( \left( D_0 - D_1 \right) \sigmad \right) \bar{\epsilon}_{+}\lambda_{-}
- \sqrt{2} \left( \left( D_0 + D_1 \right) \sigma \right) \bar{\epsilon}_{-}\lambda_{+} \bigg) \nonumber \\
 & +  \sqrt{2} \left( (D_0 - D_1)\phi^{\dagger} \right) \epsilon_{+}\psi_{-}
 + \sqrt{2} \left( (D_0 + D_1)\phi^{\dagger} \right) \epsilon_{-}\psi_{+} \nonumber \\
 & + 2ig \phi^{\dagger} \sigma \epsilon_{+}\psi_{+}
 + 2ig \phi^{\dagger} \sigmad \epsilon_{-}\psi_{-} \nonumber \\
 & -  \sqrt{2} \bar{\epsilon}_{+}\bar{\psi}_{-} (D_0 - D_1)\phi
 - \sqrt{2} \bar{\epsilon}_{-}\bar{\psi}_{+} (D_0 + D_1)\phi \nonumber \\
 & +  2ig \bar{\epsilon}_{+}\bar{\psi}_{+} \sigmad \phi
 + 2ig \bar{\epsilon}_{-}\bar{\psi}_{-} \sigma \phi \\
& =  i\; \Tr \bigg( \left( -F_{01} - ig[\sigma, \sigmad] \right) \bar{\lambda}\gamma^0\epsilon
 - i(g\phi\phi^{\dagger} - \zeta) \bar{\lambda}\gamma^1\epsilon \nonumber \\
 & +  \sqrt{2} \left( \left( D_0 - D_1 \right) \sigma \right) \bar{\lambda}P_{+}\epsilon
+ \sqrt{2} \left( \left( D_0 + D_1 \right) \sigmad \right) \bar{\lambda}P_{-}\epsilon \nonumber \\
& +  \left( F_{01} - ig[\sigma, \sigmad] \right) \bar{\epsilon}\gamma^0\lambda
 - i(g\phi\phi^{\dagger} - \zeta) \bar{\epsilon}\gamma^1\lambda \nonumber \\
 & +  \sqrt{2} \left( \left( D_0 - D_1 \right) \sigmad \right) \bar{\epsilon}P_{-}\lambda
+ \sqrt{2} \left( \left( D_0 + D_1 \right) \sigma \right) \bar{\epsilon}P_{+}\lambda \bigg) \nonumber \\
 & +  \sqrt{2} \left( D_0 \phi^{\dagger} \right) \psi\gamma^{01}\epsilon
 - \sqrt{2} \left( D_1 \phi^{\dagger} \right) \psi\epsilon \nonumber \\
 & - 2ig \phi^{\dagger} \sigma \psi\gamma^0P_{+}\epsilon
 - 2ig \phi^{\dagger} \sigmad \psi\gamma^0P_{-}\epsilon \nonumber \\
 & +  \sqrt{2} \bar{\epsilon}\gamma^{01}\bar{\psi} D_0 \phi
 + \sqrt{2} \bar{\epsilon}\bar{\psi} D_1 \phi \nonumber \\
 & + 2ig \bar{\epsilon}\gamma^0 P_{+}\bar{\psi} \sigmad \phi
 + 2ig \bar{\epsilon}\gamma^0 P_{-}\bar{\psi} \sigma \phi
\label{supercurrent}
\end{align}
where we have defined the 2d chirality projectors
\begin{equation}
P_{\pm} = \frac{1}{2}\left( I \pm \gamma^{01} \right)
\end{equation}
and all spinor contractions are of the form
$\lambda \gamma^{\mu} \epsilon = \lambda^{\alpha} \gamma^{\mu\phantom{\alpha}\beta}_{\phantom{\mu}\alpha} \epsilon_{\beta}$.

The generalization to $N_f$ fundamental chiral multiplets is straightforward,
we just replace $\phi \rightarrow \phi^i$, $\phi^{\dagger} \rightarrow \phi^{\dagger}_i$, $\psi \rightarrow \psi^i$ and $\bar{\psi} \rightarrow \bar{\psi}_i$.
With implicit contraction of the flavor indices, the supercurrent is given by
the same expression (\ref{supercurrent}). It is also straightforward to generalize to chiral multiplets in other representation of the gauge group.

For the bulk equations of motion to be consistent the boundary terms must vanish
when deriving the Euler-Lagrange equations. This leads to the following
Euler-Lagrange boundary
conditions on a boundary at fixed $x^1=0$
%\begin{eqnarray}
\begin{align}
0 &= - \Tr \left( 2 \left( F_{01} - \frac{\theta}{4\pi} \right) \delta A_{0} + 2 (D_1 \sigma)\delta \sigmad + 2 (D_1 \sigmad)\delta \sigma - i \delta \lambda \sigma^1 \bar{\lambda} - i \delta \bar{\lambda} \bar{\sigma}^1 \lambda \right) \nonumber \\
& -(\delta \phi^{\dagger}) D_1 \phi - (D_1 \phi^{\dagger}) \delta \phi
 + \frac{i}{2} \delta \psi \sigma^1 \bar{\psi} + \frac{i}{2} \delta \bar{\psi} \bar{\sigma}^1 \psi \\
&= - \Tr \left( 2 \left( F_{01} - \frac{\theta}{4\pi} \right) \delta A_{0} + 2 (D_1 \sigma)\delta \sigmad + 2 (D_1 \sigmad)\delta \sigma + i \delta \lambda \gamma^1 \bar{\lambda} + i \delta \bar{\lambda} \gamma^1 \lambda \right) \nonumber \\
&- (\delta \phi^{\dagger}) D_1 \phi - (D_1 \phi^{\dagger}) \delta \phi
 - \frac{i}{2} \delta \psi \gamma^1 \bar{\psi} - \frac{i}{2} \delta \bar{\psi} \gamma^1 \psi \label{ELbc}
\end{align}
%\end{eqnarray}

Note that
\begin{equation}
\delta \psi \sigma^1 \bar{\psi} = - \delta \psi \gamma^1 \bar{\psi} = \delta \psi_{+} \bar{\psi}_{+} - \delta \psi_{-} \bar{\psi}_{-}
\end{equation}

There are two ways to make the fermionic terms vanish while preserving the
maximum number of half the independent components. We can impose either
\begin{itemize}
\item A-type: a linear relation between the
spinor and its complex conjugate, i.e.\ a type of reality condition
\item B-type: a projection condition on the spinor.
\end{itemize}

For $\lambda$ we need to satisfy
\begin{equation}
\delta\lambda \gamma^1 \bar{\lambda} + \delta\bar{\lambda} \gamma^1 \lambda = 0 \; .
\end{equation}

%\begin{equation}
%\gamma^1 \lambda = \pm \lambda \implies \delta \lambda \gamma^1 \bar{\lambda} = 0
%\end{equation}
%while alternatively

For gauge group $U(1)$, taking A-type the most general solution is
\begin{equation}
\bar{\lambda} = e^{i \alpha} \gamma^1 \lambda \implies \delta \lambda \gamma^1 \bar{\lambda} = - \delta \bar{\lambda} \gamma^1 \lambda
\end{equation}
while for B-type the most general solution is
\begin{equation}
\left( \begin{array}{c} \lambda_1 \\ \lambda_2 \end{array} \right) =
\left( \begin{array}{cc} 0 & e^{-i\alpha} \\ e^{i\alpha} & 0 \end{array} \right)
\left( \begin{array}{c} \lambda_1 \\ \lambda_2 \end{array} \right)
\end{equation}
where $\alpha$ is an arbitrary real constant. Similar conditions must be
imposed on $\psi$. Note that these boundary conditions can be generalized in
the case of larger gauge group (for $\lambda$) or multiple flavors (for $\psi$).
This is discussed below in sections \ref{sec_Amaxproj} and \ref{sec_Bmaxproj}.

We will now see that these boundary conditions are compatible with
supersymmetry, in particular allowing preservation of half of the supersymmetry,
i.e.\ preserving two supercharges. For this we need to take either A-type for
all spinors or B-type for all spinors.

Note that we must also impose suitable Euler-Lagrange boundary conditions on
the bosonic fields in order to satisfy (\ref{ELbc}). The most obvious is to
impose either Neumann or Dirichlet boundary conditions for each field, e.g.\
for $\phi$ we could have Neumann $D_1 \phi = D_1 \phi^{\dagger} = 0$ or
Dirichlet $\delta \phi = \delta \phi^{\dagger} = 0$ so that $\phi$ takes a
constant boundary value. For the gauge field the conditions are $F_{01} = \theta / (4\pi)$
for Neumann or $\delta A_0 = 0$ for Dirichlet, which we may as well write as
$A_0 = 0$. However, more involved conditions are also possible such as a
Dirichlet condition on a linear combination of $A_0$, $\sigma$ and $\sigmad$
without any of the individual fields having to satisfy Dirichlet or Neumann
conditions. We will see an example of this in section \ref{sec_BNeupole}.

%%%%%%%%%%%%%%%%%%%%%%%%%%%%%%%%%%
\subsection{A-type boundary conditions}
%%%%%%%%%%%%%%%%%%%%%%%%%%%%%%%%%%
We can impose a condition on the spinors as follows
\begin{equation}
\label{A_fer}
\bar{\epsilon} = e^{i \vtheta} \gamma^1 \epsilon \; , \;\;
\bar{\lambda} = e^{i \alpha} \gamma^1 \lambda \; , \;\;
\bar{\psi} = e^{2 i \beta} \gamma^1 \psi \; , \;\;
\end{equation}
for some real $\alpha$, $\beta$ and $\vtheta$. This leads to identifications of
spinor bilinears such as
\begin{eqnarray}
\bar{\epsilon} \lambda & = & - \exp(i(\vtheta - \alpha)) \bar{\lambda} \epsilon \\
\bar{\epsilon} \gamma^{01} \lambda & = & - \exp(i(\vtheta - \alpha)) \bar{\lambda} \gamma^{01} \epsilon \\
\bar{\epsilon} \gamma^{0} \lambda & = & - \exp(i(\vtheta - \alpha)) \bar{\lambda} \gamma^{0} \epsilon \\
\bar{\epsilon} \gamma^{1} \lambda & = & \exp(i(\vtheta - \alpha)) \bar{\lambda} \gamma^{1} \epsilon \; .
\end{eqnarray}
So defining 
\begin{align}
c_{\pm} &= 1 \pm \exp(i(\vtheta - \alpha)), 
\end{align}
the boundary conditions for supersymmetry are
\begin{eqnarray}
c_{+}F_{01} + igc_{-}[\sigma, \sigmad] & = & 0 \\
c_{+} (g \phi \phi^{\dagger} - \zeta ) & = & 0 \\
(c_{-} D_0 - c_{+} D_1)\sigma & = & 0 \\
(c_{-} D_0 + c_{+} D_1) \sigmad & = & 0 \\
D_0 \left( \phi^{\dagger} - \exp(i(\vtheta + 2 \beta)) \phi \right) & = & 0 \\
D_1 \left( \phi^{\dagger} + \exp(i(\vtheta + 2 \beta)) \phi \right) & = & 0 \\
\phi^{\dagger}\sigma + \exp(i(\vtheta + 2 \beta)) \sigma \phi & = & 0 \\
\phi^{\dagger}\sigmad + \exp(i(\vtheta + 2 \beta)) \sigmad \phi & = & 0
\end{eqnarray}
Here we have neglected the superpotential $\mathcal{W}$, but one can be
included provided $\Im(e^{i\vtheta}\mathcal{W})$ is constant on the
boundary \cite{Hori:2000ck}.
Possibly this condition can be relaxed by including boundary degrees of freedom. 
The problem is known as the Warner problem 
\cite{Warner:1995ay, Govindarajan:1999js, Hori:2000ck, Hori:2000ic, Hellerman:2001ct, Kapustin:2002bi, Brunner:2003dc, Hori:2004zd} 
and studied for B-type boundary conditions,
but the details are not known for A-type boundary conditions and we do not consider boundary matter in this paper.

Clearly we can set $\vtheta = 0$ without loss of generality by redefining $\alpha$
and $\beta$, so we now do that. We then have two special choices of
$\alpha = 0, \pi$ giving
rise to Neumann (for $c_{-} = 0$) or Dirichlet (for $c_{+} = 0$) boundary
conditions for the gauge multiplet complex scalar $\sigma$. For the Neumann
case we also have a Neumann boundary condition for the gauge field,
$F_{01} = 0$ while for Dirichlet we see that the constant boundary values of
$\sigma$ and $\sigmad$ must commute.
If we impose gauge field boundary conditions $A_0 = 0$ for Dirichlet,
these two options satisfy the boundary conditions (\ref{ELbc}) required for the
variational principle, except that for the Neumann case the supersymmetry
and Euler-Lagrange boundary conditions are consistent only for vanishing theta
angle. For the Neumann case with $\theta \ne 0$ we would also need to impose
the Dirichlet boundary condition $\delta \left( \Tr A_0 \right) = 0$ which would completely fix
the $U(1)$ part of the gauge field on the boundary. Of course, for gauge group
$U(N_c) \cong (SU(N_c) \times U(1))/\Zb_{N_c}$ we could also impose different Dirichlet or Neumann boundary conditions for the $SU(N_c)$ and $U(1)$ gauge fields.

In general we need not choose $\alpha = 0$ or $\alpha = \pi$ but
let us first consider the basic boundary conditions, i.e. 
Neumann and Dirichlet boundary conditions
which neatly project 2d $\mathcal{N}=(2,2)$ supermultiplets onto 1d $\mathcal{N}=2_A$ supermultiplets. 

%%%%%%%%%%%%%%%%%%%%%%%%%%%%%%%%%%
\subsubsection{A-type Neumann b.c. for the vector multiplet}
%%%%%%%%%%%%%%%%%%%%%%%%%%%%%%%%%%
We start with the basic boundary conditions for the vector multiplet. 
For $\exp(i\alpha) = 1$ we find the Neumann boundary conditions for the vector multiplet:
\begin{align}
F_{01} &= 0 \label{AvmNeu1} \\
g \phi \phi^{\dagger} &= \zeta \label{phiphidagger} \\
D_1\sigma &= 0 \label{AvmNeu2} \\
D_1 \sigmad &=0 \label{AvmNeu3}
\end{align}
The condition (\ref{AvmNeu1}) would be interpreted as the Neumann boundary condition 
for the gauge field when $\theta = 0$ whereas (\ref{AvmNeu2}) and (\ref{AvmNeu3}) impose the Neumann boundary condition 
on $\sigma$ and $\sigma^{\dag}$, consistent with the Euler-Lagrange Neumann conditions.
These A-type Neumann boundary conditions for the vector multiplet admit 
the 1d $\mathcal{N}=2_A$ gauge multiplet which will be obtained from 
the 3d $\mathcal{N}=1$ gauge multiplet \cite{Gates:1983nr} or 2d $\mathcal{N}=(1,1)$ gauge multiplet \cite{Ferrara:1975nf}
and described by the $({\bf 2},{\bf 2},{\bf 0})$ supermultiplet 
whose complex bosonic scalar fields are compatible with the remaining $\sigma$ and $\sigmad$ 
obeying the Neumann boundary conditions (\ref{AvmNeu2}) and (\ref{AvmNeu3}). 

While in the absence of the chiral multiplets, 
under the conditions (\ref{AvmNeu1}), (\ref{AvmNeu2}) and (\ref{AvmNeu3}), 
the gauge symmetry could be completely preserved at the boundary, 
when the chiral multiplets are coupled to the vector multiplet, the UV boundary conditions get more complicated 
as additional conditions, including (\ref{phiphidagger}), are required. 
%It should be straightforwardly generalized if we include additional matters. 
Of course, without chiral multiplets, or with chiral multiplets with boundary conditions $\phi = 0$, we must take vanishing FI parameter in order to satisfy
(\ref{phiphidagger}).

%%%%%%%%%%%%%%%%%%%%%%%%%%%%%%%%%%
\subsubsection{A-type Dirichlet b.c. for the vector multiplet}
%%%%%%%%%%%%%%%%%%%%%%%%%%%%%%%%%%
For $\exp(i\alpha) = -1$ we have the Dirichlet boundary conditions for the vector multiplet:
\begin{align}
[\sigma, \sigmad] & = 0 \label{AvmDir1} \\
D_0 \sigma & = 0 \label{AvmDir2} \\
D_0 \sigmad &= 0 \label{AvmDir3} \; .
\end{align}
The equations (\ref{AvmDir2}) and (\ref{AvmDir3}) fix the values of $\sigma$ and $\sigma^{\dag}$ 
at the boundary. 
We also must impose the Dirichlet boundary condition for the gauge field. 
The A-type Dirichlet boundary conditions for the vector multiplet admit  
the 1d $\mathcal{N}=2_A$ scalar multiplet described by the $({\bf 1},{\bf 2},{\bf 1})$ supermultiplets 
whose bosonic physical scalar field is identified with the surviving $A_1$ component of the gauge field. 

Together with the condition (\ref{AvmDir1}), 
these boundary conditions can be solved by 
\begin{align}
\label{AvmDir4}
A_0&=0,\\
\label{AvmDir5}
\sigma&=0. 
\end{align}
The Dirichlet boundary condition for the vector multiplet 
breaks the gauge symmetry $G$ at the boundary 
and the global gauge transformation at the boundary 
would lead to the global symmetry $G_{\partial}$. 

We can also deform the condition (\ref{AvmDir5}) to 
\begin{align}
\label{AvmDir5a}
\sigma&=s
\end{align}
where $s\neq 0$ is some constant satisfying $[s, s^{\dagger}] = 0$. 
In this case, the flavor symmetry $G_{\partial}$ is broken 
to the stabilizer of $s$, that is 
the subgroup whose action on $s$ is trivial.

%%%%%%%%%%%%%%%%%%%%%%%%%%%%%%%%%%
\subsubsection{A-type b.c. for the chiral multiplet}
\label{sec_AbcCM}
%%%%%%%%%%%%%%%%%%%%%%%%%%%%%%%%%%
Consider the boundary conditions for the chiral multiplets. 
Similarly to the choice of $\alpha = 0, \pi$ for the vector multiplet, we can
take the special choices $2\beta = 0, \pi$. For $e^{2i\beta} = 1$ we have
\begin{align}
D_0 \left( \Im(\phi) \right) & = 0 \label{Acm11} \\
D_1 \left( \Re(\phi) \right) & = 0  \label{Acm12} \\
\phi^{\dagger} \sigma + \sigma \phi & = 0  \label{Acm13} \\
\phi^{\dagger} \sigmad + \sigmad \phi & = 0  \label{Acm14}
\end{align}
while for $e^{2 i\beta} = -1$ we have
\begin{align}
D_0 \left( \Re(\phi) \right) & = 0  \label{Acm21} \\
D_1 \left( \Im(\phi) \right) & =  0  \label{Acm22} \\
\phi^{\dagger} \sigma - \sigma \phi & = 0 \label{Acm23} \\
\phi^{\dagger} \sigmad - \sigmad \phi & = 0 \label{Acm24} \; .
\end{align}
Note however that the second set of boundary conditions is equivalent to the first up to a redefinition $\phi \rightarrow i \phi$.
The boundary conditions (\ref{Acm11}) and (\ref{Acm12}) 
(or (\ref{Acm21}) and (\ref{Acm14})) for the chiral multiplet are the Dirichlet and Neumann boundary conditions  
on the imaginary and real (or real and imaginary) part of $\phi$, 
which are compatible with the Lagrangian splitting of the chiral multiplet to 
the 1d $\mathcal{N}=2_A$ scalar multiplets described by the $({\bf 1},{\bf 2},{\bf 1})$ supermultiplets 
whose bosonic physical scalar field is identified with the real (or imaginary) part of $\phi$. 

When the chiral multiplets are coupled to a vector multiplet, 
we have additional conditions (\ref{Acm13}) and (\ref{Acm14}) (or (\ref{Acm23}) and (\ref{Acm24})). 
As they involve both the vector multiplet scalar field $\sigma$ and the chiral multiplet scalar $\phi$, 
the Coulomb branch and Higgs branch vevs obstruct one another. 

Let us consider the case where the vector multiplet obeys the Dirichlet boundary conditions 
where $\sigma$ is frozen according to the Dirichlet boundary conditions (\ref{AvmDir2}) and (\ref{AvmDir3}). 
%(\ref{AvmDir2}) and (\ref{AvmDir3}) and one sets $\sigma$ to zero, they are satisfied, but there are more general solutions as we now explain.
%
%Next consider the boundary conditions for the chiral multiplet. 
%When the vector multiplet obeying the singular boundary conditions (\ref{A_singular_sol1}) and (\ref{A_singular_sol2}) 
%is coupled to the chiral multiplet which has no explicit dependence on $x^0$, 
The boundary condition (\ref{Acm11}) is solved by requiring that 
$\Im \phi$ obeys the Dirichlet boundary conditions 
and that $A_0\cdot \Im \phi$ vanishes. 
They can be solved by simply setting $\im \phi$ to zero.  
On the other hand, (\ref{Acm12}) can be solved by requiring that $\Re \phi$ is
subject to the Neumann boundary condition.
Accordingly, the additional conditions (\ref{Acm13}) and (\ref{Acm14}) become 
\begin{align}
\label{AcmG7}
\Re (\sigma_+ \Re\phi)&=0, \\
\label{AcmG8}
\Im (\sigma_- \Re\phi)&=0
\end{align}
where $\sigma_{\pm} \equiv (\sigma \pm \sigmad)/\sqrt{2}$.
Noting that $\phi = \Re\phi$ and defining $\sigma_{S} = ( \sigma + \sigma^{T} ) / 2$ these equations are equivalent to
\begin{equation}
\sigma_{S} \phi = 0 = \sigma_{S}^{*} \phi = \sigma_{S}^{\dagger} \phi
\end{equation}
%For the singular solution to the boundary conditions for the vector multiplet
%this is equivalent to
%\begin{equation}
%(\mathfrak{s}_{+} + \mathfrak{s}_{+}^T) \hat{\phi} = 0 \; . % = \Re (c_2 \mathfrak{s}_{+}^T) \hat{\phi} \; .
%\end{equation}
There are two obvious solutions to this condition: $\sigma_S = 0$ or
$\phi = 0$.
When the vector multiplet satisfies Dirichlet boundary conditions, we can have
solutions with constant non-zero $\sigma$ at the boundary and this can allow
non-zero values for both
$\sigma_S$ and $\phi$ if $\sigma_S$ and
$\sigma_S^{\dagger}$ have common eigenstates with eigenvalue $0$.
%For $\mathfrak{g}=\mathfrak{su}(2)$ the only solution to this is $\phi = 0$. 
%as one can see this by considering an explicit representation
%\begin{equation}
%\mathfrak{s}_{+} = \frac{1}{\sqrt{2}} \left( \begin{array}{cc} 0 & 1 \\ 0 & 0 \end{array} \right). 
%\end{equation}

%However, in a larger gauge group we can have non-zero $\phi$. 
For example, for $G=SU(4)$ we could have
\begin{equation}
\sigma_S = \left( \begin{array}{cccc} s_{11} & s_{12} & 0 & 0 \\ s_{12} & s_{22} & 0 & 0 \\ 0 & 0 & 0 & 0 \\ 0 & 0 & 0 & 0 \end{array} \right)
\end{equation}
which allows
\begin{equation}
\phi = \left( \begin{array}{c} 0 \\ 0 \\ \phi_3 \\ \phi_4 \end{array} \right)
\end{equation}
for arbitrary real $\phi_3$ and $\phi_4$.

Instead, when the vector multiplet is subject to the Neumann boundary condition, 
the conditions (\ref{Acm13}) and (\ref{Acm14}) (or (\ref{Acm23}) and (\ref{Acm24})) 
would require the modification of the basic boundary conditions. 
In particular when the $U(1)$ vector multiplet satisfies 
the Neumann boundary conditions (\ref{AvmNeu2}) and (\ref{AvmNeu3}), we cannot find a solution 
to the constraints (\ref{Acm13}) and (\ref{Acm14}) (or (\ref{Acm23}) and (\ref{Acm24})). 
This is because the Lagrangian splitting of a charged chiral multiplet 
leads to scalar multiplets which cannot be charged under the $U(1)$ gauge symmetry. 

While it is not clear whether or how one can obtain the Neumann boundary conditions, 
we note that our boundary conditions on fermions (\ref{A_fer}) may be generalized 
when the theory has multiple matter fields or/and higher rank gauge group, as we will see in section \ref{sec_Amaxproj}. 
In such a case, one may modify the additional conditions or ordinary Neumann boundary conditions to those which break $G$ down to $H$ 
and find some consistent Lagrangian splitting of chiral multiplets so that the surviving scalar multiplets are coupled to the $H$ gauge field. 
In other words, it may be possible to obtain boundary conditions in which the
gauge group $G$ is broken to its subgroup $H$ 
so that $G$ decomposes as
\begin{align}
\mathfrak{g}&=\mathfrak{h}\oplus \mathfrak{h}^{\perp}
\end{align}
where $\mathfrak{g}$ and $\mathfrak{h}$ are the Lie algebras of $G$ and $H$ 
and $\mathfrak{h}^{\perp}$ is the orthocomplement that is not Lie algebra. 
Correspondingly, we split the adjoint-valued fields as
\begin{align}
\sigma&=\sigma^{\parallelsum}+\sigma^{\perp}
\end{align}
where $\sigma^{\parallelsum}\in \mathfrak{h}$ 
and $\sigma^{\perp}\in \mathfrak{h}^{\perp}$. 
Then we would get the boundary conditions which reduce 
the gauge group $G$ to its subgroup $H$ at the boundary. 
In addition, one can introduce extra boundary terms 
which may modify or remove the constraints encountered above. 
We leave the issue of the Neumann boundary condition for the vector multiplet coupled to chiral multiplets to future work.

%%%%%%%%%%%%%%%%%%%%%%%%%%%%%%%%%%
\subsubsection{Generic A-type boundary conditions}
%%%%%%%%%%%%%%%%%%%%%%%%%%%%%%%%%%
Above we commented on the specific cases where either $c_{+} = 0$ or $c_{-} = 0$
but in general we will have the following equations
\begin{align}
u & \equiv  \frac{1}{\sqrt{2}} \left( \frac{x^0}{c_{-}} - \frac{x^1}{c_{+}} \right) \label{AvmG0} \\
\hat{\sigma} & \equiv  c_{-} \sigma \label{AvmG1} \\
\hat{\phi} & \equiv e^{i\beta}\phi \label{AcmG0} \\
F_{u \bar{u}} & = - ig[\hat{\sigma}, \hat{\sigma}^{\dagger}]
\label{AvmG2} \\
g \hat{\phi} \hat{\phi}^{\dagger} & = \zeta 
\label{AvmG3} \\
D_u \hat{\sigma} & = 0 
\label{AvmG4} \\
D_{\bar{u}} \hat{\sigma}^{\dagger} & = 0 \label{AvmG5} \\
D_0 \Im \left( \hat{\phi} \right) & = 0 \label{AcmG1} \\
D_1 \Re \left( \hat{\phi} \right) & = 0 \label{AcmG2} \\
\hat{\phi}^{\dagger}\hat{\sigma} + \hat{\sigma} \hat{\phi} & = 0 \label{AcmG3} \\
\hat{\phi}^{\dagger} \hat{\sigma}^{\dagger} + \hat{\sigma}^{\dagger} \hat{\phi} & = 0 \label{AcmG4}
\end{align}

Defining $\hat{\sigma}_{\pm} = \hat{\sigma} \pm \hat{\sigma}^{\dagger}$, 
the equations (\ref{AcmG3}) and (\ref{AcmG4}) can be written as 
\begin{align}
\label{AcmG5}
\Re( \hat{\sigma}_{+} \hat{\phi}) & = 0 \\
\Im( \hat{\sigma}_{-} \hat{\phi}) & = 0
\label{AcmG6}
\end{align}

If we define 
\begin{align}
\mathcal{A}_{u}&:=A_{u} + e^{i\gamma} \hat{\sigma}^{\dagger}
\end{align}
for arbitrary $\gamma \in \Rb$ where 
\begin{align}
%A_{u}&:=\frac{1}{2}\left(\frac{A_0}{c_{-}}+\frac{A_{1}}{c_{+}} \right)
A_{u}&:=\frac{1}{\sqrt{2}}\left( c_{-} A_0 - c_{+} A_{1} \right) \; , \\
A_{\bar{u}}&:=- e^{-i(\theta - \alpha)} \frac{1}{\sqrt{2}}\left( c_{-} A_0 + c_{+} A_{1} \right) \; ,
\end{align}
then equations (\ref{AvmG2}), (\ref{AvmG4}) and (\ref{AvmG5}) imply that
% can be rewritten as 
\begin{align}
\label{DuDubar}
[\mathcal{D}_u, \mathcal{D}_{\overline{u}}]&=0
\end{align}
where 
\begin{align}
D_u & = \frac{1}{\sqrt{2}}(c_- D_0 - c_+ D_1), \\
D_{\overline{u}} & = - e^{-i(\theta - \alpha)} \frac{1}{\sqrt{2}}(c_- D_0 + c_+ D_1), \\
\mathcal{D}_u & = \partial_u - i g \mathcal{A}_u.
\end{align}

Note that (\ref{AvmG4}) and (\ref{AvmG5}) are mixed boundary conditions which
may be problematic as they both
contain derivatives with respect to $x^0$ and $x^1$. Hence whether we impose Dirichlet or
Neumann conditions on $\sigma$ to satisfy (\ref{ELbc}), we will also have a
constraint on the other derivative. Hence we should look for singular solutions
for the generic A-type boundary conditions, otherwise we would only expect
solutions with $\sigma$ completely fixed. 

However, although it could be interesting to explore singular solutions for the generic boundary conditions 
(\ref{AvmG0})-(\ref{AcmG4}), we cannot find them for physical theories. 
If fact, as we will discuss in section~\ref{A_D4''brane}, the brane configuration
for this case does not admit such singular boundary behavior.
As we show in appendix~\ref{AtypeSing} it is possible to have singular boundary
conditions, but only with a non-compact non-Abelian gauge group. In particular,
the specific signs
in (\ref{AvmG2}), given (\ref{AvmG4}) and (\ref{AvmG5}) mean that we have
Nahm equations which admit
singular solutions but based on the Lie algebra $\mathfrak{su}(1,1)$ rather
than $\mathfrak{su}(2)$.
It is not clear if such singular solutions can appear in any physically
relevant cases. Also they are only possible for non-Abelian
groups. If we have an Abelian theory, in the gauge $A_1 = 0$ we would have a
boundary condition $0 = F_{01} = -\partial_1 A_0$ which clearly does not admit
singular behaviour of the gauge potential. Then the boundary conditions for
$\hat{\sigma}$ also only admit regular solutions.

%%%%%%%%%%%%%%%%%%%%%%%%%%%%%%%%%%
\subsubsection{General gauge group projections and multiple matter multiplets}
\label{sec_Amaxproj}
%%%%%%%%%%%%%%%%%%%%%%%%%%%%%%%%%%
More generally, when we have a gauge group other than $U(1)$ we can impose
matrix projection conditions on $\lambda^a$, and similarly for multiple matter multiplets $\psi^i$. 
In other words, there are more general A-type boundary conditions by 
taking such matrix projection conditions on fermions. 
Specifically, we can impose the boundary conditions 
\begin{align}
(R \overline{\lambda} )^{a} = R^{ab} \overline{\lambda}^{b} & = \gamma^1 \lambda^{a} \; , &
(S \overline{\psi})^{i} = S^{ij} \overline{\psi}_{j} & = \gamma^1 \psi^{i} \\
\overline{R}^{ab} & = (\overline{R^{ab}}) \; , & \overline{S}_{ij} & = (\overline{S^{ij}}) \; . 
\end{align}
Then (\ref{ELbc}) is satisfied if $R$ and $S$ are symmetric, i.e.
\begin{align}
R^{ab} & = R^{ba} \; , & S^{ij} & = S^{ji}
\end{align}
while using $(\gamma^1)^2 = I$ implies
\begin{align}
R \overline{R} & = I \; , & S \overline{S} & = I \; .
\end{align}
Together these conditions mean that $R$ and $S$ are both unitary and symmetric.

%[{\bf To fix the notation. }]

This leads to the following identifications of
spinor bilinears
\begin{eqnarray}
\bar{\epsilon} \lambda & = & - e^{i \vtheta} R \bar{\lambda} \epsilon \\
\bar{\epsilon} \gamma^{01} \lambda & = & - e^{i \vtheta} R \bar{\lambda} \gamma^{01} \epsilon \\
\bar{\epsilon} \gamma^{0} \lambda & = & - e^{i \vtheta} R \bar{\lambda} \gamma^{0} \epsilon \\
\bar{\epsilon} \gamma^{1} \lambda & = & e^{i \vtheta} R \bar{\lambda} \gamma^{1} \epsilon \; .
\end{eqnarray}
Clearly we can absorb the factor $e^{i\vtheta}$ into $R$.
So the boundary conditions for supersymmetry are
\begin{eqnarray}
\left( (I + R)F_{01} + ig(I - R)[\sigma, \sigmad] \right) \bar{\lambda} \gamma^{0} \epsilon & = & 0 \\
\left( g \phi^i \phi^{\dagger\; \bar{i}} - \zeta \right) (I + R) \bar{\lambda} \gamma^{1} \epsilon & = & 0  \\ %\;\; \textrm{unless} \; R = -I \\
\left( ( (I - R) D_0 - (I + R) D_1)\sigma \right) \bar{\lambda} P_{+} \epsilon & = & 0 \\
\left( ( (I - R) D_0 + (I + R) D_1) \sigmad \right) \bar{\lambda} P_{-} \epsilon & = & 0
\end{eqnarray}

%[{\bf To express them by including vectors $v^I$, $w^I$ and matrix $R$}]

Similarly for multiple chiral multiplets we have
\begin{eqnarray}
\psi \epsilon & = & - e^{-i \vtheta} \bar{\epsilon} S \bar{\psi} \\
\psi \gamma^{01} \epsilon & = & - e^{-i \vtheta} \bar{\epsilon} \gamma^{01} S \bar{\psi} \\
\psi \gamma^{0} \epsilon & = & - e^{-i \vtheta} \bar{\epsilon} \gamma^{0} S \bar{\psi} \\
\psi \gamma^{1} \epsilon & = & e^{-i \vtheta} \bar{\epsilon} \gamma^{1} S \bar{\psi} \; .
\end{eqnarray}
so again we can absorb $e^{i\vtheta}$ by redefining $S^i_{\bar{j}}$,
giving supersymmetric boundary conditions
\begin{eqnarray}
\bar{\epsilon} \gamma^{01} \bar{\psi} \left( D_0 \phi - S D_0 \phi^{\dagger} \right) & = & 0 \\
\bar{\epsilon} \bar{\psi} \left( D_1 \phi + S D_1 \phi^{\dagger} \right) & = & 0 \\
\bar{\epsilon} \gamma^{0} P_{-} \bar{\psi} \left( S \phi^{\dagger}\sigma + \sigma \phi \right) & = & 0 \\
\bar{\epsilon} \gamma^{0} P_{+} \bar{\psi} \left( S \phi^{\dagger}\sigmad + \sigmad \phi \right) & = & 0
\end{eqnarray}
Note that $R$ and $S$ may depend on bosonic fields, 
which would lead to more general boundary conditions. 
Although we do not complete the analysis above in this paper, 
it would be interesting to explore such general boundary conditions.

%[{\bf To express them by including vectors $v^I$, $w^I$ and matrix $S$}]

%%%%%%%%%%%%%%%%%%%%%%%%%%%%%%%%%%
\subsection{B-type boundary conditions}
%%%%%%%%%%%%%%%%%%%%%%%%%%%%%%%%%%
The B-type conditions arise from imposing a projection condition on the
spinors.
The most general such boundary condition is
%\begin{equation}
%M \epsilon \equiv \left(a \gamma^0 + b \gamma^1 + c \gamma^{01} \right) \epsilon = \epsilon \; , \;\; a^2 = b^2 + c^2 - 1
%\end{equation}
%with the constraint imposing $M^2 = M$.
%
%For example we can impose
%\begin{equation}
%\gamma^1 \epsilon = \epsilon
%\end{equation}
%which gives the relations
%\begin{eqnarray}
%\bar{\lambda}\gamma^1\epsilon & = & \bar{\lambda}\epsilon \\
%2\bar{\lambda}P_{\pm}\epsilon & = & \bar{\lambda}\epsilon \pm \bar{\lambda}\gamma^0\epsilon \; .
%\end{eqnarray}
\begin{align}
\left( \begin{array}{c} \epsilon_1 \\ \epsilon_2 \end{array} \right) & = 
\left( \begin{array}{cc} 0 & e^{-i\vtheta} \\ e^{i\vtheta} & 0 \end{array} \right)
\left( \begin{array}{c} \epsilon_1 \\ \epsilon_2 \end{array} \right) \label{B_fer1} \\
\left( \begin{array}{c} \lambda_1 \\ \lambda_2 \end{array} \right) & = 
\left( \begin{array}{cc} 0 & e^{-i\alpha} \\ e^{i\alpha} & 0 \end{array} \right)
\left( \begin{array}{c} \lambda_1 \\ \lambda_2 \end{array} \right) \label{B_fer2} \\
\left( \begin{array}{c} \psi_1 \\ \psi_2 \end{array} \right) & = 
\left( \begin{array}{cc} 0 & e^{-i\beta} \\ e^{i\beta} & 0 \end{array} \right)
\left( \begin{array}{c} \psi_1 \\ \psi_2 \end{array} \right) \label{B_fer3}
\end{align}

The boundary conditions for supersymmetry then become
\begin{align}
\sqrt{2} D_1 \left( e^{i\vtheta} \sigma + e^{-i\alpha} \sigmad \right)
 - \left( e^{i(\vtheta - \alpha)} + 1 \right) F_{01}
 + i\left( e^{i(\vtheta - \alpha)} - 1 \right) (g \phi \phi^{\dagger} - \zeta ) & =0 \\
\sqrt{2} D_0 \left( e^{i\vtheta} \sigma - e^{-i\alpha} \sigmad \right)
 + ig \left( e^{i(\vtheta - \alpha)} + 1 \right) [\sigma, \sigmad]
 & =0 \\
\left( e^{i\beta} + e^{i\vtheta} \right) D_0 \phi^{\dagger}
 + \left( e^{i\beta} - e^{i\vtheta} \right) D_1 \phi^{\dagger}
 + \sqrt{2}ig \phi^{\dagger} \left( e^{i(\vtheta + \beta)}\sigma + \sigmad \right) & =0 \\
\left( e^{i\beta} + e^{i\vtheta} \right) D_0 \phi
 - \left( e^{i\beta} - e^{i\vtheta} \right) D_1 \phi
 - \sqrt{2}ig \left( e^{i(\vtheta + \beta)}\sigma + \sigmad \right) \phi & =0 \; .
\end{align}
%where we have defined
%$\sigma_{\pm} = \frac{1}{2} \left( \sigma \pm \sigmad \right)$.

%\begin{eqnarray}
%Tr \left( \left( \sqrt{2}\left( F_{01} - ig[\sigma, \sigmad] \right) + (D_0 - D_1)\sigma - (D_0 + D_1)\sigmad \right) \bar{\lambda}\gamma^0\epsilon \right) & = & 0 \\
%Tr \left( \left( (D_0 - D_1)\sigma + (D_0 + D_1)\sigmad - \sqrt{2}ig\phi \phi^{\dagger} \right) \bar{\lambda}\epsilon \right) & = & 0 \\
%\left( \sqrt{2} D_0 \phi^{\dagger} + ig \phi^{\dagger} (\sigma + \sigmad) \right) \psi \gamma^0 \epsilon & = & 0 \\
%\left( \sqrt{2} D_1 \phi^{\dagger} + ig \phi^{\dagger} (\sigma - \sigmad) \right) \psi\epsilon & = & 0 \; .
%\end{eqnarray}
%Now only two of these equations, one of the first two and one of the second two,
%are non-trivial since the fermion boundary conditions in order to have a
%well-defined variational principle require that we also have either
%\begin{eqnarray}
%\gamma^1 \lambda = \lambda & \implies & \bar{\lambda} \epsilon = 0 \\
%\mathrm{or} \; \gamma^1 \lambda = - \lambda & \implies & \bar{\lambda} \gamma^1 \epsilon = 0
%\end{eqnarray}
%and similarly for $\gamma^1 \psi = \pm \psi$.

%Taking into account the Hermitian conjugate of these equations, and defining
%$\sigma_{\pm} = \frac{1}{2} \left( \sigma \pm \sigmad \right)$ we have either

Here we have neglected the superpotential $\mathcal{W}$, and indeed in order to
preserve supersymmetry with B-type boundary conditions we require
$\mathcal{W}$ to be constant on the boundary \cite{Hori:2000ck, Hori:2000ic}.
This condition can be relaxed by including boundary degrees of freedom
\cite{Hori:2004zd}, but we do not consider boundary matter in this paper.

Note that if we simultaneously shift $\alpha$ and $\beta$ by $\vtheta$ and
redefine $\sigma$ by a phase factor $e^{i \vtheta}$ we can absorb all $\vtheta$
dependence in these boundary conditions. From now on we assume this has been done, which is equivalent to setting $\vtheta = 0$, leaving us with
\begin{align}
\sqrt{2} D_1 \left( \sigma + e^{-i\alpha} \sigmad \right)
 - \left( e^{-i \alpha} + 1 \right) F_{01}
 + i\left( e^{-i \alpha} - 1 \right) (g \phi \phi^{\dagger} - \zeta ) & =0 \\
\sqrt{2} D_0 \left( \sigma - e^{-i\alpha} \sigmad \right)
 + ig \left( e^{-i \alpha} + 1 \right) [\sigma, \sigmad]
 & =0 \\
\left( e^{i\beta} + 1 \right) D_0 \phi^{\dagger}
 + \left( e^{i\beta} - 1 \right) D_1 \phi^{\dagger}
 + \sqrt{2}ig \phi^{\dagger} \left( e^{i \beta}\sigma + \sigmad \right) & =0 \\
\left( e^{i\beta} + 1 \right) D_0 \phi
 - \left( e^{i\beta} - 1 \right) D_1 \phi
 - \sqrt{2}ig \left( e^{i \beta}\sigma + \sigmad \right) \phi & =0 \; .
\end{align}

If we choose $e^{i \alpha} = \pm 1$ then we can have Dirichlet type boundary conditions for
$\sigma_{+}$ and Neumann type for $\sigma_{-}$ or vice-versa, where we define
\begin{align}
\sigma_{\pm} = \frac{1}{\sqrt{2}} \left( \sigma \pm \sigmad \right).
\end{align}
We also have additional simplification as some terms vanish in this case.

Similarly, we can get either Dirichlet or Neumann type conditions for both $\phi$ and $\phi^{\dagger}$ for the choices
$e^{i \beta} = \pm 1$. These can all
happen simultaneously for the four special cases of
$e^{i \alpha} = \pm 1$ and $e^{i \beta} = \pm 1$.

As in the A-type boundary conditions, 
let us first consider the basic boundary conditions, i.e. 
Neumann and Dirichlet boundary conditions (labelled here by the type of boundary condition obeyed by the field strength $F_{01}$) which project 2d $\mathcal{N}=(2,2)$ supermultiplets 
onto 1d $\mathcal{N}=2_B$ supermultiplets. 
Some basic boundary conditions were already discussed in \cite{Honda:2013uca, Hori:2013ika}. 

%%%%%%%%%%%%%%%%%%%%%%%%%%%%%%%%%%
\subsubsection{B-type Neumann b.c. for the vector multiplet}
%%%%%%%%%%%%%%%%%%%%%%%%%%%%%%%%%%
Let us begin with the basic B-type boundary conditions for the vector multiplet. 
For $\exp(i\alpha) = 1$, 
we find the Neumann boundary conditions for the vector multiplet:
\begin{align}
F_{01}&=0,\label{BvmNeu1}\\
D_{1}\sigma_+&=0, \label{BvmNeu2}\\
D_{0}\sigma_-&=0, \label{BvmNeu3}\\
[\sigma_-,\sigma_+]&=0. \label{BvmNeu4}
\end{align}
The boundary conditions (\ref{BvmNeu2}) and (\ref{BvmNeu3}) are the 
Neumann boundary conditions for $\sigma_+$ and the Dirichlet boundary conditions for $\sigma_-$ respectively, consistent with the Euler-Lagrange boundary conditions if we also impose $A_0 \sigma_{-} = 0$ at the boundary. 
This is analogous to the Lagrangian splitting of the A-type boundary conditions for the chiral multiplets in 
section \ref{sec_AbcCM}. 
It allows for a Lagrangian submanifold $\mathcal{L}$ of the space $\mathfrak{g}_{\mathbb{C}}$ 
of the K\"{a}hler manifold labeled by $\sigma$. 
The B-type Neumann boundary conditions for the vector multiplet are compatible with  
the 1d $\mathcal{N}=2_B$ gauge multiplet described by the $({\bf 1},{\bf 2},{\bf 1})$ supermultiplets \cite{Hori:2014tda} 
whose real bosonic scalar field is identified with $\sigma_+$. 

As $\sigma_-$ obeys the Dirichlet boundary condition, 
the additional condition (\ref{BvmNeu4}) can be simply solved by setting
\begin{align}
\label{Nzero_sigma}
\sigma_-&=0. 
\end{align}
In this case, the gauge symmetry $G$ can be preserved at the boundary as we
%can take $A_1 = 0$ but
need not impose any constraints on the boundary
value of $A_0$. 
Such Neumann boundary conditions for the vector multiplet 
were studied in the context of gauged linear sigma models \cite{Herbst:2008jq, Honda:2013uca, Hori:2013ika}. 
We note that the Euler-Lagrange boundary conditions are satisfied provided
we have $\theta = 0$. For $\theta \ne 0$ if we have gauge group
$U(N_c) = SU(N_c) \times U(1)$ we could have the Neumann boundary conditions
$F_{01} = 0$ for the $SU(N_c)$ gauge fields, but if we impose (\ref{BvmNeu1}) for
the $U(1)$ gauge field we would need to impose Dirichlet boundary conditions
for the Abelian gauge field, fixing $A_0$.

When instead we choose 
\begin{align}
\label{Nnon_sigma}
\sigma_-&=t_-
\end{align}
where $t_-$ is some nonzero valued constant, 
the gauge symmetry $G$ may be explicitly broken to $H$ 
in such a way that $\sigma_+$ valued in $H$ commutes with $\sigma_-$ 
according to the condition (\ref{BvmNeu4}).

%%%%%%%%%%%%%%%%%%%%%%%%%%%%%%%%%%
\subsubsection{B-type Dirichlet b.c. for the vector multiplet}
%%%%%%%%%%%%%%%%%%%%%%%%%%%%%%%%%%
On the other hand, the Dirichlet boundary conditions for the vector multiplet can be obtained 
for $\exp(i\alpha) = -1$:
\begin{align}
D_{1} \sigma_-&=0, \label{BvmDir1}\\
D_{0} \sigma_+&=0, \label{BvmDir2}\\
g\phi\phi^{\dag}&=\zeta. \label{BvmDir3}
\end{align}
The boundary conditions (\ref{BvmDir1}) and (\ref{BvmDir2}) 
are the Neumann boundary condition for $\sigma_{-}$ and Dirichlet boundary condition for $\sigma_{+}$ respectively, satisfying the Euler-Lagrange boundary conditions provided also $A_0 \sigma_{+} = 0$.
The B-type Dirichlet boundary conditions admit 
the 1d $\mathcal{N}=2_B$ chiral multiplet described by the $({\bf 2},{\bf 2},{\bf 0})$ supermultiplet 
whose complex scalar field corresponds to the $A_1$ component of gauge field and $\sigma_-$. 

Similarly to the A-type, 
a simple solution
\begin{align}
\label{BvmDir2a}
\sigma_+&=0
\end{align}
would lead to the boundary flavor symmetry $G_{\partial}$ as the global gauge transformation of $G$. 
Its deformation 
\begin{align}
\label{BvmDir2b}
\sigma_+&=t_+
\end{align}
where $t_+$ is some nonzero real valued constant 
would break the flavor symmetry $G_{\partial}$ to the stabilizer of $t_+$. 
For example, for generic $t_+$, the flavor symmetry is broken to its maximal torus.

In the presence of the chiral multiplets coupled to the vector multiplet, 
the additional condition (\ref{BvmDir3}) is imposed. 
For the Dirichlet boundary conditions (\ref{BcmDir1}) and (\ref{BcmDir2}) 
of the chiral multiplets with gauge group $U(1)$ it can be simply solved by fixing the squared norm of $\phi$ as $\zeta/g$. 
However, the Neumann boundary conditions (\ref{BcmNeu1}) and (\ref{BcmNeu2}), together with (\ref{BvmDir1}) imply that the boundary values of
$\sigma_{-}$ and $\phi$ should not be fixed, but this is incompatible with
(\ref{BvmDir3}), (\ref{BcmNeu3}) and (\ref{BcmNeu4}) which
prevent the chiral multiplets from freely fluctuating at the boundary. 
This indicates that the pairing of the Dirichlet boundary condition for the vector multiplet 
and the Neumann boundary condition for the chiral multiplet should be generalized, 
as we will see below.

%%%%%%%%%%%%%%%%%%%%%%%%%%%%%%%%%%
\subsubsection{B-type Neumann b.c. for the chiral multiplet}
%%%%%%%%%%%%%%%%%%%%%%%%%%%%%%%%%%
For $\exp(i\beta) = -1$
we get Neumann boundary conditions for the chiral multiplet:
\begin{align}
D_{1}\phi&=0, \label{BcmNeu1}\\
D_{1}\phi^{\dag}&=0, \label{BcmNeu2}\\
\phi^{\dag} \sigma_-&=0. \label{BcmNeu3}\\
\sigma_- \phi&=0. \label{BcmNeu4}
\end{align}
The boundary conditions (\ref{BcmNeu1}) and (\ref{BcmNeu2}) are 
Neumann boundary conditions for the chiral multiplet $\phi$.
% if we also have $A_1 \phi = 0$, simply $A_1 = 0$ being the obvious choice.
The B-type Neumann boundary conditions for the chiral multiplet admit  
the 1d $\mathcal{N}=2_B$ chiral multiplet described by the $({\bf 2},{\bf 2},{\bf 0})$ supermultiplets \cite{Hori:2014tda} 
whose complex bosonic scalar field is identified with $\phi$. 

The additional conditions (\ref{BcmNeu3}) and (\ref{BcmNeu4}) 
involving the vector multiplet scalar $\sigma_-$ can be satisfied 
for the Neumann boundary condition of vector multiplet with $\sigma_{-}=0$ as in (\ref{Nzero_sigma}). 
But when $\sigma_- \neq 0$ as in (\ref{Nnon_sigma}), 
the gauge symmetry $G$ may be explicitly broken to $H$ 
and (\ref{BcmNeu3}) and (\ref{BcmNeu4}) would consistently project out the scalar $\phi$ 
so that $\phi$ takes values in $H$. 
However, for the Dirichlet boundary condition of vector multiplet with (\ref{BvmDir1}), 
they cannot be simply solved. 
Again it indicates that we need to generalize 
the pairing of the Dirichlet boundary condition for the vector multiplet 
and the Neumann boundary condition for the chiral multiplet. 

%%%%%%%%%%%%%%%%%%%%%%%%%%%%%%%%%%
\subsubsection{B-type Dirichlet b.c. for the chiral multiplet}
%%%%%%%%%%%%%%%%%%%%%%%%%%%%%%%%%%
For $\exp(i\beta) = 1$
we find Dirichlet boundary conditions for the chiral multiplet:
\begin{align}
D_{0}\phi&=0, \label{BcmDir1}\\
D_{0}\phi^{\dag}&=0, \label{BcmDir2}\\
\phi^{\dag} \sigma_+&=0. \label{BcmDir3}\\
\sigma_+ \phi&=0. \label{BcmDir4}
\end{align}
The boundary conditions (\ref{BcmDir1}) and (\ref{BcmDir2}) are 
Dirichlet boundary conditions for the chiral multiplet $\phi$ provided we also
impose $A_0 \phi = 0$.
The B-type Dirichlet boundary conditions for the chiral multiplet admit  
the 1d $\mathcal{N}=2_B$ Fermi multiplet described by the $({\bf 0},{\bf 2},{\bf 2})$ supermultiplets \cite{Hori:2014tda}. 

The additional constraints (\ref{BcmDir3}) and (\ref{BcmDir4}) can be solved 
by simply setting the boundary value of $\phi$ to zero. 
So a simple solution to (\ref{BcmDir1})-(\ref{BcmDir4}) is
\begin{align}
\phi&=0. 
\end{align}
This can lead to the maximal flavor symmetry $G_{\partial}$ 
from the global transformation of the gauge symmetry 
when the vector multiplet obeys the Dirichlet boundary condition.

%%%%%%%%%%%%%%%%%%%%%%%%%%%%%%%%%%
\subsubsection{Singular B-type boundary conditions}
%%%%%%%%%%%%%%%%%%%%%%%%%%%%%%%%%%
%We have a Neumann and a Dirichlet type condition for combinations of $\sigma$ and $\sigmad$.
We can find more general B-type boundary conditions. 

For $e^{i \alpha} = 1$ we find 
\begin{align}
D_0 \sigma_{-} & = - ig[\sigma_{-}, \sigma_{+}], 
\label{BvmNeuG1} \\
D_1 \sigma_{+} & = F_{01}, 
\label{BvmNeuG2}
\end{align}
which generalize the Neumann boundary conditions for the vector multiplet. 

For $e^{i \alpha} = -1$, we get
\begin{align}
D_0 \sigma_{+} & = 0, \label{BvmDirG1}\\
D_1 \sigma_{-} & = i (g \phi \phi^{\dagger} - \zeta ), \label{BvmDirG2} \; .
\end{align}
which contain the Dirichlet boundary conditions for the vector multiplet. 

Similarly for $e^{i \beta} = 1$ we find 
\begin{align}
D_0 \phi & =  ig \sigma_{+} \phi,  \label{BcmDirG1} \\
D_0 \phi^{\dagger} & = - ig \phi^{\dagger} \sigma_{+},  \label{BcmDirG2}
\end{align}
which contain the Dirichlet boundary conditions for the chiral multiplet. 

For $e^{i \beta} = -1$
\begin{align}
D_1 \phi & = - ig \sigma_{-} \phi 
\label{BcmNeuG1}\\
D_1 \phi^{\dagger} & = - ig \phi^{\dagger} \sigma_{-}
\label{BcmNeuG2}
\end{align}
which generalize the Neumann boundary conditions for the chiral multiplet. 

It is convenient to define
\begin{align}
\mathcal{A}_{0}&=A_{0} + \sigma_{+},& 
\mathcal{A}_{1}&=A_{1} - \sigma_{-}, 
\end{align}
so that we have the generalized covariant derivative operators
\begin{align}
\label{Bdiff}
\mathcal{D}_{0}&=D_{0} - ig \sigma_+,&
\mathcal{D}_{1}&=D_{1} + ig \sigma_-.
\end{align}
The chiral multiplet boundary conditions (\ref{BcmDirG1}) and (\ref{BcmNeuG1})
can then be written as $\mathcal{D}_{0} \phi = 0$ and
$\mathcal{D}_{1} \phi = 0$ respectively.
This also allows us to rewrite the boundary conditions (\ref{BvmNeuG1}) and (\ref{BvmNeuG2}) 
for the vector multiplet as a single equation 
\begin{align}
[\mathcal{D}_{0}, \mathcal{D}_{1}]&=0 \; ,
\end{align}
noting that we can extract two separate equations from the Hermitian and
anti-Hermitian parts of $[\mathcal{D}_{0}, \mathcal{D}_{1}]$ since
$\mathcal{A}_1$ is neither Hermitian nor anti-Hermitian.

To get solutions of the general B-type boundary conditions, we first consider the boundary conditions 
(\ref{BcmNeuG1}) and (\ref{BcmNeuG2}) for the chiral multiplets, 
which generalize the Neumann boundary conditions 
(\ref{BcmNeu1}) and (\ref{BcmNeu2}). 
They play a key role to find singular solutions. 
In fact, when we consider the chiral multiplet $\phi=\phi_1+i\phi_2$ transforming in the adjoint representation, 
the equations (\ref{BcmNeuG1}) and (\ref{BcmNeuG2}) together with (\ref{BvmDirG2}), 
which generalizes the Dirichlet boundary condition for the vector multiplet, 
are lifted to the well-known Nahm pole boundary conditions which admit singularity 
after identifying $\phi_1, \phi_2$ and $\sigma_-$ with three real scalar fields. 
In terms of the generalized differential operators (\ref{Bdiff}), 
the boundary conditions (\ref{BcmNeuG1}) and (\ref{BcmNeuG2}) for the chiral multiplets can be expressed as 
\begin{align}
\label{BcmNeuG3}
\mathcal{D}_{1}\phi&=0. 
\end{align}
This would imply that the chiral multiplets are covariantly constant 
and therefore a gauge invariant polynomial in $\phi$ should take the same values 
at any value of $x^1$. 

The equation (\ref{BcmNeuG3}) generalizes the Neumann boundary conditions for the chiral multiplet. 
In the axial gauge we find a singular solution to the boundary conditions (\ref{BcmNeuG1}) and (\ref{BcmNeuG2}) with the form
\begin{align}
\label{Bsol1}
\sigma_{-}&\sim \frac{\mathfrak{t}}{x^1}
\end{align}
where $\mathfrak{t}$ is some constant element of the Lie algebra of the gauge group, 
which we take hermitian.

Given the singular configuration (\ref{Bsol1}) of $\sigma_-$ near the boundary, 
we further consider the boundary conditions for other fields in the vector multiplet. 
To find the solutions, we set $A_1=0$ by gauge transformation in the following. 

%%%%%%%%%%%%%%%%%%%%%%%%%%%%%%%%%%
\subsubsection{Neumann b.c. for the vector multiplet with singularity}
%%%%%%%%%%%%%%%%%%%%%%%%%%%%%%%%%%
\label{sec_BNeupole}

%U(1)
Consider the singular solution (\ref{Bsol1}) along with the set of boundary conditions 
(\ref{BvmNeuG1}) and (\ref{BvmNeuG2}), 
which generalize the Neumann boundary conditions for the vector multiplet. 
For the basic Neumann boundary conditions for the vector multiplet that freezes $\sigma_-$, 
the regular value (\ref{Nnon_sigma}) of $\sigma_-$ may break the gauge symmetry. 
But for the singular solution (\ref{Bsol1}) of $\sigma_-$, 
it turns out that the scaling behavior of the scalar $\phi$ is affected. 

For the $U(1)$ vector multiplet, we find a solution
\begin{align}
\label{Bsol2u1}
\partial_{1}\sigma_{+}&=0,& A_{0}&=0, & A_{1}&=0. 
\end{align}
Accordingly, the Neumann boundary condition for $A_0$ flips to the Dirichlet boundary condition due to the singular configuration (\ref{Bsol1}). 
Given the solutions (\ref{Bsol1}) and (\ref{Bsol2u1}) to the boundary conditions for the $U(1)$ vector multiplet, 
we can find a solution to the boundary condition (\ref{BcmNeuG3}),  
which generalizes the Neumann boundary conditions for the chiral multiplet. 
If we have 
\begin{align}
 - ig\sigma_{-}& = \frac{\mathfrak{t}}{x^1}, 
\end{align}
we find a solution
\begin{align}
\label{BN_u1cm_sol}
\phi = (x^1)^{\mathfrak{t}} \phi_0
\end{align}
where $\phi_0$ is some constant value. 
When $\mathfrak{t}$ is positive, $\phi$ vanishes at the boundary, 
which essentially realizes the Dirichlet boundary condition. 
When $\mathfrak{t}$ is negative, $\phi$ has poles at the boundary of order $|\mathfrak{t}|$. 
Therefore the singularity (\ref{Bsol1}) in $\sigma_-$ dramatically alters the boundary conditions for $\phi$ 
in such a way that the scaling behavior of $\phi$ is controlled by $\mathfrak{t}$.

%non-Abelian
For the vector multiplet of non-Abelian gauge group $G$, 
the commutator in the equation (\ref{BvmNeuG1}) allows the $A_0$ component of
the gauge field to have non-zero value. 
Now, we still need to satisfy (\ref{ELbc}), and the contribution from the
vector multiplet now requires
\begin{align}
\Tr \left( F_{01} \delta A_0 + (D_1 \sigma_{+}) \delta \sigma_{+} - (D_1 \sigma_{-}) \delta \sigma_{-} \right) & = 0 \; .
\label{ELbc_vm_bos}
\end{align}
One option is to take Dirichlet boundary conditions for $\sigma_{-}$, i.e.\ in
(\ref{ELbc_vm_bos}) we have $\delta \sigma_{-} = 0$. Then
the boundary condition (\ref{BvmNeuG1}) % and (\ref{BvmNeuG2}) 
requires that $A_0$ and $\sigma_+$ take values in the same element of
the Lie algebra, up to a term which commutes with $\sigma_{-}$,
which modifies the ordinary Neumann boundary condition (\ref{BvmNeuG2}).
Specifically,
\begin{align}
A_0 & = \mp \sigma_{+} + \Sigma_{-}
\end{align}
where we require $[\sigma_{-}, \Sigma_{-}] = 0$. Then (\ref{BvmNeuG2})
reduces to
\begin{align}
\partial_0 A_1 & = D_1 \Sigma_{-}
\end{align}
and (\ref{ELbc_vm_bos}) becomes simply
\begin{align}
\Tr \left( F_{01} \delta \Sigma_{-} \right) & = 0 \; .
\end{align}
This is solved by imposing Dirichlet boundary conditions on $\Sigma_{-}$ so it
is fixed like $\sigma_{-}$.

As in the case of $U(1)$, the boundary condition (\ref{BcmNeuG3}) for the chiral multiplet 
leads to a non-trivial profile as a solution to a matrix differential equation
\begin{align}
\label{diffeq_phi1}
\frac{\partial \phi}{\partial x^1}&=\frac{\mathfrak{t}}{x^1}\phi. 
\end{align}
For example, for $G=SU(4)$ when we choose
\begin{align}
\label{t_su4a}
\mathfrak{t}&=
\left(
\begin{matrix}
0&1&0&0\\
1&0&0&0\\
0&0&1&0\\
0&0&0&-1\\
\end{matrix}
\right),
\end{align}
we find a solution
\begin{align}
\phi&=
\left(
\begin{matrix}
\frac{\alpha_1}{x^1}+\alpha_2 x^1\\
-\frac{\alpha_1}{x^1}+\alpha_2 x^1\\
\alpha_3 x^1\\
\frac{\alpha_4}{x^1}\\
\end{matrix}
\right). 
\end{align}
where $\alpha_1$, $\alpha_2$, $\alpha_3$ and $\alpha_4$ are some constant values. 
In this case, the gauge symmetry is completely broken 
and the Neumann boundary condition for $\phi$ changes so that 
all the components depend on $x^1$. 

Alternatively, when we pick 
\begin{align}
\mathfrak{t}&=
\left(
\begin{matrix}
0&1&0&0\\
1&0&0&0\\
0&0&0&0\\
0&0&0&0\\
\end{matrix}
\right),
\end{align}
the gauge symmetry is broken to $SU(2)$ 
and we find a solution
\begin{align}
\phi&=\left(
\begin{matrix}
\frac{\alpha_1}{x^1}+\alpha_2 x^1\\
-\frac{\alpha_1}{x^1}+\alpha_2 x^1\\
\phi_3\\
\phi_4\\
\end{matrix}
\right)
\end{align}
where $\phi_3$ and $\phi_4$ obey the Neumann boundary condition. 
We see that 
the non-zero values in $\mathfrak{t}$ modify 
the Dirichlet boundary condition of $\sigma_-$ 
(or equivalently the Neumann boundary condition for the vector multiplet) 
and the Neumann boundary condition on $\phi$ changes 
in such a way that the two components have nontrivial dependence on $x^1$. 

As can easily be seen in the above example, in general the solutions are
linear combinations of the eigenvectors of $\mathfrak{t}$. In particular if
$\mathfrak{t}$ has eigenvectors $v_i$ with eigenvalues $t_i$, the general
solution takes the form
\begin{equation}
\phi = \sum_i C_i (x^1)^{t_i} v_i. 
\end{equation}
Consequently, the gauge symmetry $G$ is generically broken to some $H$ 
depending on $\mathfrak{t}$. 

In particular, when the chiral multiplet scalar field has a simple pole
\begin{align}
\phi&=\frac{\mathfrak{a}}{x^1}
\end{align}
where $\mathfrak{a}$ is the eigenvector of $\mathfrak{t}$ 
with eigenvalue $-1$ according to the equation (\ref{diffeq_phi1}). 
Thus we have 
\begin{align}
\det (\mathbb{I}+\mathfrak{t})&=0. 
\end{align}

%%%%%%%%%%%%%%%%%%%%%%%%%%%%%%%%%%
\subsubsection{Dirichlet b.c. for the vector multiplet with singularity}
%%%%%%%%%%%%%%%%%%%%%%%%%%%%%%%%%%
\label{sec_BDirpole}

Consider the other set of boundary conditions 
(\ref{BvmDirG1}) and (\ref{BvmDirG2}), 
which generalizes the Dirichlet boundary conditions for the vector multiplet, 
together with the singular configuration (\ref{Bsol1}). 
For the $U(1)$ vector multiplet, working in the gauge $A_1 = 0$, from equation
(\ref{BvmDirG1}) we get
\begin{align}
\label{BDD_other}
\partial_0\sigma_+&=0,& A_0&=0,
\end{align}
which breaks the gauge symmetry. 
Then, according to the other condition (\ref{BvmDirG2}), we see that $\phi$ must have a singularity with the form
\begin{align}
\label{BDD_phi}
\phi&=\frac{c}{g x^1},& 
\phi^{\dag}&=\frac{c^*}{g x^1}
\end{align}
where $c\in \mathbb{C}$ is some constant. 
Plugging (\ref{BDD_phi}) into the condition (\ref{BcmNeuG1}), 
we find that $\sigma_{-} = \mp i / (g x^1)$, so we see that
(\ref{ELbc}), or more specifically (\ref{ELbc_vm_bos}), is satisfied with
$\sigma_{-}$ obeying Dirichlet boundary conditions.
%$\mathfrak{t}_1=i/\sqrt{2}g$. 
Together with the condition (\ref{BvmDirG2}), 
this gives rise to $|c|^2= 1$, i.e.\ $c$ is an arbitrary phase. 
Recall that 
the ordinary Dirichlet boundary conditions for the $U(1)$ vector multiplet 
together with the ordinary Neumann boundary conditions for the chiral multiplet 
have no simple solution. 
In the absence of boundary terms, 
the only consistent condition seems to be 
the singular configurations (\ref{Bsol1}) and (\ref{BDD_phi}) which resolve this obstruction.

%which is inconsistent. 
%Therefore the generalized Dirichlet b.c. for the $U(1)$ vector multiplet 
%would not admit the singular configurations (\ref{Bsol1}) and (\ref{BDD_phi}). 

%{\bf [To check the above. Analysis is correct, assuming no sign errors in
%(\ref{supercurrent}) -- still need to double check signs in supercurrent.] }

%{\bf [Is it possible to have any poles for generalized Dirichlet for non-Abelian? 

For the non-Abelian case, we also find solutions (\ref{BDD_other}) 
and the gauge symmetry is completely broken. 
However, $c$ is a vector $\mathfrak{c}$ rather than a constant: 
\begin{align}
\phi&=\frac{\mathfrak{c}}{g x^1},& 
\phi^{\dag}&=\frac{\mathfrak{c}^{\dag}}{g x^1}
\label{BDD_NA_phi}
\end{align}
Then (\ref{BcmNeuG1}) requires it to be an eigenvector of $\mathfrak{t}$ with
eigenvalue $-1$, where $\sigma_{-} = \pm i \mathfrak{t} / (g x^1)$,
and consequently (\ref{BvmDirG2}) requires that
\begin{align}
\mathfrak{t} & = - \mathfrak{c} \mathfrak{c}^{\dagger} \; . \label{BDD_t_cc}
\end{align}
%specific constant times $c c^{\dagger}$. However, although $c$ is then clearly
%an eigenvector of $\mathfrak{t}_1 \sim c c^{\dagger}$, for it to have the
%correct eigenvalue requires $c^{\dagger} c = -1/g^2$.
Clearly $\mathfrak{c}$ is automatically an eigenvector of $\mathfrak{t}$, and it has
eigenvalue $-1$ provided $\mathfrak{c}$ is normalized so that
$\mathfrak{c}^{\dagger} \mathfrak{c} = 1$. Note that this means that
$\Tr \mathfrak{t} \ne 0$ so such solutions are possible with gauge group $U(N_c)$
but not $SU(N_c)$.

In fact, the above gives the leading order behaviour at the boundary even
without the specific assumptions in (\ref{BDD_phi}) or (\ref{BDD_NA_phi}) if
we simply assume that $\phi$ can be expressed as a Taylor series in $x^1$, with
non-zero constant term, multiplied by an arbitrary non-zero power of $x^1$.
That is because in this case (\ref{BcmNeuG1}) requires the leading behaviour
of $\sigma_{-}$ to be a simple pole in $x^1$ and consequently (\ref{BvmDirG2})
requires the leading order term for $\phi$ to also be a simple pole. We could
solve (\ref{BcmNeuG1}) by including a more singular term in $\sigma_{-}$
provided it annihilated $\phi$ but that would not be consistent with
(\ref{BvmDirG2}). The only other
alternative is that $\phi$ is a regular Taylor series, in which case
(\ref{BcmNeuG1}) shows that $\sigma_{-}$ is also regular at the boundary. Then
the boundary values of $\phi$ and $\sigma_{-}$ are arbitrary, with
(\ref{BvmDirG2}) and (\ref{BcmNeuG1}) determining the boundary values of their
normal derivatives.

If we have multiple flavors, we can find more general solutions. In this case
each flavor $\phi^i$ satisfies (\ref{BcmNeuG1}) but (\ref{BvmDirG2}) is
generalized to
\begin{align}
D_1 \sigma_{-} & = \pm i (g \phi^i \phi_i^{\dagger} - \zeta ), \label{BvmDirNfG2} \; .
\end{align}
Now, as for the single flavor case above, we could have a regular solution for
$\sigma_{-}$ and all $\phi^i$. Otherwise, again $\sigma_{-}$ must have a simple
pole as must at least one of the $\phi^i$. Now we can have the situation that
some of the $\phi^i$ have poles while some are regular. We can solve
(\ref{BcmNeuG1}) with the condition that the constant values of the regular
$\phi^i$ are annihilated by the singular part of $\sigma_{-}$. Then the
singular part of the solution is described as above with
$\mathfrak{c} \rightarrow \mathfrak{c}^i$ and now
$\mathfrak{t} = -\mathfrak{c}^i \mathfrak{c}^{\dagger}_i$ where we include only
values of $i$ corresponding to singular $\phi^i$.
Again each $\mathfrak{c}^i$ must be an eigenvector of $\mathfrak{t}$ with
eigenvalue $-1$. The most general solution is
$\mathfrak{c}^i = \omega^i \mathfrak{u}^{I(i)}$ where the $\{ \mathfrak{u}^I \}$
form a set of orthonormal vectors and $\omega^i \in \Cb$. The constants
$\omega^i$ are constrained to satisfy
$\sum_{i : I(i) = J} |\omega^i|^2 = 1$ for each $J$.

However, we can use the flavor symmetry to do a field redefinition so that for
each $J$ we use a unitary transformation acting on the $\phi^i$ with
$I(i) = J$ so that only one scalar has a singular part. Hence we will split the
$N_f$ flavors $\phi^i$ into $N_s$ with a singular boundary condition.
Specifically, including the subleading regular terms we have $N_s$ scalars
\begin{equation}
\phi^I = \frac{\mathfrak{c}^I}{g x^1} + \phi_0^I + x^1 \phi_1^I \label{vDcNsrS1}
\end{equation}
with orthonormal $\{ \mathfrak{c}^I \}$ and the remaining $N_f - N_s$ will
have regular boundary conditions
\begin{equation}
\phi^{\Ih} = \phi_0^{\Ih} + x^1 \phi_1^{\Ih} \label{vDcNsrS2}
\end{equation}
where we have assumed a power series expansion of the fields near the boundary
and included only those terms relevant for the boundary conditions.
Similarly, writing the boundary condition
\begin{equation}
\sigma_{-} = \frac{\pm i}{g x^1}\mathfrak{t} \pm 2i \mathfrak{t}_0 \mp 2i x^1 \mathfrak{t}_1 \; , \label{vDcNsrS3}
\end{equation}
where $\mathfrak{t}$, $\mathfrak{t}_0$ and $\mathfrak{t}_1$ are Hermitian,
the boundary conditions (\ref{BvmDirG2}) and (\ref{BcmNeuG1}) are equivalent to
the following conditions
\begin{align}
\mathfrak{t} & = - \mathfrak{c}^I \mathfrak{c}^{\dagger}_I \label{vDcNsr1} \\
\mathfrak{t} \phi_0^I & = \mathfrak{t}_0 \mathfrak{c}^I \label{vDcNsr2} \\
\mathfrak{t} \phi_0^{\Ih} & = 0 \label{vDcNsr3} \\
(I - \mathfrak{t}) \phi_1^i & = - g \mathfrak{t}_0 \phi_0^i \label{vDcNsr4} \\
\mathfrak{c}^I \phi_{0I}^{\dagger} + \phi_0^I \mathfrak{c}_I^{\dagger} & = 0 \label{vDcNsr5} \\
\mathfrak{t}_{1} & = \mathfrak{c}^I \phi_{1I}^{\dagger} + \phi_1^I \mathfrak{c}_I^{\dagger} + g \phi_0^i \phi_{0i}^{\dagger} - \zeta \; . \label{vDcNsr6}
\end{align}
These can be simplified by defining the $N_s$-dimensional vector space
$V_s$ with orthonormal basis $\{\mathfrak{c}^I\}$, and $(N_c - N_s)$-dimensional
vector space $V_r$, so that the $\phi^i$ belong
to the $N_c$-dimensional vector space $V_s \oplus V_r$. Then (\ref{vDcNsr1})
states that $\mathfrak{t}$ projects onto $V_s$ and acts as minus the identity,
i.e. $-I_s$, on that space. Then (\ref{vDcNsr2}) shows that $\mathfrak{t}_0$
maps $V_s \rightarrow V_s$ and hence (since it is Hermitian) also
$V_r \rightarrow V_r$. Now, if we act on (\ref{vDcNsr5}) from the left and from
the right with $\mathfrak{t}$ and use
$\mathfrak{t}\mathfrak{c}^I = - \mathfrak{c}^I$ along with (\ref{vDcNsr2}) we
see that
\begin{align}
\left\{ \mathfrak{t}, \mathfrak{t}_0 \right\} & = 0
\end{align}
so that (since $\mathfrak{t}$ acts as $-I_s$) in fact $\mathfrak{t}_0$
annihilates all vectors in $V_s$. Hence (\ref{vDcNsr2}) and (\ref{vDcNsr3})
combine to give simply
\begin{align}
\mathfrak{t} \phi_0^{i} & = 0
\end{align}
which means that all $\phi_0^{i} \in V_r$. Then, noting that
$(I - \mathfrak{t})^{-1} = I + \frac{1}{2}\mathfrak{t}$ since
$\mathfrak{t}^2 = - \mathfrak{t}$, (\ref{vDcNsr4}) becomes
\begin{align}
\phi_1^i & = - g \mathfrak{t}_0 \phi_0^i \label{vDcNsrS4}
\end{align}
showing that also all $\phi_1^i \in V_r$. This simplifies (\ref{vDcNsr6}) to
\begin{align}
\mathfrak{t}_{1} & = g \phi_0^i \phi_{0i}^{\dagger} - \zeta \; . \label{vDcNsrS5}
\end{align}

So, to summarize the solutions take the form of a superposition of singular
solutions and regular solutions. For the singular solutions we choose a
non-negative integer $N_s \le \min(N_c, N_f)$ for gauge group $U(N_c)$ and $N_f$
flavors. If $N_s > 0$ we pick an $N_s$-dimensional orthonormal basis 
$\{\mathfrak{c}^I\}$ and this determines the singular part of the solution
(\ref{vDcNsrS1}) and (\ref{vDcNsrS3}). We can then choose the regular part of
the solution by specifying $N_f$ arbitrary vectors $\phi_0^i$ and an arbitrary
Hermitian mapping $\mathfrak{t}_0 : \; V_r \rightarrow V_r$ which also acts as
$\mathfrak{t}_0 \mathfrak{c}^I = 0$. Then $\phi_1^i$ and $\mathfrak{t}_1$ are
fully determined by (\ref{vDcNsrS4}) and (\ref{vDcNsrS5}), giving the complete
solutions (\ref{vDcNsrS1})-(\ref{vDcNsrS3}).

Note that the singular and regular parts of the solutions are completely
independent other than the determination of the splitting into orthogonal
subspaces $V_s$ and $V_r$. In particular, if we choose $\phi_0^i = 0$ and
$\mathfrak{t}_0 = 0$ we have only singular terms in the solution, except for
the FI parameter contribution to $\partial_1 \sigma_{-}$ through
$\mathfrak{t}_1$.

Now we also consider the symmetries preserved by such boundary conditions.
Since we have $A_0 = 0$ the gauge symmetry will become a global symmetry
$U(N_c)$ on the boundary. For $N_s > 0$ this is explicitly broken to
$U(N_s) \times U(N_c - N_s)$ simply by the choice of $V_s$. Without any regular
terms (i.e.\ with $\phi_0^i = 0$ and $\mathfrak{t}_0 = 0$) the form of
$\sigma_{-}$ given in (\ref{vDcNsrS3}) is invariant, even with an FI parameter.
However, the $\phi^I$ given by (\ref{vDcNsrS1}) are not invariant under the
$U(N_s)$ transformations. Nevertheless, we can compensate using the explicitly
broken $U(N_s)$ subgroup of the flavor symmetry. So, together with the
unbroken $U(N_f - N_s)$ flavor symmetry we have the global symmetry
$U(N_c - N_s) \times U(N_s) \times U(N_f - N_s)$. Now, introducing non-zero
regular terms into the solutions will generically completely break the
$U(N_c - N_s)$ and $U(N_f - N_s)$ symmetries, but since the regular terms are
related to $V_r$ only, the form of $\phi^{\Ih}$ in (\ref{vDcNsrS2}) and of
$\sigma_{-}$ in (\ref{vDcNsrS3}) will not break the $U(N_s)$ global symmetry.
So, the $U(N_s)$ symmetry is only broken by the regular terms in $\phi^{I}$
in (\ref{vDcNsrS1}), since these are invariant under the $U(N_s)$ from the gauge
symmetry but not under the compensating transformation from the flavor
symmetry. Therefore, if $\widetilde{N}_s$ of the $\phi_0^I = 0$ a
$U(\widetilde{N}_s)$ symmetry will be preserved, as from (\ref{vDcNsrS4}) also
the corresponding $\phi_1^I = 0$.

We note that the above analysis is for chiral multiplets in the fundamental
representation. It can easily be generalized to other representations. For
example if we had a single chiral multiplet in the adjoint representation we
would have the Nahm equations (deformed by the FI term) for the three
fields $\sigma_{-}$ and the Hermitian and anti-Hermitian parts of the scalar
$\phi$.

%{\bf [Could include analysis of the subleading terms, i.e. regular terms for
%$\phi^i$.]}
%{\bf [Explain that more generally we can solve (\ref{BcmNeuG1}) with
%$\sigma_{-} \sim \mathfrak{t}/x^1$ finding singular solution for $\phi$ is
%we have eigenvalue $-1$. More singular solutions for $\phi$ cannot solve
%(\ref{BvmDirG2}) so eigenvectors with more negative eigenvalue cannot be
%included. Not clear if we can include terms with zero eigenvalues, i.e.
%constant terms for $\phi$ while satisfying (\ref{BvmDirG2}).
%
%Need to look at multiple flavors to generalize $\mathfrak{t}$ to have more
%than one $-1$ eigenvalue.]}

%We could also have chosen the condition
%\begin{equation}
%\gamma^1 \epsilon = - \epsilon
%\end{equation}
%but this gives the same results up to a redefinition
%$\sigma \rightarrow - \sigma$.

%%%%%%%%%%%%%%%%%%%%%%%%%%%%%%%%%%
\subsubsection{General gauge group projections and multiple matter multiplets}
\label{sec_Bmaxproj}
%%%%%%%%%%%%%%%%%%%%%%%%%%%%%%%%%%
Like for the A-type constraints we can generalize the B-type boundary conditions 
when we have a gauge group other than $U(1)$, 
or/and multiplet matter multiplets 
by imposing matrix projection conditions on fermions. 
In this case we get
\begin{align}
\lambda^a_2 & = R^{ab}\lambda^b_1 \; , & \psi^i_2 & = S^{i}_{\phantom{i}j} \psi^j_1 \; .
\end{align}
Now (\ref{ELbc}) is satisfied if $R$ and $S$ are unitary.
The supersymmetric boundary conditions then require
\begin{eqnarray}
\left( \left( - F_{01} + ig[\sigma, \sigmad] \right) \left( I + e^{-i\theta}R \right)
 + i(g\phi^i\phi^{\dagger}_i - \zeta) \left( I + e^{-i\theta}R \right) \right. & & \nonumber \\
\left. + \sqrt{2} \left( D_0 + D_1 \right) \sigma R
 - \sqrt{2} e^{-i\theta} \left( D_0 - D_1 \right) \sigmad \right) \bar{\epsilon}_{-}\lambda_{-} & = & 0 \\
\bar{\epsilon}_{-}\psib_{-} \left( \left( e^{-i\theta} + S^{\dagger} \right) D_0 \phi
 + \left( - e^{-i\theta} + S^{\dagger} \right) D_1 \phi
 - 2ig e^{-i\theta} S^{\dagger} \sigmad \phi
 - 2ig \sigma \phi \right) & = & 0 \hspace{1cm}
\end{eqnarray}
Although we do not pursue the detail here, 
we can find certain combinations of the boundary conditions  we discussed above 
for the theory with higher rank gauge group or multiplet flavors 
by choosing appropriate matrices $R$ and $S$. 
It would be also interesting to explore more general boundary conditions 
by allowing $R$ and $S$ to depend on the scalar fields $\sigma$ and/or $\phi$.

%%%%%%%%%%%%%%%%%%%%%%%%%%%%%%%%%%
%%%%%%%%%%%%%%%%%%%%%%%%%%%%%%%%%%
\section{Quarter-BPS boundary conditions}
\label{sec_quarter}
%%%%%%%%%%%%%%%%%%%%%%%%%%%%%%%%%%
%%%%%%%%%%%%%%%%%%%%%%%%%%%%%%%%%%
We can impose the quarter-BPS boundary conditions by writing each spinor and
its complex conjugate in terms of a single spinor. In order to satisfy
\begin{equation}
\delta\lambda \gamma^1 \bar{\lambda} + \delta\bar{\lambda} \gamma^1 \lambda = 0 \; .
\end{equation}
the most general condition we can have is
\begin{eqnarray}
\lambda_2 & = & e^{i\alpha_1}\lambda_1 \\
\bar{\lambda}_1 & = & e^{2i\alpha_2}\lambda_1 \\
\bar{\lambda}_2 & = & e^{i(2\alpha_2 - \alpha_1)}\lambda_1
\end{eqnarray}
and similarly for $\psi$ we have
\begin{eqnarray}
\psi_2 & = & e^{i\beta_1}\psi_1 \\
\bar{\psi}_1 & = & e^{2i\beta_2}\psi_1 \\
\bar{\psi}_2 & = & e^{i(2\beta_2 - \beta_1)}\psi_1
\end{eqnarray}
We can also parameterize $\epsilon$ in terms of a single real Grassmann
parameter $\epsilon_0 = \bar{\epsilon}_0$ as
\begin{equation}
\epsilon_1 = a \epsilon_0 \; , \;\; \epsilon_2 = b \epsilon_0 \; , \;\;
\bar{\epsilon}_1 = \bar{a} \epsilon_0 \; , \;\; \bar{\epsilon}_2 = \bar{b} \epsilon_0
\end{equation}
Then the conditions to preserve 1d $\mathcal{N}=1$ supersymmetry reduce to two equations
\begin{align}
\label{quarter_vm}
& - \Re \left( a e^{i\alpha_2} + b e^{i(\alpha_2 - \alpha_1)} \right) F_{01} \nonumber \\
&+ g \Im \left( a e^{i\alpha_2} + b e^{i(\alpha_2 - \alpha_1)} \right) [\sigma, \sigmad]
+ \Im \left( a e^{i\alpha_2} - b e^{i(\alpha_2 - \alpha_1)} \right) ( g \phi \phi^{\dagger} - \zeta )  \nonumber \\
&+ \sqrt{2} D_0 \Re \left( (-b e^{i\alpha_2} + \bar{a} e^{i(\alpha_1 - \alpha_2)} ) \sigma \right)
+ \sqrt{2} D_1 \Re \left( (b e^{i\alpha_2} + \bar{a} e^{i(\alpha_1 - \alpha_2)} ) \sigma \right)
  =  0
\end{align}
and
\begin{align}
\label{quarter_cm}
&D_0 \Im \left( \left( \bar{b} e^{i\beta_2} + \bar{a} e^{i(\beta_2 - \beta_1)} \right) \phi \right)
 + D_1 \Im \left( \left( -\bar{b} e^{i\beta_2} + \bar{a} e^{i(\beta_2 - \beta_1)} \right) \phi \right) \nonumber \\
& - \sqrt{2} g \Re \left( \left( \bar{a} e^{i\beta_2} \sigmad + \bar{b} e^{i(\beta_2 - \beta_1)} \sigma \right) \phi \right)
 =  0
\end{align}

The first equation (\ref{quarter_vm}) is the quarter-BPS boundary conditions for the vector multiplet 
and the second equation (\ref{quarter_cm}) is that for the chiral multiplet. 
The parameters $a$ and $b$ fix a choice of the supercharge. 
There are four basic boundary conditions for each of vector and chiral multiplets 
so that totally we find sixteen types.

%%%%%%%%%%%%%%%%%%%%%%%%%%%%%%%%%%
\subsection{Simple examples}
%%%%%%%%%%%%%%%%%%%%%%%%%%%%%%%%%%
We first consider the case with $a=b=1$ 
where the supersymmetric boundary conditions have fewer terms. 
Although the following conditions would be required to preserve $\mathcal{N}=1$ supersymmetry, 
one needs to impose additional boundary conditions to preserve locality of the bulk fields, 
which do not show up in the projection of supercurrents. 
In the following we group the resulting quarter-BPS boundary conditions into four different sets of boundary conditions for the vector multiplet, 
which we call N$'$N$''$, N$'$D$''$, D$'$N$''$ and D$'$D$''$. 

%%%%%%%%%%%%%%%%%%%%%%%%%%%%%%%%%%
\subsubsection{N$'$N$''$ boundary conditions}
%%%%%%%%%%%%%%%%%%%%%%%%%%%%%%%%%%
There are four distinct boundary conditions 
which are compatible with A-type Neumann boundary conditions and B-type Neumann boundary conditions for the vector multiplet. 

For $\alpha_1=0$, $\alpha_2=0$, $\beta_1=0$ and $\beta_2=0$ we find 
\begin{align}
F_{01} - D_1 \sigma_+&=0, \\
D_{0}(\Im \phi) - g (\Re \phi) \sigma_+&=0. 
\end{align}
For $\alpha_1=0$, $\alpha_2=0$, $\beta_1=0$ and $\beta_2=\frac{\pi}{2}$ we find 
\begin{align}
F_{01} - D_1 \sigma_+&=0, \\
D_{0}(\Re \phi) + g (\Im \phi) \sigma_+&=0. 
\end{align}
For $\alpha_1=0$, $\alpha_2=0$, $\beta_1=\pi$ and $\beta_2=0$ we find 
\begin{align}
F_{01} - D_1 \sigma_+&=0, \\
D_{1}(\Im \phi) - i g (\Im \phi) \sigma_{-}&=0
\end{align}
For $\alpha_1=0$, $\alpha_2=0$, $\beta_1=\pi$ and $\beta_2=\frac{\pi}{2}$ we find 
\begin{align}
F_{01} - D_1 \sigma_+&=0, \\
D_{1}(\Re \phi) - i g (\Re \phi) \sigma_{-}&=0. 
\end{align}
The boundary condition for the vector multiplet take the same form as the B-type boundary condition 
(\ref{BvmNeuG2}) and involves 
the Neumann boundary condition for gauge field and that for $\sigma_+$. 
While the B-type boundary conditions also need the condition (\ref{BvmNeuG1}), 
it is not necessary for quarter of supersymmetry.

%%%%%%%%%%%%%%%%%%%%%%%%%%%%%%%%%%
\subsubsection{N$'$D$''$ boundary conditions}
%%%%%%%%%%%%%%%%%%%%%%%%%%%%%%%%%%
There are four boundary conditions 
which are compatible with A-type Dirichlet boundary condition and B-type Neumann boundary condition for the vector multiplet. 

For $\alpha_1=0$, $\alpha_2=\frac{\pi}{2}$, $\beta_1=0$ and $\beta_2=0$ we find 
\begin{align}
g[\sigma,\sigma^{\dag}]-iD_0 \sigma_-&=0, \\
D_{0}(\Im \phi) - g (\Re \phi) \sigma_+&=0
\end{align}
For $\alpha_1=0$, $\alpha_2=\frac{\pi}{2}$, $\beta_1=0$ and $\beta_2=\frac{\pi}{2}$ we find 
\begin{align}
g[\sigma,\sigma^{\dag}]-iD_0 \sigma_-&=0, \\
D_{0}(\Re \phi) + g (\Im \phi) \sigma_+&=0
\end{align}
For $\alpha_1=0$, $\alpha_2=\frac{\pi}{2}$, $\beta_1=\pi$ and $\beta_2=0$ we find 
\begin{align}
g[\sigma,\sigma^{\dag}]-iD_0 \sigma_-&=0, \\
D_{1}(\Im \phi) - i g (\Im \phi) \sigma_{-}&=0
\end{align}
For $\alpha_1=0$, $\alpha_2=\frac{\pi}{2}$, $\beta_1=\pi$ and $\beta_2=\frac{\pi}{2}$ we find 
\begin{align}
g[\sigma,\sigma^{\dag}]-iD_0 \sigma_-&=0, \\
D_{1}(\Re \phi) - i g (\Re \phi) \sigma_{-}&=0
\end{align}
The boundary condition for the vector multiplet 
corresponds to (\ref{BvmNeuG1}) in the B-type boundary conditions 
and contains the Dirichlet boundary condition for $\sigma_-$ together with the condition $[\sigma, \sigma^{\dag}]=0$. 

%%%%%%%%%%%%%%%%%%%%%%%%%%%%%%%%%%
\subsubsection{D$'$N$''$ boundary conditions}
%%%%%%%%%%%%%%%%%%%%%%%%%%%%%%%%%%
There are four boundary conditions 
which are compatible with A-type Neumann boundary condition and B-type Dirichlet boundary condition for the vector multiplet. 
They can be obtained from the N$'$D$''$ boundary conditions by 
replacing the value of $\alpha_1$ with $\pi$.  

For $\alpha_1=\pi$, $\alpha_2=\frac{\pi}{2}$, $\beta_1=0$ and $\beta_2=0$ we find 
\begin{align}
g\phi\phi^{\dag} - \zeta +iD_1 \sigma_-&=0, \\
D_{0}(\Im \phi) - g (\Re \phi) \sigma_+&=0
\end{align}
For $\alpha_1=\pi$, $\alpha_2=\frac{\pi}{2}$, $\beta_1=0$ and $\beta_2=\frac{\pi}{2}$ we find 
\begin{align}
g\phi\phi^{\dag} - \zeta +iD_1 \sigma_-&=0,  \\
D_{0}(\Re \phi) + g (\Im \phi) \sigma_+&=0
\end{align}
For $\alpha_1=\pi$, $\alpha_2=\frac{\pi}{2}$, $\beta_1=\pi$ and $\beta_2=0$ we find 
\begin{align}
g\phi\phi^{\dag} - \zeta +iD_1 \sigma_-&=0,  \\
D_{1}(\Im \phi) - i g (\Im \phi) \sigma_{-}&=0
\end{align}
For $\alpha_1=\pi$, $\alpha_2=\frac{\pi}{2}$, $\beta_1=\pi$ and $\beta_2=\frac{\pi}{2}$ we find 
\begin{align}
g\phi\phi^{\dag} - \zeta +iD_1 \sigma_-&=0,  \\
D_{1}(\Re \phi) - i g (\Re \phi) \sigma_{-}&=0
\end{align}
The boundary condition for the vector multiplet 
corresponds to (\ref{BvmDirG2}) in the B-type boundary conditions 
and contains the Neumann boundary condition for $\sigma_-$ together with the constraint $g\phi\phi^{\dag}=0$.

%%%%%%%%%%%%%%%%%%%%%%%%%%%%%%%%%%
\subsubsection{D$'$D$''$ boundary conditions}
%%%%%%%%%%%%%%%%%%%%%%%%%%%%%%%%%%
There are four boundary conditions 
which are compatible with A-type Dirichlet boundary condition and B-type Dirichlet boundary condition for the vector multiplet. 
They can be obtained from the N$'$N$''$ boundary conditions by 
replacing the value of $\alpha_1$ with $\pi$.  

For $\alpha_1=\pi$, $\alpha_2=0$, $\beta_1=0$ and $\beta_2=0$ we get
\begin{align}
D_0\sigma_+&=0, \\
D_{0}(\Im \phi) - g (\Re \phi) \sigma_+&=0
\end{align}
For $\alpha_1=\pi$, $\alpha_2=0$, $\beta_1=0$ and $\beta_2=\frac{\pi}{2}$ we find 
\begin{align}
D_0\sigma_+&=0, \\
D_{0}(\Re \phi) + g (\Im \phi) \sigma_+&=0
\end{align}
For $\alpha_1=\pi$, $\alpha_2=0$, $\beta_1=\pi$ and $\beta_2=0$ we find 
\begin{align}
D_0\sigma_+&=0, \\
D_{1}(\Im \phi) - i g (\Im \phi) \sigma_{-}&=0
\end{align}
For $\alpha_1=\pi$, $\alpha_2=0$, $\beta_1=\pi$ and $\beta_2=\frac{\pi}{2}$ we find 
\begin{align}
D_0\sigma_+&=0, \\
D_{1}(\Re \phi) - i g (\Re \phi) \sigma_{-}&=0
\end{align}
The boundary condition for the vector multiplet 
is identified with (\ref{BvmDirG1}) in the B-type boundary conditions, 
which is the Dirichlet boundary condition on $\sigma_+$. 
While the B-type boundary conditions also require the condition (\ref{BvmDirG2}), 
we do not need the latter to preserve one supercharge. 

Therefore, the N$'$N$''$, N$'$D$''$, D$'$N$''$ and D$'$D$''$ boundary conditions for the vector multiplet 
is obtained by simply picking up one of the B-type boundary conditions, that is 
(\ref{BvmNeuG2}), (\ref{BvmNeuG1}), (\ref{BvmDirG2}) and (\ref{BvmDirG1}) respectively. 
However, the four distinct quarter-BPS boundary conditions for the chiral multiplet 
cannot be obtained by naively choosing the half-BPS boundary conditions. 

%%%%%%%%%%%%%%%%%%%%%%%%%%%%%%%%%%
\subsubsection{Singular solutions}
%%%%%%%%%%%%%%%%%%%%%%%%%%%%%%%%%%
We can also find singular solutions of the quarter-BPS boundary conditions. 
Note that as in the B-type boundary conditions, the two kinds of boundary conditions 
\begin{align}
D_{1}(\Re \phi) - i g (\Re \phi) \sigma_{-}&=0
\end{align}
and 
\begin{align}
D_{1}(\Im \phi) - i g (\Im \phi) \sigma_{-}&=0
\end{align}
for the chiral multiplet can be solved by postulating the simple pole for $\sigma_-$ in the axial gauge: 
\begin{align}
\label{quarter_pole1}
\sigma_-&\sim \frac{\mathfrak{u}}{x^1}
\end{align}
where $\mathfrak{u}$ is some constant valued in the Lie algebra. 

Given the singular profile (\ref{quarter_pole1}), 
it is now straightforward to get solutions. 
For the N$'$N$''$, N$'$D$''$ and D$'$D$''$ boundary conditions one can solve them by following the previous discussion 
for the generalized B-type Neumann boundary conditions 
whereas for the D$'$N$''$ boundary conditions we can find solutions by following the discussion 
for the generalized B-type Dirichlet boundary conditions.

%%%%%%%%%%%%%%%%%%%%%%%%%%%%%%%%%%
\subsection{Other cases}
%%%%%%%%%%%%%%%%%%%%%%%%%%%%%%%%%%

%%%%%%%%%%%%%%%%%%%%%%%%%%%%%%%%%%
\subsubsection{Mixed cases}
%%%%%%%%%%%%%%%%%%%%%%%%%%%%%%%%%%
Next consider the case with $a=0$ and $b=1$. 

For fixed $\alpha_1$ and $\alpha_2$ one finds four different types of supersymmetric boundary conditions of the vector multiplet: 
\begin{align}
F_{01} + (D_0 -D_1) \sigma_+&=0,\\
g(\phi \phi^{\dag}-[\sigma, \sigma^{\dag}]-\zeta)
+i (D_0 -D_1) \sigma_- &=0,\\
g(\phi \phi^{\dag}-[\sigma, \sigma^{\dag}]-\zeta)
-i (D_0 -D_1) \sigma_- &=0,\\
F_{01} - (D_0 -D_1) \sigma_+&=0. 
\end{align}

Similarly, there are four different types of the boundary conditions for the chiral multiplet for fixed $\beta_1$ and $\beta_2$: 
\begin{align}
(D_0-D_1)\Im \phi - \Re (\phi \sigma)&=0,\\
(D_0-D_1)\Re \phi + \Im (\phi\sigma)&=0,\\
(D_0-D_1)\Re \phi - \Im (\phi\sigma)&=0,\\
(D_0-D_1)\Im \phi + \Re (\phi \sigma)&=0,
\end{align}
These boundary conditions are mixed in that 
each of the conditions contain the both Neumann and Dirichlet boundary conditions on the bosonic fields 
in a single equation as encountered in the A-type generic boundary condition.

%%%%%%%%%%%%%%%%%%%%%%%%%%%%%%%%%%
\subsubsection{Corner}
%%%%%%%%%%%%%%%%%%%%%%%%%%%%%%%%%%
Another interesting situation with 
a quarter of supersymmetry is a corner configuration of the 2d $\mathcal{N}=(2,2)$ gauge theory 
placed on a quadrant $\mathbb{R}_+\times \mathbb{R}_+$. 
It should realize 0d $\mathcal{N}=1$ supersymmetry. 
In order to preserve supersymmetry at a corner, say at $x^0=x^1=0$ 
we should further impose the condition by demanding that the component $J^0$ of the supercurrent vanishes at $x^0=0$, 
in addition to the vanishing component $J^1$ at $x^1=0$. 
To maintain the equation of motion by employing the A-type and B-type boundary conditions on fermions 
(see (\ref{A_fer}) and (\ref{B_fer1})-(\ref{B_fer3})), 
one finds that the only consistent conditions consist of the A-type boundary conditions along the $x^1$ 
and the B-type boundary conditions along the $x^0$ or vice versa. 
The resulting conditions imposed at a corner would be stronger than 
the quarter-BPS boundary conditions that we discussed in the above. 
We defer a study of the quarter-BPS corner conditions to future work.

%%%%%%%%%%%%%%%%%%%%%%%%%%%%%%%%%%
%%%%%%%%%%%%%%%%%%%%%%%%%%%%%%%%%%
\section{Brane setup in Type IIA string theory}
\label{sec_IIA}
%%%%%%%%%%%%%%%%%%%%%%%%%%%%%%%%%%
%%%%%%%%%%%%%%%%%%%%%%%%%%%%%%%%%%
In this section, we construct the BPS-boundary conditions 
in terms of branes in Type IIA string theory. We start with brane configurations
producing 2d $\mathcal{N} = (2,2)$ gauge theories and then add additional branes
to give either A-type of B-type boundary conditions. We also show that the
M-theory lift of these brane configurations is the same for both A-type and
B-type -- the difference being simply which direction is compactified to reduce
to Type IIA string theory.

%%%%%%%%%%%%%%%%%%%%%%%%%%%%%%%%%%
%%%%%%%%%%%%%%%%%%%%%%%%%%%%%%%%%%
\subsection{2d $\mathcal{N}=(2,2)$ supersymmetric gauge theories}
\label{sec_IIAbrane}
%%%%%%%%%%%%%%%%%%%%%%%%%%%%%%%%%%
%%%%%%%%%%%%%%%%%%%%%%%%%%%%%%%%%%
First, let us briefly review the Hanany-Hori construction \cite{Hanany:1997vm} 
of 2d $\mathcal{N}=(2,2)$ gauge theory. 
We start from the brane configuration consisting of the following branes in Type IIA string theory: 
\begin{align}
\label{22_brane}
\begin{array}{c|cccccccccc}
&0&1&2&3&4&5&6&7&8&9 \\ \hline 
\textrm{D2}&\circ&\circ&-&-&-&-&\circ&-&-&- \\ 
\textrm{NS5}&\circ&\circ&\circ&\circ&\circ&\circ&-&-&-&- \\ 
\widetilde{\textrm{NS5}}&\circ&\circ&\circ&\circ&-&-&-&-&\circ&\circ \\ 
\textrm{D4}&\circ&\circ&-&-&-&-&-&\circ&\circ&\circ \\ 
\widetilde{\textrm{D4}}&\circ&\circ&-&-&\circ&\circ&-&\circ&-&- \\ 
\end{array}
\end{align}
These configurations break space-time symmetry down to $SO(1,1)_{01}\times SO(2)_{23}\times SO(2)_{45}\times SO(2)_{89}$. 
Let $\epsilon_{L}$ and $\epsilon_{R}$ be the supersymmetry parameters of Type IIA string theory 
associated with the left and right moving supercharges $Q_L$ and $Q_R$. 
They are chiral and antichiral 
\begin{align}
\Gamma \epsilon_{L}&=\epsilon_{L},& 
\Gamma \epsilon_{R}&=-\epsilon_{R}
\end{align}
where $\Gamma=\Gamma_{01\cdots 9}$. 
The supercharges preserved in the brane configuration (\ref{22_brane}) satisfy the conditions (with some consistent convention chosen for the signs)
\begin{align}
\label{superch_brane}
\begin{array}{cc}
\textrm{D2}:&\Gamma_{016} \left(\begin{matrix} \epsilon_L\\ \epsilon_R\\ \end{matrix}\right)= \left(\begin{matrix} \epsilon_R\\ \epsilon_L\\ \end{matrix}\right) \\
\textrm{NS5}:& \Gamma_{012345} \left(\begin{matrix} \epsilon_L\\ \epsilon_R\\ \end{matrix}\right) =  \left(\begin{matrix} \epsilon_L\\ \epsilon_R\\ \end{matrix}\right)\\
\widetilde{\textrm{NS5}}:& \Gamma_{012389} \left(\begin{matrix} \epsilon_L\\ \epsilon_R\\ \end{matrix}\right) =  \left(\begin{matrix} \epsilon_L\\ \epsilon_R\\ \end{matrix}\right)\\
\textrm{D4}:&\Gamma_{01789} \left(\begin{matrix} \epsilon_L\\ \epsilon_R\\ \end{matrix}\right)= \left(\begin{matrix} \epsilon_R\\ -\epsilon_L\\ \end{matrix}\right) \\
\widetilde{\textrm{D4}}:&\Gamma_{01457} \left(\begin{matrix} \epsilon_L\\ \epsilon_R\\ \end{matrix}\right)= \left(\begin{matrix} \epsilon_R\\ - \epsilon_L\\ \end{matrix}\right) \\
\end{array}
\end{align}
There are four preserved supercharges obeying the projection conditions (\ref{superch_brane}).

We consider $N_c$ D4-branes suspended between the NS5-brane say at $x^6=$ $x^7=$ $x^8=$ $x^9=0$ 
and $\widetilde{\textrm{NS5}}$-brane at $x^6=L$, $x^4=$ $x^5=0$ 
and $N_f$ D4-branes which have the same $x^6$ position as the $\widetilde{\textrm{NS5}}$-brane 
in the upper-half space $x^7>0$ 
and $N_a$ D4-branes in the lower-half space $x^7<0$. 
Similarly one can also introduce $\widetilde{\textrm{D4}}$-branes which
have the same $x^6$ position as the NS5-brane. 

Since the D2-branes have finite extent along $x^6$, 
the low-energy effective theory on the world-volume of the D2-branes is a two-dimensional field theory 
along $(x^0,x^1)$ preserving 2d $\mathcal{N}=(2,2)$ supersymmetry. 

The D2-D2 strings yield a 2d $\mathcal{N}=(8,8)$ $U(N_c)$ vector multiplet.
However, six of eight scalar fields are frozen due to the boundary condition 
required at the end points of NS5-brane and $\widetilde{\textrm{NS5}}$-brane. 
The surviving two of eight scalar fields describe the motion of the D2-branes along the $(x^2, x^3)$ directions 
and correspond to the complex scalar field $\sigma$ in the vector
multiplet.\footnote{In terms of the string length $l_{\textrm{st}}$, the vector multiplet scalar is given by $\sigma=l_{\textrm{st}}^{-2}(x^3+ix^4)$. }
The vector multiplet scalar $\sigma$ has charge $+2$ under the $U(1)_{23}$. 

The open strings stretched between the D2-branes and upper-half D4-branes 
give rise to the fundamental chiral multiplets $\Phi^i$, $i=1,\cdots, N_f$. 
Similarly, the open strings between the D2-branes and lower-half D4-branes 
lead to the antifundamental chiral multiplets $\widetilde{\Phi}^{j}$, $j=1,\cdots, N_a$. 
These matter multiplets carry charge $+1$ under the $U(1)_{89}$. 

The $U(1)_{23}$ and $U(1)_{45}$ are 
the $U(1)_{A}$ and $U(1)_V$ R-symmetries respectively. 
The $U(1)_{89}$ is the axial $U(1)$ part of the $U(N_f)\times U(N_a)$ symmetry of the D4-branes, 
which breaks down to the $S\left[U(N_f)\times U(N_a)\right]$ flavor symmetry 
since the vector $U(1)$ part of them is gauged. 

When $N_f=N_a$, the D4-branes can combine to form $N_f$ infinite D4-branes 
and the flavor symmetry is broken to $SU(N_f)$. 
For $N_f\ge N_c$ there are $\left( \begin{smallmatrix} N_f\\N_c \end{smallmatrix} \right)$ 
supersymmetric ground states 
and supersymmetry is broken for $N_f< N_c$ as a consequence of the s-rule. 

The parameters in the 2d $\mathcal{N}=(2,2)$ gauge theories are realized as the positions of the D4-branes and NS5-branes. 
The gauge coupling is given by the distance $L$ between the NS5- and $\widetilde{\textrm{NS5}}$-branes along the $x^6$ direction. 
\begin{align}
\label{g_brane}
\frac{1}{g^2}&=\frac{Ll_{\textrm{st}}}{g_{\textrm{st}}}
\end{align}
where $g_{\textrm{st}}$ is the Type IIA string coupling constant. 

The FI parameter for the $U(1)$ factor of the $U(N_c)$ gauge symmetry 
is realized by the $x^7$ position of $\widetilde{\textrm{NS5}}$-brane
\begin{align}
\label{FI_brane}
-\zeta&=\frac{x^7(\widetilde{\textrm{NS5}})}{g_{\textrm{st}} l_{\textrm{st}}}
\end{align}

The mass parameter is given by the positions in the $(x^4, x^5)$ directions of 
joined $i$-th upper-half D4-brane and $j$-th lower-half D4-brane
\begin{align}
\label{m_brane}
m_{ij}&=x^4(\textrm{D4}_{ij})+ix^5(\textrm{D4}_{ij}). 
\end{align}

The twisted mass parameter $\widetilde{m}_f^{(i)}$ for the $i$-th fundamental chiral multiplet 
and the twisted mass parameter $\widetilde{m}_a$ for the $j$-th antifundamental chiral multiplet 
are realized by the $(x^2, x^3)$ positions of the upper-half and lower-half D4-branes respectively
\begin{align}
\label{tm_brane}
\widetilde{m}_f^{(i)}&=\frac{x^2(\textrm{D4}_i)+ix^3(\textrm{D4}_i)}{l_{\textrm{st}}^2},& 
\widetilde{m}_a^{(j)}&=\frac{x^2(\textrm{D4}_j)+ix^3(\textrm{D4}_j)}{l_{\textrm{st}}^2}. 
\end{align}

The theta parameter is most simply described in the M-theory lift,
corresponding to the separation of the M5-brane and $\widetilde{\textrm{M5}}$-brane 
along the $x^{10}$ direction where the M5-brane and $\widetilde{\textrm{M5}}$-brane 
are lifted from the NS5-brane and $\widetilde{\textrm{NS5}}$-brane respectively.
The D4 and $\widetilde{\textrm{D4}}$ brane also lift to M5-branes while the D2-branes
become M2-branes.
These M-theory brane configurations describe M2-branes ending on intersecting
M5-branes.

The classical Coulomb branch of the theory 
corresponds to the $(x^2, x^3)$ positions of the D2-branes 
suspended between NS5- and $\widetilde{\textrm{NS5}}$-branes. 
The classical Higgs branch of the theory 
is parametrized by the $(x^7, x^8, x^9)$ positions of D2-branes 
stretched between the D4-branes.

We note that there is another type of D4-brane, which we call $\widetilde{\textrm{D4}}$-brane. 
When we have both D4-branes and $\widetilde{\textrm{D4}}$-branes, the theory may have superpotential terms \cite{Aharony:1997ju}.

The 2d $\mathcal{N}=(2,2)$ supersymmetric gauge theories 
with orthogonal and symplectic gauge group 
studied in \cite{Hori:2011pd} can be constructed by using orientifold planes \cite{Bergman:2018vqe}. For simplicity we do not include orientifold planes and so
focus only on unitary gauge groups.

%%%%%%%%%%%%%%%%%%%%%%%%%%%%%%%%%%
%%%%%%%%%%%%%%%%%%%%%%%%%%%%%%%%%%
\subsection{A-type boundaries}
%%%%%%%%%%%%%%%%%%%%%%%%%%%%%%%%%%
%%%%%%%%%%%%%%%%%%%%%%%%%%%%%%%%%%
%%%%%%%%%%%%%%%%%%%%%%%%%%%%%%%%%%
\subsubsection{A-type boundary conditions}
%%%%%%%%%%%%%%%%%%%%%%%%%%%%%%%%%%
Now we would like to find the brane construction of the A-type boundary conditions. 
We further introduce the NS5$''$-branes or/and D4$''$-branes to the configuration (\ref{22_brane}): 
\begin{align}
\label{1dN2A_brane}
\begin{array}{c|cccccccccc}
&0&1&2&3&4&5&6&7&8&9 \\ \hline 
\textrm{D2}&\circ&\circ&-&-&-&-&\circ&-&-&- \\ 
\textrm{NS5}&\circ&\circ&\circ&\circ&\circ&\circ&-&-&-&- \\ 
\widetilde{\textrm{NS5}}&\circ&\circ&\circ&\circ&-&-&-&-&\circ&\circ \\ 
\textrm{D4}&\circ&\circ&-&-&-&-&-&\circ&\circ&\circ \\ 
\widetilde{\textrm{D4}}&\circ&\circ&-&-&\circ&\circ&-&\circ&-&- \\ \hline
\textrm{NS5$''$}&\circ&-&\circ&\circ&\circ&-&\circ&-&-&\circ \\ 
\widetilde{\textrm{NS5$''$}}&\circ&-&\circ&\circ&-&\circ&\circ&-&\circ&- \\ 
\textrm{D4$''$}&\circ&-&-&-&-&\circ&\circ&\circ&\circ&- \\ 
\widetilde{\textrm{D4$''$}}&\circ&-&-&-&\circ&-&\circ&\circ&-&\circ \\ 
\end{array}
\end{align}
The supercharges preserved in the brane configuration (\ref{1dN2A_brane}) satisfy the conditions (\ref{superch_brane}) and 
\begin{align}
\label{superch_braneA}
\begin{array}{cc}
\textrm{NS5}'':& \Gamma_{023469} \left(\begin{matrix} \epsilon_L\\ \epsilon_R\\ \end{matrix}\right) =  \left(\begin{matrix} \epsilon_L\\ \epsilon_R\\ \end{matrix}\right)\\
\widetilde{\textrm{NS5}}'':& \Gamma_{023568} \left(\begin{matrix} \epsilon_L\\ \epsilon_R\\ \end{matrix}\right) =  \left(\begin{matrix} \epsilon_L\\ \epsilon_R\\ \end{matrix}\right)\\
\textrm{D4}'':&\Gamma_{05678} \left(\begin{matrix} \epsilon_L\\ \epsilon_R\\ \end{matrix}\right)= \left(\begin{matrix} -\epsilon_R\\ \epsilon_L\\ \end{matrix}\right) \\
\widetilde{\textrm{D4}}'':&\Gamma_{04679} \left(\begin{matrix} \epsilon_L\\ \epsilon_R\\ \end{matrix}\right)= \left(\begin{matrix} -\epsilon_R\\ \epsilon_L\\ \end{matrix}\right) \\
\end{array}
\end{align}
It follows that the configuration (\ref{1dN2A_brane}) preserves two supercharges. 
We identify this with 1d $\mathcal{N}=2_A$ supersymmetry along the $x^0$ direction 
as the continuous R-symmetry is classically broken. 

We consider the configuration with the NS5-brane at $x^6=x^7=x^8=x^9=0$, 
the $\widetilde{\mathrm{NS5}}$ at $x^6=L$, $x^4=x^5=x^7=0$ 
and the D4-branes at $x^6=L$, $x^2=x^3=x^4=x^5=0$ 
so that FI parameters, mass parameters and twisted mass parameters are set to zero.

%%%%%%%%%%%%%%%%%%%%%%%%%%%%%%%%%%
\subsubsection{NS5$''$-brane}
%%%%%%%%%%%%%%%%%%%%%%%%%%%%%%%%%%
When a D2-brane is stretched between the NS5-brane and $\widetilde{\textrm{NS5}}$-brane along the $x^6$ direction, 
it admits the 2d $\mathcal{N}=(2,2)$ $U(1)$ vector multiplet. 
The boundary condition coming from the NS5$''$-brane at $x^1=x^5=x^7=x^8=0$ 
projects out the $A_1$ component of the gauge field 
while the complex scalar field $\sigma$ describing the motion of D2-branes along the $(x^2,x^3)$ directions can fluctuate 
and the $A_0$ component of gauge field survives. 
Thus the NS5$''$-brane provides the A-type Neumann boundary condition for the vector multiplet
\begin{align}
F_{01}&=0, \label{NS5Abc_vm1}\\
\partial_1 \sigma&=0 \label{NS5Abc_vm2},
\end{align}
which correspond to (\ref{AvmNeu1}) and (\ref{AvmNeu2}) for $\alpha=0$. 
However, for the bulk supersymmetry to be unbroken, 
the D4-brane must be introduced. 

When an upper-half D4-brane emanating from the $\widetilde{\textrm{NS5}}$-brane at $x^6=L$ is further added, 
the effective D2-brane theory has a chiral multiplet of charge $+1$ 
which arises from D2-D4 strings. 
As the D4-brane intersects with the $\widetilde{\textrm{NS5}}$-brane at $x^6=L$,
the complex scalar field in the chiral multiplet would correspond to the fluctuations of the D2-brane along the $(x^8,x^9)$ directions. 
The boundary condition arising from the NS5$''$-brane classically fixes the $x^8$ position of the D2-brane. 
This would split the complex scalar fields into real scalar fields obeying the Neumann and Dirichlet boundary conditions 
\begin{align}
\mathrm{Im} \phi&=0, \label{NS5Abc_cm1} \\
\partial_{1}(\mathrm{Re} \phi)&=0, \label{NS5Abc_cm2}
\end{align}
which can be obtained from (\ref{Acm11}) and (\ref{Acm12}) for $\beta=0$ in the A-type boundary condition for the chiral multiplet. 

Thus the NS5$''$-like boundary conditions may naturally correspond to the case of $\alpha=0$ and $\beta=0$, 
equipped with the fermionic boundary conditions
\begin{align}
\gamma^1\lambda&=\overline{\lambda}, \label{NS5Abc_fm} \\
\gamma^1\psi&=\overline{\psi}. 
\end{align}

Similarly, the $\widetilde{\textrm{NS5}}''$-brane at $x^1=x^4=x^7=x^9=0$ may naturally lead to 
the A-type Neumann boundary conditions (\ref{NS5Abc_vm1}) and (\ref{NS5Abc_vm2}) for the $U(1)$ vector multiplet 
and the A-type boundary conditions for the chiral multiplet opposite to the conditions (\ref{NS5Abc_cm1}) and (\ref{NS5Abc_cm2}), 
which are realized when $\alpha=0$ and $\beta=\pi$. 

However, we found from the field theory analysis that the additional conditions 
(\ref{Acm13}) and (\ref{Acm14}) (or (\ref{Acm23}) and (\ref{Acm24})) 
cannot be simply solved together with the Neumann boundary conditions for the vector multiplet. 
At this stage it is not clear to see such an obstruction associated to the NS5$''$-brane 
(or $\widetilde{\textrm{NS5}}''$-brane) from the brane picture.

%%%%%%%%%%%%%%%%%%%%%%%%%%%%%%%%%%
\subsubsection{D4$''$-brane}
\label{A_D4''brane}
%%%%%%%%%%%%%%%%%%%%%%%%%%%%%%%%%%
On the other hand, the D4$''$-brane at $x^1=x^2=x^3=x^4=x^9$ on which the D2-brane ends projects out the $A_0$ component of the gauge field 
and the motion of D2-branes along the $(x^2,x^3)$ directions described by complex scalar field $\sigma$ 
whereas $A_1$ is not frozen. 
Hence the D4$''$-brane yields the A-type Dirichlet boundary conditions for the vector multiplet. 
\begin{align}
A_0&=0, \label{D4Abc_vm1} \\
\sigma&=0, \label{D4Abc_vm2}
\end{align}
which can be found from the conditions (\ref{AvmDir2}) and (\ref{AvmDir3}) for $\alpha=\pi$. 
We note that, unlike for the B-type configurations we will discuss later, this
boundary condition, fixing $\sigma$, excludes any singular boundary behaviour
for the D2-branes. So, for A-type boundary conditions we do not get any Nahm pole like behaviour.

When an upper-half D4-brane coinciding with the $\widetilde{\textrm{NS5}}$-brane is introduced, 
the theory has a charged multiplet coupled to the $U(1)$ vector multiplet 
and the D4$''$-brane then fixes the $x^9$ position of the D2-brane. 
This requires a splitting of the complex scalar field into two real scalar fields obeying the boundary conditions 
\begin{align}
\mathrm{Re} \phi&=0, \label{D4Abc_cm1} \\
D_{1}(\mathrm{Im} \phi)&=0, \label{D4Abc_cm2}
\end{align}
which are different from the boundary conditions (\ref{NS5Abc_cm1}) and (\ref{NS5Abc_cm2}) imposed by the NS5$''$-brane.  
Therefore the D4$''$-brane leads to the A-type boundary condition with $\alpha=\pi$ and $\beta=\pi$. 
Correspondingly the fermionic boundary conditions for the D4$''$-brane are
\begin{align}
\gamma^1\lambda&=-\overline{\lambda}, \label{D4Abc_fm}\\
\gamma^1\psi&=-\overline{\psi}. 
\end{align}

Analogously, the $\widetilde{\textrm{D4}}''$-brane at $x^1=x^2=x^3=x^5=x^8$ also introduces the A-type Dirichlet boundary condition 
for the $U(1)$ vector multiplet and the A-type boundary conditions (\ref{NS5Abc_cm1}) and (\ref{NS5Abc_cm2}) for the chiral multiplets. 
This corresponds to the A-type boundary conditions with $\alpha=\pi$ and $\beta=0$.

In summary, any of these additional four branes,  
NS5$''$, $\widetilde{\textrm{NS5}}''$, D4$''$, $\widetilde{\textrm{D4}}''$, placed at $x^1=0$ 
would naturally correspond to the basic A-type boundary conditions in 2d $\mathcal{N}=(2,2)$ gauge theories. 
The angle $\alpha$ that defines the boundary conditions (\ref{A_fer}) on the gaugino $\lambda$ 
controls a ratio of the 5-branes and the 4-branes for each system. 
The angle $\beta$ that determines the boundary condition (\ref{A_fer}) on the matter fermion $\psi$ 
describes the ratio for two types of 5-brane, NS5$''$- and $\widetilde{\textrm{NS5}}''$-branes 
and that for two types of 4-branes, D4$''$- and $\widetilde{\textrm{D4}}''$-branes.

%%%%%%%%%%%%%%%%%%%%%%%%%%%%%%%%%%
\subsubsection{M-theory configuration}
%%%%%%%%%%%%%%%%%%%%%%%%%%%%%%%%%%
Consider the M-theory lift of the D2-NS5-$\widetilde{\textrm{NS5}}$-NS5$''$ configuration. 
It is again recognized as the M2-M5 system, 
however, unlike the D2-NS5-$\widetilde{\textrm{NS5}}$ system of the 2d $\mathcal{N}=(2,2)$ gauge theories, 
all the types of NS5-branes and D4-branes can become a single wrapped M5-brane. 
This M5-brane fills the $(x^0, x^2, x^3)$ directions and wraps a special Lagrangian submanifold of a Calabi-Yau three-fold 
and the M2-brane has a boundary on this special Lagrangian cycle. 
The M5-brane will lead to a 3d $\mathcal{N}=2$ field theory in the $(x^0,x^2,x^3)$ directions 
while the M2-brane will be understood as a charged particle in the theory. 

In fact the complete A-type system can be realized in M-theory as an M5-brane
wrapped on the product of a holomorphic curve in $\Cb^2$, having complex
coordinates $x^2 + i x^3$ and $x^7 + i x^{10}$, with a special
Lagrangian 3-cycle in $\Cb^3$, having complex coordinates $x^1 + i x^6$,
$x^4 - i x^8$ and $x^5 + i x^9$. In particular, reducing to Type IIA, branes
wrapping $x^7 + i x^{10}$ become D4-branes while those wrapping $x^2 + i x^3$
become NS5-branes.

In general an M5-brane wrapping a special Lagrangian 3-cycle will preserve
one eighth on the supersymmetry, i.e.\ 4 supercharges. We see the four types of
NS5-branes and D4-branes arising from M5-branes wrapping the real 3-cycles in
the $145$, $189$, $469$ and $568$ directions. For the lifts of the NS5-banes we
have the projection conditions
\begin{align}
\Gamma_{023}\Gamma_{145} \epsilon & = \epsilon \\
\Gamma_{023}\Gamma_{189} \epsilon & = \epsilon \\
\Gamma_{023}\Gamma_{568} \epsilon & = \epsilon \\
\Gamma_{023}\Gamma_{469} \epsilon & = \epsilon
\end{align}
but it is easy to see that only 3 conditions are independent, and indeed we can
express the conditions as
\begin{align}
\Gamma_{012345} \epsilon & = \epsilon \\
\Gamma_{4589} \epsilon & = - \epsilon \\
\Gamma_{1468} \epsilon & = - \epsilon \; .
\end{align}

A general holomorphic curve in $\Cb^2$ will preserve half of the supersymmetry
and in this case is described by the additional projection condition
\begin{align}
\Gamma_{237(10)} \epsilon & = - \epsilon
\end{align}
resulting in 2 preserved supercharges. It can be quickly checked that these four
independent conditions imply the projection conditions for an M2-brane in the
$016$ directions
\begin{align}
\Gamma_{016} \epsilon & = \epsilon
\end{align}
so that overall the system we considered indeed preserves 2 supercharges.
Note that because the M5-brane wraps a special Lagrangian 3-cycle where one of
the complex coordinates is $x^1 + i x^6$, an M2-brane will always have a
boundary on (or codimension-one intersection with) the M5-brane.

%%%%%%%%%%%%%%%%%%%%%%%%%%%%%%%%%%
\subsubsection{$\mathcal{N}=2_A$ line operators}
%%%%%%%%%%%%%%%%%%%%%%%%%%%%%%%%%%
Let us further introduce the following additional D2$''$-branes or/and D2$'''$-branes:  
\begin{align}
\label{1dN2A_brane2}
\begin{array}{c|cccccccccc}
&0&1&2&3&4&5&6&7&8&9 \\ \hline 
\textrm{D2}&\circ&\circ&-&-&-&-&\circ&-&-&- \\ 
\textrm{NS5}&\circ&\circ&\circ&\circ&\circ&\circ&-&-&-&- \\ 
\widetilde{\textrm{NS5}}&\circ&\circ&\circ&\circ&-&-&-&-&\circ&\circ \\ 
\textrm{D4}&\circ&\circ&-&-&-&-&-&\circ&\circ&\circ \\ 
\widetilde{\textrm{D4}}&\circ&\circ&-&-&\circ&\circ&-&\circ&-&- \\ \hline
\textrm{NS5$''$}&\circ&-&\circ&\circ&\circ&-&\circ&-&-&\circ \\ 
\widetilde{\textrm{NS5$''$}}&\circ&-&\circ&\circ&-&\circ&\circ&-&\circ&- \\ 
\textrm{D4$''$}&\circ&-&-&-&-&\circ&\circ&\circ&\circ&- \\ 
\widetilde{\textrm{D4$''$}}&\circ&-&-&-&\circ&-&\circ&\circ&-&\circ \\ \hline 
\textrm{D2}''&\circ&-&-&-&\circ&-&-&-&\circ&- \\ 
\widetilde{\textrm{D2}}''&\circ&-&-&-&-&\circ&-&-&-&\circ \\ 
\end{array}
\end{align}
The supercharges preserved in the brane configuration (\ref{1dN2A_brane2}) satisfy the conditions (\ref{superch_brane}), (\ref{superch_braneA}) and 
\begin{align}
\label{superch_braneA2}
\begin{array}{cc}
\textrm{D2}'':&\Gamma_{048} \left(\begin{matrix} \epsilon_L\\ \epsilon_R\\ \end{matrix}\right)= \left(\begin{matrix} -\epsilon_R\\ -\epsilon_L\\ \end{matrix}\right) \\
\widetilde{\textrm{D2}}'':&\Gamma_{059} \left(\begin{matrix} \epsilon_L\\ \epsilon_R\\ \end{matrix}\right)= \left(\begin{matrix} \epsilon_R\\ \epsilon_L\\ \end{matrix}\right) \\
\end{array}
\end{align}
One can check that 
the configuration (\ref{1dN2A_brane2}) preserves 1d $\mathcal{N}=2_A$ supersymmetry along the $x^0$ direction 
without further breaking supersymmetry. 

When the D2$''$-branes or/and $\widetilde{\textrm{D2}}''$-branes at $x^1=0$ are added, 
they would realize domain walls or line operators compatible with the A-type boundary condition in 2d $\mathcal{N}=(2,2)$ gauge theories. 
Although we do not pursue here, 
it would be interesting to figure out the field theory interpretation. 
We also note that the M-theory lift could in general describe a single M2-brane wrapping
a holomorphic cycle in $\Cb^3$.

%%%%%%%%%%%%%%%%%%%%%%%%%%%%%%%%%%
\subsection{B-type boundaries}
%%%%%%%%%%%%%%%%%%%%%%%%%%%%%%%%%%

%%%%%%%%%%%%%%%%%%%%%%%%%%%%%%%%%%
\subsubsection{B-type boundary conditions}
%%%%%%%%%%%%%%%%%%%%%%%%%%%%%%%%%%
Now let us consider the brane construction of the B-type boundary conditions. 
We take another set of additional NS5$'$-branes or/and D4$'$-branes as follows: 
\begin{align}
\label{1dN2B_brane}
\begin{array}{c|cccccccccc}
&0&1&2&3&4&5&6&7&8&9 \\ \hline 
\textrm{D2}&\circ&\circ&-&-&-&-&\circ&-&-&- \\ 
\textrm{NS5}&\circ&\circ&\circ&\circ&\circ&\circ&-&-&-&- \\ 
\widetilde{\textrm{NS5}}&\circ&\circ&\circ&\circ&-&-&-&-&\circ&\circ \\ 
\textrm{D4}&\circ&\circ&-&-&-&-&-&\circ&\circ&\circ \\ 
\widetilde{\textrm{D4}}&\circ&\circ&-&-&\circ&\circ&-&\circ&-&- \\ \hline
\textrm{NS5$'$}&\circ&-&\circ&-&-&-&\circ&\circ&\circ&\circ \\ 
\widetilde{\textrm{NS5$'$}}&\circ&-&\circ&-&\circ&\circ&\circ&\circ&-&- \\ 
\textrm{D4$'$}&\circ&-&-&\circ&\circ&\circ&\circ&-&-&- \\ 
\widetilde{\textrm{D4$'$}}&\circ&-&-&\circ&-&-&\circ&-&\circ&\circ \\ 
\end{array}
\end{align}
The supercharges preserved in the brane configuration (\ref{1dN2B_brane}) satisfy the conditions (\ref{superch_brane}) and 
\begin{align}
\label{superch_braneB}
\begin{array}{cc}
\textrm{NS5}':& \Gamma_{026789} \left(\begin{matrix} \epsilon_L\\ \epsilon_R\\ \end{matrix}\right) =  \left(\begin{matrix} \epsilon_L\\ \epsilon_R\\ \end{matrix}\right)\\
\widetilde{\textrm{NS5}}':& \Gamma_{024567} \left(\begin{matrix} \epsilon_L\\ \epsilon_R\\ \end{matrix}\right) =  \left(\begin{matrix} \epsilon_L\\ \epsilon_R\\ \end{matrix}\right)\\
\textrm{D4}':&\Gamma_{03456} \left(\begin{matrix} \epsilon_L\\ \epsilon_R\\ \end{matrix}\right)= \left(\begin{matrix} -\epsilon_R\\ \epsilon_L\\ \end{matrix}\right) \\
\widetilde{\textrm{D4}}':&\Gamma_{03689} \left(\begin{matrix} \epsilon_L\\ \epsilon_R\\ \end{matrix}\right)= \left(\begin{matrix} -\epsilon_R\\ \epsilon_L\\ \end{matrix}\right) \\
\end{array}
\end{align}
It can be checked that 
the configuration (\ref{1dN2B_brane}) preserves two supercharges. 
In this case the space-time symmetry is classically broken to $SO(2)_{45}\times SO(2)_{89}$. 
We identify the corresponding supersymmetry with 1d $\mathcal{N}=2_B$ supersymmetry 
along $x^0$. 

Again we take the configuration with the NS5-brane at $x^6=x^7=x^8=x^9=0$, 
the $\widetilde{\mathrm{NS5}}$ at $x^6=L$, $x^4=x^5=x^7=0$ 
and the D4-branes at $x^6=L$, $x^2=x^3=x^4=x^5=0$ 
so that FI parameters, mass parameters and twisted mass parameters are turned off.

%%%%%%%%%%%%%%%%%%%%%%%%%%%%%%%%%%
\subsubsection{NS5$'$-brane}
%%%%%%%%%%%%%%%%%%%%%%%%%%%%%%%%%%
For the theory of a 2d $\mathcal{N}=(2,2)$ $U(1)$ vector multiplet 
arising from a D2-brane suspended between the NS5-brane and $\widetilde{\textrm{NS5}}$-brane along the $x^6$ direction, 
the NS5$'$-brane at $x^1=x^3=x^4=x^5=0$ on which the D2-brane terminates fixes the motion of the D2-brane along the $x^3$ direction and the $A_1$ component of gauge field, 
while the motion of the D2-brane along $x^2$ is not fixed. 
Accordingly, the NS5$'$-brane realizes the B-type Neumann boundary conditions for the $U(1)$ vector multiplet
\begin{align}
F_{01}&=0, \label{NS5Bbc_vm1} \\
\partial_{1} \sigma_+&=0, \label{NS5Bbc_vm2}\\
\sigma_-&=0, \label{NS5Bbc_vm3}
\end{align}
where $\sigma_+$ and $\sigma_-$ correspond to the $x^2$ and $x^3$ positions of the D2-brane. 
These can be found from the equations (\ref{BvmNeu1}), (\ref{BvmNeu2}) and (\ref{BvmNeu3}) for $\alpha=0$. 

In the presence of an upper-half D4-brane, 
which leads to a charged chiral multiplet, 
when the D2-brane ends on the NS5$'$-brane, 
the motion of the D2-brane along the $(x^8,x^9)$ directions can still fluctuate. 
This corresponds to the B-type Neumann boundary conditions for the chiral multiplet
\begin{align}
\partial_1&\phi=0, \label{NS5Bbc_cm1}\\
\partial_1&\phi^{\dagger}=0, \label{NS5Bbc_cm2}
\end{align}
which are the conditions (\ref{BcmNeu1}) and (\ref{BcmNeu2}) obtained for $\beta=\pi$. 
One can then identify the NS5$'$-like boundary conditions with the B-type boundary conditions 
with $\alpha=0$ and $\beta=\pi$. 
Whereas these basic boundary conditions will correspond to the NS5$'$-brane at $x^1=x^3=x^4=x^5=0$, 
it would be interesting to shift the transverse positions of the NS5$'$-brane to find the generalized boundary conditions, 
as we argued for the deformation of the Neumann boundary conditions for the vector multiplet 
with a singular profile of $\sigma_-$ in the field theory analysis in section \ref{sec_BNeupole}.

%, which imposes the fermionic boundary conditions
%\begin{align}
%\gamma^1\lambda&=\lambda, \label{NS5Bbc.fm}\\
%\gamma^1\psi&=\psi. 
%\end{align}

On the other hand, the B-type Neumann boundary conditions (\ref{NS5Bbc_vm1}), (\ref{NS5Bbc_vm2}) and (\ref{NS5Bbc_vm3}) 
for the $U(1)$ vector multiplet can be also given by the $\widetilde{\textrm{NS5}}'$-brane at $x^1=x^3=x^8=x^9=0$. 
However, in this case the B-type boundary conditions for the chiral multiplets are the Dirichlet boundary conditions, 
rather than the Neumann boundary conditions (\ref{NS5Bbc_cm1}) and (\ref{NS5Bbc_cm2}). 
Thus the $\widetilde{\textrm{NS5}}'$-brane would lead to the B-type boundary conditions with $\alpha=0$ and $\beta=0$.

%%%%%%%%%%%%%%%%%%%%%%%%%%%%%%%%%%
\subsubsection{D4$'$-brane}
%%%%%%%%%%%%%%%%%%%%%%%%%%%%%%%%%%
When the D2-brane terminates on a D4$'$-brane at $x^1=x^2=x^7=x^8=x^9=0$, 
the motion of the D2-brane along the $x^2$ direction and the $A_0$ component of gauge field are frozen, 
while the D2-brane can still free move along the $x^3$ direction. 
Thus the D4$'$-brane realizes the B-type Dirichlet boundary conditions for the $U(1)$ vector multiplet
\begin{align}
A_0&=0, \label{D4Bbc_vm1}\\
\sigma_+&=0, \label{D4Bbc_vm2}\\
D_{1} \sigma_-&=0, \label{D4Bbc_vm3}
\end{align}
which we found in the equations (\ref{BvmDir1}), (\ref{BvmDir2}) for $\alpha=\pi$. 

In the presence of an upper-half D4-brane, the effective theory has a charged chiral multiplet. 
When the D2-brane ends on the D4$'$-brane, the fluctuations of the D2-brane along the $(x^8,x^9)$ directions are projected out. 
This gives rise to the B-type Dirichlet boundary conditions for the chiral multiplet 
\begin{align}
\phi&=0,\label{D4Bbc_cm1}\\
\phi^{\dag}&=0, \label{D4Bbc_cm2}
\end{align}
which can be obtained from the equations (\ref{BcmDir1}) and (\ref{BcmDir2}) for $\beta=0$. 
Hence the D4$'$-like boundary condition is identified with the B-type boundary condition 
with $\alpha=\pi$ and $\beta=0$. 
%, which imposes the fermionic boundary conditions
%\begin{align}
%\gamma^1\lambda&=-\lambda,\label{D4Bbc_fm}\\
%\gamma^1\psi&=\psi.
%\end{align}

So far, we have argued that the NS5$'$-branes, the $\widetilde{\textrm{NS5}}'$ and the D4$'$-branes introduced at $x^1=0$ 
can produce the basic B-type boundary conditions in 2d $\mathcal{N}=(2,2)$ gauge theories 
in such a way that the phase $\alpha$ in the boundary conditions (\ref{B_fer2}) on the gaugino $\lambda$ 
determines a ratio of 5-branes and the 4-branes, 
while the phase $\beta$ in the boundary conditions (\ref{B_fer3}) on the matter fermion $\psi$ 
encodes the ratio for two types of 5-branes and that for two types of 4-branes. 

Although one may expect that a $\widetilde{\textrm{D4}}'$-brane at $x^1=x^2=x^4=x^5=x^7=0$ also realizes the B-type Dirichlet boundary conditions 
(\ref{D4Bbc_vm1}), (\ref{D4Bbc_vm2}) and (\ref{D4Bbc_vm3}) for the $U(1)$ vector multiplet, 
as well as the B-type Neumann-type boundary conditions for the chiral multiplet, 
it does not seem to be the case 
because we found from the field theory analysis that such a combination of the basic boundary conditions 
for vector and chiral multiplets cannot be simply solved 
due to the obstruction from the conditions (\ref{BvmDir3}), (\ref{BcmNeu3}) and (\ref{BcmNeu4}).
Instead, we found in section \ref{sec_BDirpole} 
that the corresponding B-type boundary conditions are generalized so that 
they allow for the consistent singular solutions 
where $\sigma_-$ and $\phi$ have a simple pole at the boundary. 
Indeed this might be expected from the brane configuration as the
D2-$\widetilde{\mathrm{D4}}'$ branes are T-dual to the familiar D1-D3 system described by the
Nahm equation \cite{Diaconescu:1996rk, Tsimpis:1998zh}.

\begin{figure}
\begin{center}
\includegraphics[width=6.5cm]{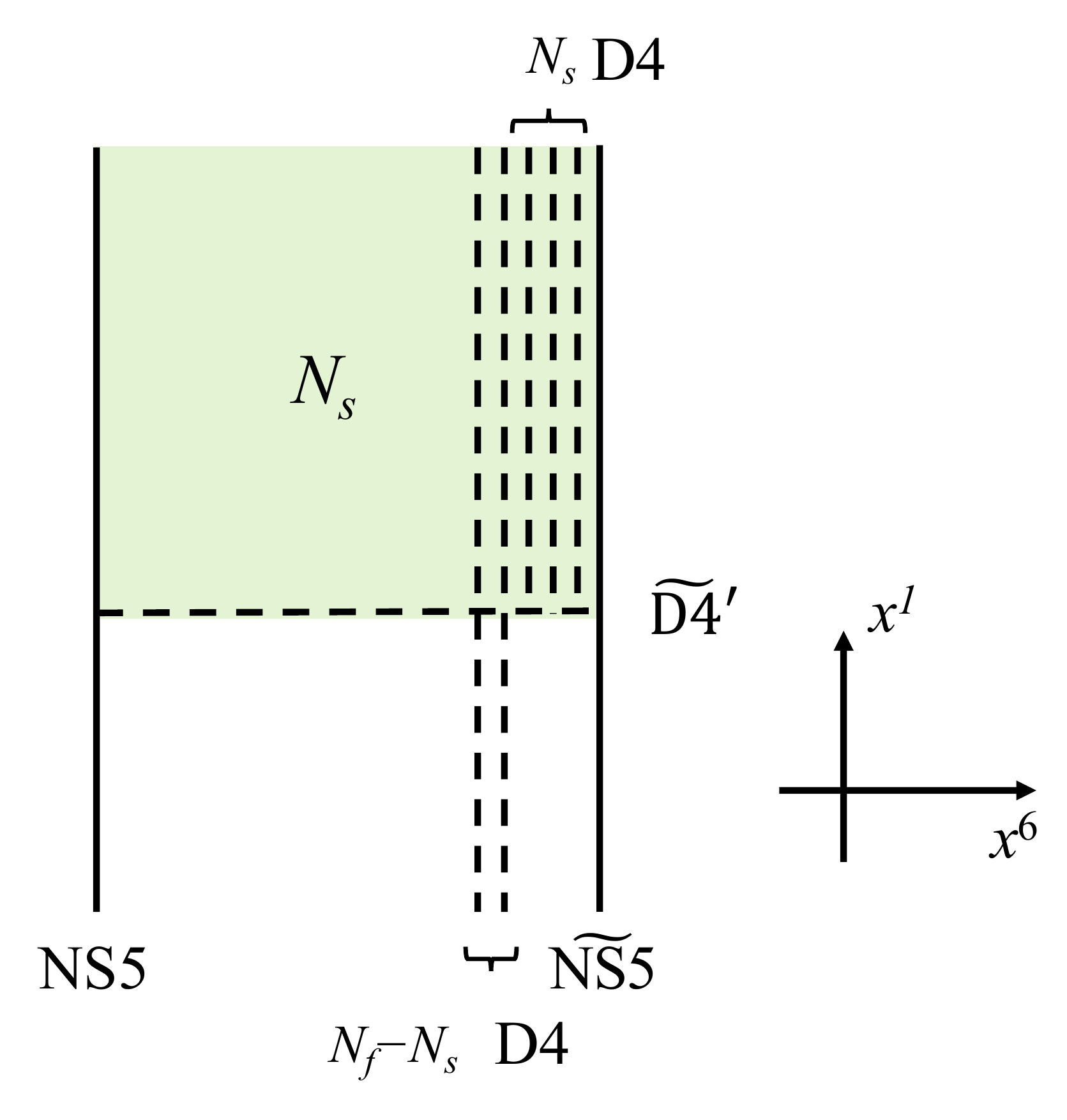}
\caption{
The B-type Dirichlet boundary conditions for the $U(N_s)$ vector multiplet coupled 
to $N_f$ chiral multiplets with singularity 
realized by the $\widetilde{\textrm{D4}}'$-brane at $x^1=0$ on which 
$N_s$ D2-branes end 
and $N_s$ of $N_f$ D4-branes end. 
}
\label{figpole}
\end{center}
\end{figure}

As the residue $\mathfrak{t}$ at the pole (\ref{Bsol1}) for $\sigma_-$ 
typically breaks the boundary global symmetry $G_{\partial}=U(N_c)$ down to $U(N_c-N_s)$, 
this will correspond to the brane configuration 
where $N_s$ of the $N_c$ D2-branes end on a single $\widetilde{\textrm{D4}}'$-brane 
in a similar manner as the regular Nahm pole boundary condition of rank $r$ 
realized by the D5-brane on which $r$ D3-branes end \cite{Gaiotto:2008sa}. 
In fact, when we rotate the $\widetilde{\textrm{NS5}}$-brane to be parallel to the NS5-brane 
and T-dualize the system along $x^2$, the D2-branes ending on the $\widetilde{\textrm{D4}}'$-brane 
become the D3-branes ending on the D5-brane. 
However, in our case the fundamental scalar field $\phi$ arising from the D2-D4 strings 
also contains a pole (\ref{BDD_NA_phi}) 
whose residue $\mathfrak{c}$ satisfies the relation $\mathfrak{t}=-\mathfrak{c}\mathfrak{c}^{\dag}$, 
which may also break the $U(N_f)$ flavor symmetry down to $U(N_f-N_s)$ 
as we discussed in section \ref{sec_BDirpole}.  
In the brane configuration, such a symmetry breaking may occur 
when some of $N_f$ flavor D4-branes terminate on the single $\widetilde{\textrm{D4}}'$-brane at $x^1=0$. 
On the other hand, we argued for 
the regular terms appearing in the solutions.
However, it is not clear from the brane configuration. 
We illustrate the case with a maximal rank of the pole 
with $N_c=N_s$ where all the D2-branes and $N_s$ D4-branes terminate on the 
$\widetilde{\textrm{D4}}'$-brane in Figure \ref{figpole}.

%
%
%
%
%

%%%%%%%%%%%%%%%%%%%%%%%%%%%%%%%%%%
\subsubsection{M-theory configuration}
%%%%%%%%%%%%%%%%%%%%%%%%%%%%%%%%%%
As for the A-type brane configuration, we can lift the IIA configuration to
M-theory. Interestingly, we then see that in eleven dimensions the A-type and B-type brane
configurations are equivalent, i.e.\ they can both be described as an M5-brane
wrapping the product of a holomorphic curve in $\Cb^2$ with a special
Lagrangian 3-cycle in $\Cb^3$. The only difference is our identification of the
coordinates, and in particular when reducing to ten dimensions with the
M-theory circle in the $\Cb^2$ we get A-type while if it is in the $\Cb^3$
we get B-type.

Specifically, for A-type we saw that the complex coordinates in $\Cb^2$ were
$x^2 + i x^3$ and $x^7 + i x^{10}$, while for B-type we have
$x^4 + i x^5$ and $x^8 + i x^9$. For the $\Cb^3$ we had coordinates
$x^1 + i x^6$, $x^4 - i x^8$ and $x^5 + i x^9$ for A-type while for B-type we
have instead $x^1 + i x^6$, $x^2 + i x^{10}$ and $x^3 + i x^7$. In both cases
we preserve 2 supercharges from the wrapped M5-brane, the M2-branes
spanning $016$ always have a boundary on the M5-brane since $x^1 + i x^6$ is a
complex coordinate in $\Cb^3$ in both cases, and the M2-branes do not break any
further supersymmetry. For B-type one set of four independent projection
conditions is
\begin{align}
\Gamma_{012345} \epsilon & = \epsilon \\
\Gamma_{237(10)} \epsilon & = - \epsilon \\
\Gamma_{1367} \epsilon & = \epsilon \\
\Gamma_{4589} \epsilon & = - \epsilon \; .
\end{align}

In summary, the A-type and B-type boundary conditions can be realized in M-theory setup as follows: 
\begin{align}
\label{Mthy_setup}
\begin{matrix}
\textrm{space-time:}&\mathbb{R}&\times&\mathbb{C}^2&\times&\mathbb{C}^3 \\
&&&\cup&&\cup \\
\textrm{M5:}&\mathbb{R}&\times&\Sigma&\times&M_3 \\
\textrm{M2:}&\mathbb{R}&\times&&&C \\
\end{matrix}
\end{align}
where the M5-brane is wrapped on 
the special Lagrangian 3-cycle $M_3$ in $\mathbb{C}^3$ 
and the holomorphic 2-cycle $\Sigma$ in $\mathbb{C}^2$ 
while the M2-branes wrap
the holomorphic 2-cycle $C$ in $\mathbb{C}^3$ whose boundary is in $M_3$. 

It may be interesting to explore the supergravity description of such M-brane
configurations. We are not aware of any such solutions but solutions for the
M5-brane wrapping $\Sigma \times M_3$ have been described \cite{Gauntlett:2001jj}. It may be possible to understand some aspects of the field theory using M-brane probes in these backgrounds, such as calculating central charges similar to
\cite{SanchezLoureda:2006ifg}. It would be particularly interesting to investigate supergravity solutions including M2-branes giving rise to $AdS_2$ geometry
as duals of superconformal QM. Examples without M2-branes were found in
\cite{Gauntlett:2002rv, Gauntlett:2001jj}.

%%%%%%%%%%%%%%%%%%%%%%%%%%%%%%%%%%
\subsubsection{$\mathcal{N}=2_B$ line operators}
%%%%%%%%%%%%%%%%%%%%%%%%%%%%%%%%%%
We also note that there are other objects preserving $\mathcal{N}=2_B$ supersymmetry along $x^0$. 
Let us further introduce the fundamental strings (F1) and D2$'$-branes: 
\begin{align}
\label{1dN2B_brane2}
\begin{array}{c|cccccccccc}
&0&1&2&3&4&5&6&7&8&9 \\ \hline 
\textrm{D2}&\circ&\circ&-&-&-&-&\circ&-&-&- \\ 
\textrm{NS5}&\circ&\circ&\circ&\circ&\circ&\circ&-&-&-&- \\ 
\widetilde{\textrm{NS5}}&\circ&\circ&\circ&\circ&-&-&-&-&\circ&\circ \\ 
\textrm{D4}&\circ&\circ&-&-&-&-&-&\circ&\circ&\circ \\ 
\widetilde{\textrm{D4}}&\circ&\circ&-&-&\circ&\circ&-&\circ&-&- \\ \hline
\textrm{NS5$'$}&\circ&-&\circ&-&-&-&\circ&\circ&\circ&\circ \\ 
\widetilde{\textrm{NS5$'$}}&\circ&-&\circ&-&\circ&\circ&\circ&\circ&-&- \\ 
\textrm{D4$'$}&\circ&-&-&\circ&\circ&\circ&\circ&-&-&- \\ 
\widetilde{\textrm{D4$'$}}&\circ&-&-&\circ&-&-&\circ&-&\circ&\circ \\ 
\textrm{F1}&\circ&-&\circ&-&-&-&-&-&-&- \\ 
\textrm{D2}'&\circ&-&-&\circ&-&-&-&\circ&-&- \\ 
\end{array}
\end{align}
The supercharges preserved in the brane configuration (\ref{1dN2B_brane2}) satisfy the additional conditions (\ref{superch_brane}), (\ref{superch_braneB}) and 
\begin{align}
\label{superch_braneB2}
\begin{array}{cc}
\textrm{F1}:& \Gamma_{02} \left(\begin{matrix} \epsilon_L\\ \epsilon_R\\ \end{matrix}\right) =  \left(\begin{matrix} \epsilon_L\\ -\epsilon_R\\ \end{matrix}\right)\\
\textrm{D2}':&\Gamma_{037} \left(\begin{matrix} \epsilon_L\\ \epsilon_R\\ \end{matrix}\right)= \left(\begin{matrix} \epsilon_R\\ \epsilon_L\\ \end{matrix}\right) \\
\end{array}
\end{align}
The configuration (\ref{1dN2B_brane2}) preserves 1d $\mathcal{N}=2_B$ supersymmetry along the $x^0$ direction 
without further breaking supersymmetry. 

The fundamental strings along the $(x^0,x^2)$ directions or/and D2$'$-branes 
would realize domain walls or line operators supported along the boundary 
which are compatible with the B-type boundary condition in 2d $\mathcal{N}=(2,2)$ gauge theories. In the M-theory lift, the F1 and D2' become M2-branes, and
as for the A-type case, we can have a single M2-brane wrapping a holomorphic cycle in $\Cb^3$.

%%%%%%%%%%%%%%%%%%%%%%%%%%%%%%%%%%
\subsubsection{Dualities}
%%%%%%%%%%%%%%%%%%%%%%%%%%%%%%%%%%
As we commented above, the A-type and B-type configurations are equivalent when lifted to M-theory, the only difference being the choice of direction to compactify on to reduce to Type IIA string theory, along with a relabelling of some
of the coordinates. One choice of mapping of coordinates is
\begin{align}
\label{AB_coords}
\begin{array}{c|ccccccccccc}
\mathrm{A-type} & 0 & 1 & 2 & 3 & 4 & 5 & 6 & 7 & 8 & 9 & 10 \\
\mathrm{B-type} & 0 & 1 & 4 & 5 & 2 & 3 & 6 & 8 & 10 & 7 & 9
\end{array}
\end{align}
which results in the following mapping of branes
\begin{align}
\label{AB_branemap}
\begin{array}{c|cccccccccc}
\mathrm{A-type} & \mathrm{NS5} & \widetilde{\mathrm{NS5}} & \mathrm{D4} & \widetilde{\mathrm{D4}} & \mathrm{NS5''} & \widetilde{\mathrm{NS5''}} & \mathrm{D4''} & \widetilde{\mathrm{D4''}} & \mathrm{D2''} & \widetilde{\mathrm{D2''}} \\
\mathrm{B-type} & \mathrm{NS5} & \widetilde{\mathrm{D4}} & \mathrm{D4} & \widetilde{\mathrm{NS5}} & \widetilde{\mathrm{NS5'}} & \mathrm{D4'} & \widetilde{\mathrm{D4'}} & \mathrm{NS5'} & \mathrm{F1} & \mathrm{D2'}
\end{array}
\end{align}
Now we also have several mappings of the coordinates which preserve the A-type or B-type configurations. We consider only those which map D2-branes to D2-branes and do not result in new orientations of branes we have not considered. The mappings which satisfy this condition and have a non-trivial effect on some
branes are for A-type any combination of
\begin{itemize}
\item $x^2 \leftrightarrow x^7$ and $x^3 \leftrightarrow x^{10}$
\item $x^4 \leftrightarrow x^5$ and $x^8 \leftrightarrow x^9$
\item $x^4 \leftrightarrow x^8$ and $x^5 \leftrightarrow x^9$
\end{itemize}
and for B-type any combination of
\begin{itemize}
\item $x^4 \leftrightarrow x^8$ and $x^5 \leftrightarrow x^9$ 
\item $x^2 \leftrightarrow x^3$ and $x^7 \leftrightarrow x^{10}$
\item $x^2 \leftrightarrow x^{10}$ and $x^3 \leftrightarrow x^7$ .
\end{itemize}
This gives the following mapping of branes
\begin{align}
\label{AB_fullbranemap}
\begin{array}{c|c|cccccccccc}
\mathrm{A0} & \mathrm{A-type} & \mathrm{NS5} & \widetilde{\mathrm{NS5}} & \mathrm{D4} & \widetilde{\mathrm{D4}} & \mathrm{NS5''} & \widetilde{\mathrm{NS5''}} & \mathrm{D4''} & \widetilde{\mathrm{D4''}} & \mathrm{D2''} & \widetilde{\mathrm{D2''}} \\
\mathrm{A1} & \mathrm{A-type} & \widetilde{\mathrm{D4}} & \mathrm{D4} & \widetilde{\mathrm{NS5}} & \mathrm{NS5} & \widetilde{\mathrm{D4''}} & \mathrm{D4''} & \widetilde{\mathrm{NS5''}} & \mathrm{NS5''} & \mathrm{D2''} & \widetilde{\mathrm{D2''}} \\
\mathrm{A2} & \mathrm{A-type} & \mathrm{NS5} & \widetilde{\mathrm{NS5}} & \mathrm{D4} & \widetilde{\mathrm{D4}} & \widetilde{\mathrm{NS5''}} & \mathrm{NS5''} & \widetilde{\mathrm{D4''}} & \mathrm{D4''} & \widetilde{\mathrm{D2''}} & \mathrm{D2''} \\
\mathrm{A3} & \mathrm{A-type} & \widetilde{\mathrm{D4}} & \mathrm{D4} & \widetilde{\mathrm{NS5}} & \mathrm{NS5} & \mathrm{D4''} & \widetilde{\mathrm{D4''}} & \mathrm{NS5''} & \widetilde{\mathrm{NS5''}} & \widetilde{\mathrm{D2''}} & \mathrm{D2''} \\
\mathrm{A4} & \mathrm{A-type} & \widetilde{\mathrm{NS5}} & \mathrm{NS5} & \widetilde{\mathrm{D4}} & \mathrm{D4} & \widetilde{\mathrm{NS5''}} & \mathrm{NS5''} & \widetilde{\mathrm{D4''}} & \mathrm{D4''} & \mathrm{D2''} & \widetilde{\mathrm{D2''}} \\
\mathrm{A5} & \mathrm{A-type} & \mathrm{D4} & \widetilde{\mathrm{D4}} & \mathrm{NS5} & \widetilde{\mathrm{NS5}} & \mathrm{D4''} & \widetilde{\mathrm{D4''}} & \mathrm{NS5''} & \widetilde{\mathrm{NS5''}} & \mathrm{D2''} & \widetilde{\mathrm{D2''}} \\
\mathrm{A6} & \mathrm{A-type} & \widetilde{\mathrm{NS5}} & \mathrm{NS5} & \widetilde{\mathrm{D4}} & \mathrm{D4} & \mathrm{NS5''} & \widetilde{\mathrm{NS5''}} & \mathrm{D4''} & \widetilde{\mathrm{D4''}} & \widetilde{\mathrm{D2''}} & \mathrm{D2''} \\
\mathrm{A7} & \mathrm{A-type} & \mathrm{D4} & \widetilde{\mathrm{D4}} & \mathrm{NS5} & \widetilde{\mathrm{NS5}} & \widetilde{\mathrm{D4''}} & \mathrm{D4''} & \widetilde{\mathrm{NS5''}} & \mathrm{NS5''} & \widetilde{\mathrm{D2''}} & \mathrm{D2''} \\
\mathrm{B0} & \mathrm{B-type} & \mathrm{NS5} & \widetilde{\mathrm{D4}} & \mathrm{D4} & \widetilde{\mathrm{NS5}} & \widetilde{\mathrm{NS5'}} & \mathrm{D4'} & \widetilde{\mathrm{D4'}} & \mathrm{NS5'} & \mathrm{F1} & \mathrm{D2'} \\
\mathrm{B1} & \mathrm{B-type} & \widetilde{\mathrm{NS5}} & \mathrm{D4} & \widetilde{\mathrm{D4}} & \mathrm{NS5} & \mathrm{NS5'} & \widetilde{\mathrm{D4'}} & \mathrm{D4'} & \widetilde{\mathrm{NS5'}} & \mathrm{F1} & \mathrm{D2'} \\
\mathrm{B2} & \mathrm{B-type} & \mathrm{NS5} & \widetilde{\mathrm{D4}} & \mathrm{D4} & \widetilde{\mathrm{NS5}} & \mathrm{D4'} & \widetilde{\mathrm{NS5'}} & \mathrm{NS5'} & \widetilde{\mathrm{D4'}} & \mathrm{D2'} & \mathrm{F1} \\
\mathrm{B3} & \mathrm{B-type} & \widetilde{\mathrm{NS5}} & \mathrm{D4} & \widetilde{\mathrm{D4}} & \mathrm{NS5} & \widetilde{\mathrm{D4'}} & \mathrm{NS5'} & \widetilde{\mathrm{NS5'}} & \mathrm{D4'} & \mathrm{D2'} & \mathrm{F1} \\
\mathrm{B4} & \mathrm{B-type} & \widetilde{\mathrm{D4}} & \mathrm{NS5} & \widetilde{\mathrm{NS5}} & \mathrm{D4} & \mathrm{D4'} & \widetilde{\mathrm{NS5'}} & \mathrm{NS5'} & \widetilde{\mathrm{D4'}} & \mathrm{F1} & \mathrm{D2'} \\
\mathrm{B5} & \mathrm{B-type} & \mathrm{D4} & \widetilde{\mathrm{NS5}} & \mathrm{NS5} & \widetilde{\mathrm{D4}} & \widetilde{\mathrm{D4'}} & \mathrm{NS5'} & \widetilde{\mathrm{NS5'}} & \mathrm{D4'} & \mathrm{F1} & \mathrm{D2'} \\
\mathrm{B6} & \mathrm{B-type} & \widetilde{\mathrm{D4}} & \mathrm{NS5} & \widetilde{\mathrm{NS5}} & \mathrm{D4} & \widetilde{\mathrm{NS5'}} & \mathrm{D4'} & \widetilde{\mathrm{D4'}} & \mathrm{NS5'} & \mathrm{D2'} & \mathrm{F1} \\
\mathrm{B7} & \mathrm{B-type} & \mathrm{D4} & \widetilde{\mathrm{NS5}} & \mathrm{NS5} & \widetilde{\mathrm{D4}} & \mathrm{NS5'} & \widetilde{\mathrm{D4'}} & \mathrm{D4'} & \widetilde{\mathrm{NS5'}} & \mathrm{D2'} & \mathrm{F1}
\end{array}
\end{align}

Another possible duality is to T-dualize to Type IIB along $x^2$, then perform
S-duality before T-dualizing back to Type IIA, again along $x^2$. This would
result in new orientations of branes, but if we then exchange
$x^3 \leftrightarrow x^7$
we get the same type of branes back.
However, this is already included in the above mappings via M-theory, e.g.\
mapping A0 to A1 or B0 to B4. It is
worth noting that this TST duality does not exchange A-type with B-type, but as we see above this is possible via M-theory.

There will also be interesting Seiberg-like dualities arising from Hanany-Witten brane
rearrangements.
It would be interesting to explore some of these dualities and find interpretations in the field theory, but we leave that for future work. In particular we
expect to find dualities of boundary conditions related to 2d mirror symmetry
and through T-duality this should be closely related to 3d mirror symmetry
\cite{Aganagic:2001uw, Chen:2013pha, Aharony:2017adm, Karch:2018mer, Jockers:2018sfl}.

%%%%%%%%%%%%%%%%%%%%%%%%%%%%%%%%%%
%%%%%%%%%%%%%%%%%%%%%%%%%%%%%%%%%%
\subsection{Quarter-BPS boundaries}
%%%%%%%%%%%%%%%%%%%%%%%%%%%%%%%%%%
%%%%%%%%%%%%%%%%%%%%%%%%%%%%%%%%%%

%%%%%%%%%%%%%%%%%%%%%%%%%%%%%%%%%%
\subsubsection{$\mathcal{N}=1$ quarter BPS boundary conditions}
%%%%%%%%%%%%%%%%%%%%%%%%%%%%%%%%%%
When we consider the configuration in which the NS5$''$- and D4$''$-branes 
and the NS5$'$- and D4$'$-branes exist
\begin{align}
\label{1dN1_brane}
\begin{array}{c|cccccccccc}
&0&1&2&3&4&5&6&7&8&9 \\ \hline 
\textrm{D2}&\circ&\circ&-&-&-&-&\circ&-&-&- \\ 
\textrm{NS5}&\circ&\circ&\circ&\circ&\circ&\circ&-&-&-&- \\ 
\widetilde{\textrm{NS5}}&\circ&\circ&\circ&\circ&-&-&-&-&\circ&\circ \\ 
\textrm{D4}&\circ&\circ&-&-&-&-&-&\circ&\circ&\circ \\ 
\widetilde{\textrm{D4}}&\circ&\circ&-&-&\circ&\circ&-&\circ&-&- \\ \hline
\textrm{NS5$'$}&\circ&-&\circ&-&-&-&\circ&\circ&\circ&\circ \\ 
\widetilde{\textrm{NS5$'$}}&\circ&-&\circ&-&\circ&\circ&\circ&\circ&-&- \\ 
\textrm{NS5$''$}&\circ&-&\circ&\circ&\circ&-&\circ&-&-&\circ \\ 
\widetilde{\textrm{NS5$''$}}&\circ&-&\circ&\circ&-&\circ&\circ&-&\circ&- \\ 
\textrm{D4$'$}&\circ&-&-&\circ&\circ&\circ&\circ&-&-&- \\ 
\widetilde{\textrm{D4$'$}}&\circ&-&-&\circ&-&-&\circ&-&\circ&\circ \\ 
\textrm{D4$''$}&\circ&-&-&-&-&\circ&\circ&\circ&\circ&- \\ 
\widetilde{\textrm{D4$''$}}&\circ&-&-&-&\circ&-&\circ&\circ&-&\circ \\ 
\end{array}
\end{align}
one can check that there remains $\mathcal{N}=1$ supersymmetry. 
In addition, one can also introduce the fundamental strings and three kinds of D2-branes in 
the configurations (\ref{1dN2A_brane2}) and (\ref{1dN2B_brane2}), 
corresponding to the line operators keeping $\mathcal{N}=1$ supersymmetry. 

According to the identification of the parameters in the half-BPS boundary conditions, 
we can identify the phases $\alpha_1$, $\alpha_2$, $\beta_1$ and $\beta_2$ 
introduced in the field theory analysis with parameters which 
characterize the different types of boundary branes. 
The angle parameter $\alpha_1$ distinguishes two kinds of branes for the B-type 
so that the case with $\alpha_1=0$ (resp.\ $\pi$) describes the 4-brane (resp.\ 5-brane). 
The parameter $\alpha_2-\alpha_1$ distinguishes the two kinds of branes for the A-type 
so that one finds the same kinds of branes in A-type for $\alpha_2-\alpha_1$ $=0$ $\mod \pi$ 
and the opposite kinds of branes in A-type for $\alpha_2-\alpha_1=\pi/2$ $\mod \pi$. 
The angle parameters $\beta_1$ and $\beta_2$ further characterize the 
rotation of each of the 4-branes (and 5-branes) in B-type and A-type respectively.

As shown in Table \ref{tab_angles}, sixteen combinations of 
A-type and B-type boundary branes correspond to the basic quarter-BPS boundary conditions discussed in section 
\ref{sec_quarter}. 
We see that the N$'$N$''$ boundary conditions are realized for 
$(\alpha_1,\alpha_2)=(0,0)$, 
the N$'$D$''$ boundary conditions for 
$(\alpha_1,\alpha_2)=(0,\frac{\pi}{2})$, 
the D$'$D$''$ boundary conditions for 
$(\alpha_1,\alpha_2)=(\pi,0)$ 
and the D$'$N$''$ boundary conditions for 
$(\alpha_1,\alpha_2)=(\pi,\frac{\pi}{2})$. 

In particular, we can again check that 
$\sigma_+$ which corresponds to the $x^2$ positions of D2-branes
is fixed by a pair of A-type and B-type D4-branes 
while it can still freely move for a pair of A-type and B-type NS5-branes 
as we found in the D$'$D$''$ boundary conditions and N$'$N$''$ boundary conditions. 

\begin{table}
\begin{center}
\begin{tabular}{|c|c|c|c|c|c|} \hline
a pair of branes&boundary condition &$\alpha_1$ & $\alpha_2$ & $\beta_1$& $\beta_2$\\ \hline
$\widetilde{\textrm{NS5}}'$-NS$''$ &N$'$N$''$&$0$&$0$&$0$&$0$ \\
$\widetilde{\textrm{NS5}}'$-$\widetilde{\textrm{NS5}}''$ &N$'$N$''$&$0$&$0$&$0$&$\frac{\pi}{2}$ \\
NS5$'$-NS5$''$ &N$'$N$''$&$0$&$0$&$\pi$&$\frac{\pi}{2}$ \\
NS5$'$-$\widetilde{\textrm{NS5}}''$ &N$'$N$''$&$0$&$0$&$\pi$&$0$ \\
$\widetilde{\textrm{NS5}}'$-$\widetilde{\textrm{D4}}''$ &N$'$D$''$&$0$&$\frac{\pi}{2}$&$0$&$0$ \\
$\widetilde{\textrm{NS5}}'$-D4$''$ &N$'$D$''$&$0$&$\frac{\pi}{2}$&$0$&$\frac{\pi}{2}$ \\ 
NS5$'$-D4$''$ &N$'$D$''$&$0$&$\frac{\pi}{2}$&$\pi$&$0$ \\
NS5$'$-$\widetilde{\textrm{D4}}''$ &N$'$D$''$&$0$&$\frac{\pi}{2}$&$\pi$&$\frac{\pi}{2}$ \\
$\widetilde{\textrm{D4}}'$-$\widetilde{\textrm{D4}}''$&D$'$D$''$&$\pi$&$0$&$0$&$0$ \\
$\widetilde{\textrm{D4}}'$-D4$''$ &D$'$D$''$&$\pi$&$0$&$0$&$\frac{\pi}{2}$ \\ 
D4$'$-$\widetilde{\textrm{D4}}''$ &D$'$D$''$&$\pi$&$0$&$\pi$&$\frac{\pi}{2}$ \\
D4$'$-D4$''$ &D$'$D$''$&$\pi$&$0$&$\pi$&$0$ \\ 
$\widetilde{\textrm{D4}}'$-NS$''$ &D$'$N$''$&$\pi$&$\frac{\pi}{2}$&$0$&$0$ \\
$\widetilde{\textrm{D4}}'$-$\widetilde{\textrm{NS5}}''$ &D$'$N$''$&$\pi$&$\frac{\pi}{2}$&$0$&$\frac{\pi}{2}$ \\
D4$'$-$\widetilde{\textrm{NS5}}''$ &D$'$N$''$&$\pi$&$\frac{\pi}{2}$&$\pi$&$0$ \\ 
D4$'$-NS5$''$ &D$'$N$''$&$\pi$&$\frac{\pi}{2}$&$\pi$&$\frac{\pi}{2}$ \\ \hline
\end{tabular}
\caption{The phases $(\alpha_1, \alpha_2, \beta_1, \beta_2)$ appearing in the fermionic quarter BPS boundary 
conditions for $a=b=1$ and the corresponding pairs of A-type and B-type branes in Type IIA string theory. }
\label{tab_angles}
\end{center}
\end{table}

We have not yet understood the brane construction of the mixed quarter-BPS boundary conditions 
with $a=0$, $b=1$. 
It would be intriguing to explore the additional objects such as fluxes which realize them while maintaining the 1d $\mathcal{N}=1$ supersymmetry. 

Finally, we note that the quarter-BPS configurations have a simple M-theory description as a single M5-brane wrapping a special Lagrangian 5-cycle in
$\Cb^5$ with complex coordinates $x^1 + i x^6$, $x^2 + i x^{10}$, $x^3 + i x^7$,
$x^4 - i x^8$ and $x^5 + i x^9$. This specialises to the half-BPS 
configurations, noting that in that case the holomorphic curve in $\Cb^2$
can equivalently be described as a special Lagrangian 2-cycle with a different
complex structure. As for the half-BPS configurations, we are not aware of any
supergravity solutions for the M-brane system, but for the M5-brane alone
wrapping a special Lagrangian 5-cycle, some solutions are described in
\cite{Gauntlett:2000ng}.

%%%%%%%%%%%%%%%%%%%%%%%%%%%%%%%%%%%
\subsection*{Acknowledgements}
The authors would like to thank Mauricio Romo for useful discussions and comments. 
This work is supported by STFC Consolidated Grant ST/P000371/1.
%%%%%%

\appendix

%%%%%%%%%%%%%%%%%%%%%%%%%%%%%%%%%%
%%%%%%%%%%%%%%%%%%%%%%%%%%%%%%%%%%
\section{Reduction from 4d $\mathcal{N}=1$ to 2d $\mathcal{N}=(2,2)$}
\label{sec_4dN1}
Here we summarize some of the conventions used, particularly for spinors and
superfields. We also include some details of the reduction from 4d
$\mathcal{N}=1$ to 2d $\mathcal{N}=(2,2)$ as particularly for the supercurrent
much of the calculation can be carried out more conveniently in 4d and then the
result reduced to 2d.

\subsection{Spinor conventions}
\label{sec_spinorconv}
%{\bf [To compare our conventions with 
%\cite{Witten:1993yc} for 2d $\mathcal{N}=(2,2)$ supersymmetric field theories 
%\cite{Wess:1992cp} for 4d $\mathcal{N}=1$ supersymmetric field theories.
%} 

Our spinor conventions in 4d are those used in \cite{Wess:1992cp}. For convenience we list some of these here.
%\footnote{Some good sources of conventions are Wess \& Bagger, Witten hep-th/9301042, \cite{Hanany:1997vm}.}
\begin{itemize}
\item We use 4d spacetime indices $m \in \{0, 1, 2, 3\}$
with Minkowski metric
$\eta_{mn}$, with $\eta_{00} = -1$.

\item We use standard $\alpha, \dot{\alpha}$ notation in 4d, lowering and
raising indices with $\epsilon_{\alpha\beta}$ and $\epsilon^{\alpha\beta}$
where $\epsilon^{12} = -\epsilon_{12} = 1$. Specifically,
$\psi^{\alpha} = \epsilon^{\alpha\beta}\psi_{\beta}$ and
$\psi_{\alpha} = \epsilon_{\alpha\beta}\psi^{\beta}$. We also refer to the
spinor indices $(1, 2)$ as $(-, +)$.

\item The 4d gamma-matrices can be written in terms of
$\sigma^m_{\alpha\dot{\alpha}}$, the three Pauli sigma matrices together with
$\sigma^0 = -\sigma_0 = -I$. The matrices $\bar{\sigma}^m$ are defined by
raising the spinor indices on $\sigma^m$,
$\bar{\sigma}^{m\dot{\alpha}\alpha} = \epsilon^{\dot{\alpha}\dot{\beta}} \epsilon^{\alpha\beta} \sigma^m_{\alpha\dot{\beta}}$.

\item We contract spinor indices as follows:
$\lambda \psi = \lambda^{\alpha} \psi_{\alpha}$,
$\bar{\lambda} \bar{\psi} = \bar{\lambda}_{\dot{\alpha}} \bar{\psi}^{\dot{\alpha}}$,
$\theta \sigma^m \bar{\theta} = \theta^{\alpha} \sigma^m_{\alpha \dot{\alpha}} \bar{\theta}^{\dot{\alpha}}$
and $\bar{\theta} \bar{\sigma}^m \theta = \bar{\theta}_{\dot{\alpha}} (\bar{\sigma}^m)^{\dot{\alpha} \alpha} \theta_{\alpha}$.
%$\epsilon\gamma^{\mu}\bar{\lambda} = \epsilon^{\alpha}\gamma^{\mu}_{\alpha\dot{\alpha}}\lambda^{\dot{\alpha}}$
%and $\bar{\epsilon}\gamma^{\mu}\lambda = \bar{\epsilon}^{\dot{\alpha}}\gamma^{\mu}_{\dot{\alpha}\alpha}\lambda^{\alpha}$.

\item Antisymmetric products of
sigma matrices are defined as
\begin{align}
\sigma^{mn} & = \frac{1}{4} \left( \sigma^m \sigmab^n - \sigma^n \sigmab^m \right) \\
\overline{\sigma}^{mn} & = \frac{1}{4} \left( \sigmab^m \sigma^n - \sigmab^n \sigma^m \right)
\end{align}

\end{itemize}

The following summarizes our conventions in 2d, and the reduction from 4d to 2d,
matching those in \cite{Witten:1993yc} for 2d $\mathcal{N}=(2,2)$ supersymmetric field theories.
\begin{itemize}
\item We denote the 2d Minkowski metric $\eta_{\mu\nu}$ with $\eta_{00} = -1$
and in the reduction from 4d we identify Lorentz indices
$m=0$ with $\mu =0$ but $m=3$ with $\mu = 1$.

\item The 4d vector, chiral and anti-chiral multiplets all reduce to the
corresponding 2d multiplets.
The 4d gauge potential $v_m$ becomes the 2d gauge potential $A_{\mu}$
with $A_0 = v_0$, $A_1 = v_3$ and a complex scalar
$\sigma = \frac{1}{\sqrt{2}}(v_1 - iv_2)$.

\item The 2d spinor indices are the same as those in 4d except we no longer
distinguish $\dot{\alpha}$ from $\alpha$ and all index contractions are top-left to bottom-right. This can lead to some changes of sign
compared to 4d expressions, e.g.\ $\bar{\lambda} \bar{\psi} = \bar{\lambda}_{\dot{\alpha}} \bar{\psi}^{\dot{\alpha}} \rightarrow \bar{\lambda}_{\alpha} \bar{\psi}^{\alpha} = - \bar{\lambda}^{\alpha} \bar{\psi}_{\alpha} = - \bar{\lambda} \bar{\psi}$.

\item The 2d gamma-matrices $(\gamma^{\mu})_{\alpha}^{\phantom{\alpha} \beta}$
are defined by raising the second spinor index of $\sigma^0$ and $\sigma^3$,
with the explicit expressions
\begin{equation}
\left( \gamma^{0 \phantom{\alpha} \beta}_{\phantom{0} \alpha} \right) = \left( \begin{array}{cc} 0 & 1 \\ -1 & 0 \end{array} \right) \; , \;\;\;
\left( \gamma^{1 \phantom{\alpha} \beta}_{\phantom{0} \alpha} \right) = \left( \begin{array}{cc} 0 & -1 \\ -1 & 0 \end{array} \right) \; .
\end{equation}
We contract indices with gamma matrices in this way, e.g.\
$\bar{\epsilon}\gamma^{\mu}\lambda = \bar{\epsilon}^{\alpha}\gamma^{\mu \phantom{\alpha} \beta}_{\phantom{\mu} \alpha}\lambda_{\beta}$.

\end{itemize}

%[{\bf Can we summarize them w/o itemize?}]

%%%%%%%%%%%%%%%%%%%%%%%%%%%%%%%%%%
\subsection{Supermultiplet}
%%%%%%%%%%%%%%%%%%%%%%%%%%%%%%%%%%

%%%%%%%%%%%%%%%%%%%%%%%%%%%%%%%%%%
\subsubsection{Vector multiplet}
%%%%%%%%%%%%%%%%%%%%%%%%%%%%%%%%%%
We work in WZ gauge where the vector multiplet $V$ is a real superfield with
component superfield expansion
\begin{eqnarray}
V(x, \theta) & = & -\theta \sigma^m \bar{\theta} v_m + i \theta \theta \bar{\theta} \bar{\lambda} - i \bar{\theta} \bar{\theta} \theta \lambda + \frac{1}{2} \theta \theta \bar{\theta} \bar{\theta} D \\
 & = & -\theta \gamma^{\mu} \bar{\theta} A_{\mu} - \sqrt{2} \theta^{-} \bar{\theta}^{+} \sigma - \sqrt{2} \theta^{+} \bar{\theta}^{-} \sigmad \nonumber \\
 & & + i \theta \theta \bar{\theta} \bar{\lambda} - i \bar{\theta} \bar{\theta} \theta \lambda + \frac{1}{2} \theta \theta \bar{\theta} \bar{\theta} D
\end{eqnarray}

The supersymmetric transformations of these components are
\begin{eqnarray}
\delta A_{\mu} & = & i \bar{\epsilon}\gamma_{\mu}\lambda + i \epsilon\gamma_{\mu}\bar{\lambda} \\
\delta \sigma & = & -i\sqrt{2} \bar{\epsilon}_{+}\lambda_{-} -i\sqrt{2} \epsilon_{-}\bar{\lambda}_{+} \\
\delta \sigmad & = & -i\sqrt{2} \epsilon_{+}\bar{\lambda}_{-} -i\sqrt{2} \bar{\epsilon}_{-}\lambda_{+} \\
\delta \lambda_{+} & = & i \epsilon_{+} D + \sqrt{2}(D_0 + D_1)
\sigmad \epsilon_{-} - F_{01}\epsilon_{+} \\
\delta \lambda_{-} & = & i \epsilon_{-} D + \sqrt{2}(D_0 - D_1)
\sigma \epsilon_{+} + F_{01}\epsilon_{-} \\
\delta \bar{\lambda}_{+} & = & - i \bar{\epsilon}_{+} D + \sqrt{2}(D_0 + D_1)
\sigma \bar{\epsilon}_{-} - F_{01}\bar{\epsilon}_{+} \\
\delta \bar{\lambda}_{-} & = & - i \bar{\epsilon}_{-} D + \sqrt{2}(D_0 - D_1)
\sigmad \bar{\epsilon}_{+} + F_{01}\bar{\epsilon}_{-} \\
\delta D & = & - \bar{\epsilon}_{+}(D_0 - D_1)\lambda_{+}
 - \bar{\epsilon}_{-}(D_0 + D_1)\lambda_{-} \nonumber \\
 & & + \epsilon_{+}(D_0 - D_1)\bar{\lambda}_{+}
 + \epsilon_{-}(D_0 + D_1)\bar{\lambda}_{-}
\end{eqnarray}

%%%%%%%%%%%%%%%%%%%%%%%%%%%%%%%%%%
\subsubsection{Chiral and anti-chiral multiplets}
%%%%%%%%%%%%%%%%%%%%%%%%%%%%%%%%%%
A chiral superfield $\Phi$ is defined by $\bar{D}_{\dot{\alpha}} \Phi = 0$
while and its conjugate $\bar{\Phi}$ is automatically an anti-chiral superfield,
in general defined by $D_{\alpha} \bar{\Phi} = 0$. If we define
\begin{eqnarray}
y^m & = & x^m + i \theta \sigma^m \bar{\theta} \\
\bar{y}^m & = & x^m - i \theta \sigma^m \bar{\theta}
\end{eqnarray}
the general solutions take the form
\begin{eqnarray}
\Phi(x, \theta) & = & \phi(y) + \sqrt{2}\theta\psi(y) + \theta \theta F(y) \\
\bar{\Phi}(x, \theta) & = & \bar{\phi}(\bar{y}) + \sqrt{2}\bar{\theta}\bar{\psi}(\bar{y}) + \bar{\theta} \bar{\theta} \bar{F}(\bar{y})
\end{eqnarray}
Expanding to write all fields as functions of $x^m$ rather than $y^m$ we have
\begin{eqnarray}
\Phi(x, \theta) & = & \phi + i \theta \sigma^m \bar{\theta} \partial_m \phi
 + \frac{1}{4} \theta \theta \bar{\theta} \bar{\theta} \Box \phi \nonumber \\
 & & + \sqrt{2}\theta\psi
 - \frac{i}{\sqrt{2}}\theta \theta (\partial_m \psi) \sigma^m \bar{\theta}
 + \theta \theta F
\end{eqnarray}

For gauge theories, replace partial derivatives with gauge covariant
derivatives $D_m = \partial_m + ig v_m$.

To reduce to 2d we just drop derivatives wrt.\ the two compactified dimensions.
This gives
\begin{eqnarray}
\Phi(x, \theta) & = & \phi + i \theta \gamma^{\mu} \bar{\theta} D_{\mu} \phi
 - \sqrt{2} g \theta^{-} \bar{\theta}^{+} \sigma \phi
 - \sqrt{2} g \theta^{+} \bar{\theta}^{-} \sigmad \phi \nonumber \\
 & & + \sqrt{2}\theta\psi
 - \frac{i}{\sqrt{2}}\theta \theta (D_{\mu} \psi) \gamma^{\mu} \bar{\theta}
 + \theta \theta F \nonumber \\
 & & + g \theta \theta \left(\sigma \psi^{-} \bar{\theta}^{+} + \sigmad \psi^{+} \bar{\theta}^{-} \right) \nonumber \\
 & & + \frac{1}{4} \theta \theta \bar{\theta} \bar{\theta} (\Box \phi + \cdots)
\end{eqnarray}

The supersymmetric transformations are
\begin{eqnarray}
\delta \phi & = & \sqrt{2} \left( \epsilon_{+}\psi_{-} - \epsilon_{-} \psi_{+} \right) \\
\delta \psi_{+} & = & i\sqrt{2}(D_0 + D_1)\phi \bar{\epsilon}_{-}
	+ \sqrt{2}\epsilon_{+} F - 2 \sigmad\phi \bar{\epsilon}_{+} \\
\delta \psi_{-} & = & - i\sqrt{2}(D_0 - D_1)\phi \bar{\epsilon}_{+} 
        + \sqrt{2}\epsilon_{-} F + 2 \sigmad\phi \bar{\epsilon}_{-} \\
\delta F & = & - i\sqrt{2} \bar{\epsilon}_{+} (D_0 - D_1)\psi_{+}
	- i\sqrt{2} \bar{\epsilon}_{-} (D_0 + D_1)\psi_{-} \nonumber \\
	& & + 2(\bar{\epsilon}_{+}\sigmad\psi_{-} + \bar{\epsilon}_{-}\sigmad\psi_{+})
        + 2i(\bar{\epsilon}_{-}\bar{\lambda}_{+} - \bar{\epsilon}_{+}\bar{\lambda}_{-} ) \phi
\end{eqnarray}

%%%%%%%%%%%%%%%%%%%%%%%%%%%%%%%%%%
\subsubsection{Twisted chiral and anti-chiral multiplets}
%%%%%%%%%%%%%%%%%%%%%%%%%%%%%%%%%%
A twisted chiral superfield $\widetilde{\Phi}$ is defined in 2d by
$\bar{D}_{-} \widetilde{\Phi} = 0 = D_{+} \widetilde{\Phi}$
while and its conjugate $\bar{\widetilde{\Phi}}$ is automatically a twisted
anti-chiral superfield, in general defined by
$\bar{D}_{+} \bar{\widetilde{\Phi}} = 0 = D_{-} \bar{\widetilde{\Phi}}$.
If we define
\begin{eqnarray}
z^0 & = & x^0 + i \theta \gamma^1 \bar{\theta} \\
z^1 & = & x^0 - i \theta \gamma^0 \bar{\theta}
\end{eqnarray}
where in 2d notation $\theta \gamma^{\mu} \bar{\theta} = \theta^{\alpha} \gamma^{\mu\phantom{\alpha}\beta}_{\phantom{\mu}\alpha} \bar{\theta}_{\beta}$
the general solution for a twisted chiral superfield takes the form
\begin{eqnarray}
\widetilde{\Phi}(x, \theta) & = & \widetilde{\phi}(z) + \sqrt{2}\theta^{-}\widetilde{\psi}_{-}(z) + \sqrt{2}\bar{\theta}^{+}\overline{\widetilde{\psi}}_{+}(z) + 2 \bar{\theta}^{R} \theta^{L} \widetilde{F}(z)
%\bar{\Phi}(x, \theta) & = & \bar{\phi}(\bar{y}) + \sqrt{2}\bar{\theta}\bar{\psi}(\bar{y}) + \bar{\theta} \bar{\theta} \bar{F}(\bar{y})
\end{eqnarray}
%Expanding to write all fields as functions of $x^m$ rather than $z^m$ we have
%\begin{eqnarray}
%\widetilde{\Phi}(x, \theta) & = & \phi + i \theta \sigma^m \bar{\theta} \partial_m \phi
% + \frac{1}{4} \theta \theta \bar{\theta} \bar{\theta} \Box \phi \nonumber \\
% & & + \sqrt{2}\theta\psi
% - \frac{i}{\sqrt{2}}\theta \theta (\partial_m \psi) \sigma^m \bar{\theta}
% + \theta \theta F
%\end{eqnarray}
%
%For gauge theories, replace partial derivatives with gauge covariant
%derivatives $D_m = \partial_m + ig v_m$.
%
%To reduce to 2d we just drop derivatives wrt.\ the two compactified dimensions.
%This gives
%\begin{eqnarray}
%\widetilde{\Phi}(x, \theta) & = & %\phi + i \theta \gamma^{\mu} \bar{\theta} D_{\mu} \phi
% - \sqrt{2} g \theta^{-} \bar{\theta}^{+} \sigma \phi
% - \sqrt{2} g \theta^{+} \bar{\theta}^{-} \sigmad \phi \nonumber \\
% & & + \sqrt{2}\theta\psi
% - \frac{i}{\sqrt{2}}\theta \theta (D_{\mu} \psi) \gamma^{\mu} \bar{\theta}
% + \theta \theta F \nonumber \\
% & & + g \theta \theta \left(\sigma \psi^{-} \bar{\theta}^{+} + \sigmad \psi^{+} \bar{\theta}^{-} \right) \nonumber \\
% & & + \frac{1}{4} \theta \theta \bar{\theta} \bar{\theta} (\Box \phi + \cdots)
%\end{eqnarray}

%{\bf [To complete the above]}

Compared to the chiral superfield we have the following replacements
\begin{align}
\theta^{+} \leftrightarrow \bar{\theta}^{+} & , \;\;
\theta_{-} \leftrightarrow \bar{\theta}_{-} \\
\epsilon^{+} \leftrightarrow \bar{\epsilon}^{+} & , \;\;
\epsilon_{-} \leftrightarrow \bar{\epsilon}_{-} \\
\psi_{+} \rightarrow \overline{\widetilde{\psi}}_{+} & .
\end{align}
with the other components of $\theta$ and $\epsilon$ unchanged, and other fields trivially gaining a tilde.
Therefore we can map all the result for the chiral superfield. In particular,
the supersymmetric transformations are
\begin{eqnarray}
\delta \widetilde{\phi} & = & \sqrt{2} \left( \epsilon_{+}\widetilde{\psi}_{-} - \bar{\epsilon}_{-} \bar{\widetilde{\psi}}_{+} \right) \\
\delta \bar{\widetilde{\psi}}_{+} & = & i\sqrt{2}(D_0 + D_1)\widetilde{\phi} \epsilon_{-}
        + \sqrt{2}\epsilon_{+} \widetilde{F} - 2 \sigmad\widetilde{\phi} \bar{\epsilon}_{+} \\
\delta \widetilde{\psi}_{-} & = & - i\sqrt{2}(D_0 - D_1)\widetilde{\phi} \bar{\epsilon}_{+}
        + \sqrt{2}\epsilon_{-} \widetilde{F} + 2 \sigmad\widetilde{\phi} \epsilon_{-} \\
\delta \widetilde{F} & = & - i\sqrt{2} \bar{\epsilon}_{+} (D_0 - D_1)\bar{\widetilde{\psi}}_{+}
        - i\sqrt{2} \epsilon_{-} (D_0 + D_1)\widetilde{\psi}_{-} \nonumber \\
        & & + 2(\bar{\epsilon}_{+}\sigmad\widetilde{\psi}_{-} + \epsilon_{-}\sigmad\bar{\widetilde{\psi}}_{+})
        + 2i(\epsilon_{-}\bar{\lambda}_{+} - \bar{\epsilon}_{+}\bar{\lambda}_{-} ) \widetilde{\phi}
\end{eqnarray}

%%%%%%%%%%%%%%%%%%%%%%%%%%%%%%%%%%
\subsection{Supercurrent}
%%%%%%%%%%%%%%%%%%%%%%%%%%%%%%%%%%
In WZ gauge the action for the gauge field and chiral multiplets are given by
\begin{eqnarray}
\mathcal{L}_{\textrm{gauge}} & = & \frac{1}{2} \Tr \left( W^{\alpha}W_{\alpha}\vert_{\theta\theta} + h.c. \right) \\
 & = & -\frac{1}{2} \Tr (v_{mn}v^{mn}) - 2i \Tr(\bar{\lambda}\bar{\sigma}^mD_m\lambda) + \Tr(DD) \nonumber \\
 & & + i \partial_m \Tr(\bar{\lambda}\bar{\sigma}^m\lambda) \\
\mathcal{L}_{\textrm{chiral}} & = & \Phi^{\dagger} e^V \Phi \vert_{\theta\theta\bar{\theta}\bar{\theta}} \\
 & = & g \phi^{\dagger}D\phi - (D_m\phi^{\dagger})(D^m \phi)
 - i \bar{\psi} \bar{\sigma}^m D_m \psi + F^{\dagger} F \nonumber \\
 & & + i \sqrt{2} g \left(\phi^{\dagger}\lambda\psi - \bar{\psi}\bar{\lambda}\phi \right) + \frac{i}{2}\partial_m(\bar{\psi}\bar{\sigma}^m\psi)
\end{eqnarray}
where the final terms in each Lagrangian are required for the Lagrangians to be
real in the presence of a boundary.

The supersymmetric transformations in 4d are
\begin{eqnarray}
\delta \phi & = & \sqrt{2}\epsilon \psi \\
\delta \psi & = & i \sqrt{2}\sigma^m \bar{\epsilon} D_m \phi + \sqrt{2}\epsilon F \\
\delta F & = & i \sqrt{2} \bar{\epsilon} \bar{\sigma}^m D_m \psi
	+ i 2g \bar{\epsilon} \bar{\lambda} \phi \\
\delta v_m & = & -i \bar{\lambda} \bar{\sigma}^m \epsilon + i \bar{\epsilon} \bar{\sigma}^m \lambda \\
\delta \lambda & = & \sigma^{mn} \epsilon v_{mn} + i \epsilon D \\
\delta D & = & -\epsilon \sigma^m D_m \bar{\lambda} - (D_m \lambda) \sigma^m \bar{\epsilon}
\end{eqnarray}

If we vary the Lagrangians with constant $\epsilon$ we find a total derivative
so the action is invariant with suitable boundary conditions. If we let
$\epsilon$ depend on the spacetime coordinates we find additional terms
of the form $J^m \partial_m \epsilon + \bar{J}^m \partial_m \bar{\epsilon}$
which define the supercurrents $J$ and $\bar{J}$. Note, in our conventions
this defines $\bar{J}^m$ to be the conjugate of $J^m$. Explicitly 
we have
\begin{eqnarray}
\delta \mathcal{L}_{\textrm{gauge}} & = & \partial_m \Tr \left( -\bar{\lambda}\bar{\sigma}^m \epsilon D - i \bar{\lambda}\bar{\sigma}_n \epsilon v^{mn} - \bar{\epsilon} \bar{\sigma}_n \lambda \tilde{v}^{mn} \right) \nonumber \\
 & & + i \left( (\partial_m\bar{\epsilon})\bar{\sigma}_n\lambda - \bar{\lambda}\bar{\sigma}_n (\partial_m \epsilon) \right) v^{mn} + \bar{\lambda}\bar{\sigma}_n(\partial_m \epsilon) \tilde{v}^{mn} + (\partial_m \bar{\epsilon})\bar{\sigma}_n \lambda \tilde{v}^{mn} \\
\delta \mathcal{L}_{\textrm{chiral}} & = & \partial_m \left( g\phi^{\dagger}(\bar{\lambda}\bar{\sigma}^m \epsilon + \bar{\epsilon}\bar{\sigma}^m \lambda)\phi - \sqrt{2}i\bar{\psi}\bar{\sigma}^m \epsilon F - \sqrt{2}\bar{\epsilon}\bar{\psi}D^m\phi -2\sqrt{2}\phi^{\dagger}\epsilon\sigma^{mn} D_n\psi \right) \nonumber \\
 & & + \sqrt{2} \bar{\psi}\bar{\sigma}^m\sigma^n(D_n \phi) (\partial_m \bar{\epsilon})
 + \sqrt{2} (\partial_m \epsilon) \left( 2\phi^{\dagger} \sigma^{mn}D_n\psi - (D^m \phi^{\dagger})\psi \right) \nonumber \\
  & & - g \phi^{\dagger} \bar{\lambda} \bar{\sigma}^m(\partial_m \epsilon) \phi
 - g \phi^{\dagger} (\partial_m \bar{\epsilon}) \bar{\sigma}^m \lambda \phi \nonumber \\
 & = & \partial_m \left( g\phi^{\dagger}(\bar{\lambda}\bar{\sigma}^m \epsilon + \bar{\epsilon}\bar{\sigma}^m \lambda)\phi - \sqrt{2}i\bar{\psi}\bar{\sigma}^m \epsilon F - \sqrt{2}\bar{\epsilon}\bar{\psi}D^m\phi + 2\sqrt{2}(D_n \phi^{\dagger})\epsilon\sigma^{mn} \psi \right) \nonumber \\
 & & + \sqrt{2} \bar{\psi}\left( 2\bar{\sigma}^{mn} - \eta^{mn} \right) (D_n \phi) (\partial_m \bar{\epsilon})
 - \sqrt{2} (\partial_m \epsilon) (D_n \phi^{\dagger}) \left( 2\sigma^{mn} + \eta^{mn} \right) \psi \nonumber \\
  & & - g \phi^{\dagger} \bar{\lambda} \bar{\sigma}^m(\partial_m \epsilon) \phi
 - g \phi^{\dagger} (\partial_m \bar{\epsilon}) \bar{\sigma}^m \lambda \phi
\end{eqnarray}
where we have defined
\begin{equation}
\tilde{v}^{mn} = \frac{1}{2}\epsilon^{mnpq} v_{pq}
\end{equation}

Extracting the supercurrents from each part we find for the gauge multiplet
\begin{eqnarray}
J^m & = & -(\tilde{v}^{mn} - i v^{mn})\sigma_n \bar{\lambda} \\
\bar{J}^m & = & (\tilde{v}^{mn} + i v^{mn})\bar{\sigma}_n \lambda \; .
\end{eqnarray}
Similarly for the fundamental chiral multiplet we have
\begin{eqnarray}
J^m & = & -2\sqrt{2}(D_n \phi^{\dagger}) \sigma^{mn} \psi - \sqrt{2}(D^m\phi^{\dagger})\psi + g \phi^{\dagger} \sigma^m\bar{\lambda} \phi \\
\bar{J}^m & = & -2\sqrt{2}\bar{\sigma}^{mn}\bar{\psi}(D_n\phi) - \sqrt{2}\bar{\psi}(D^m\phi) - g \phi^{\dagger} \bar{\sigma}^m\lambda \phi
\end{eqnarray}

\section{More on A-type boundary conditions}
\label{AtypeSing}
%%%%%%%%%%%%%%%%%%%%%%%%%%%%%%%%%%
%%%%%%%%%%%%%%%%%%%%%%%%%%%%%%%%%%
We note that (\ref{AvmG2}), (\ref{AvmG4}) and (\ref{AvmG5}) 
take a form similar to Hitchin's equations \cite{Hitchin:1986vp}
\begin{align}
F_A&=[\Phi, \overline{\Phi}], \label{Hitchin_eq1}\\
\overline{D}_{A}\Phi&=0 \label{Hitchin_eq2}. 
\end{align}
%The moduli space of solutions to the Hitchin's equations is a hyperk\"{a}hler manifold with three complex structures. 
The singular solution of the Hitchin's equation was mathematically studied in 
\cite{MR1159261} and it was used to define the surface operator in 4d $\mathcal{N}=4$ SYM theory \cite{Gukov:2006jk,Gukov:2008sn}. 
The surface operator was constructed by postulating the rotational invariant singular configurations
\begin{align}
A&=a(r)d\theta+f(r)\frac{dr}{r},\\
\Phi&=b(r)\frac{dr}{r}-c(r)d\theta
\end{align}
where $x^0+ix^1=re^{i\theta}$ are the coordinates on $\mathbb{R}^2$. 
$f(r)$ can be set to zero by gauge transformation. 
When we introduce a new variable $s=-\log r$, 
the Hitchin's equations (\ref{Hitchin_eq1}) and (\ref{Hitchin_eq2}) takes the form of the Nahm's equations
\begin{align}
\label{Nahm_eq}
\frac{da}{ds}&=[b,c],& 
\frac{db}{ds}&=[c,a],& 
\frac{dc}{ds}&=[a,b]. 
\end{align}
The superconformal invariant solution which has no dependence on $s$ can be obtained 
by setting $a,b$ and $c$ to a constant elements $\alpha, \beta$ and $\gamma$ 
of the Lie algebra $\mathfrak{t}$ of a maximal torus $\mathbb{T}$ of the gauge group. 
Then one finds the singular solution to the Hitchin's equation with the form 
\begin{align}
A&=\alpha d\theta,\\
\Phi&=\beta \frac{dr}{r}-\gamma d\theta
\end{align}
where the Higgs field $\Phi$ has a pole at the origin. 

Now let us go back to the boundary conditions. 
To find the solutions of the boundary conditions (\ref{AvmG2}), (\ref{AvmG4}) and (\ref{AvmG5}) 
for the vector multiplet, 
we first fix the gauge so that $A_u$ and $A_{\overline{u}}$ commute. 
We then take the ansatz
\begin{align}
\label{A_singular_sol1}
A_{u}&=c_1\frac{\mathfrak{s}_3}{x^1} + \cdots ,& 
A_{\overline{u}}&=c_1^*\frac{\mathfrak{s}_3}{x^1} + \cdots , \\
\label{A_singular_sol2}
\hat{\sigma}&=c_2\frac{\mathfrak{s}_+}{x^1} + \cdots , & \hat{\sigma}^{\dagger}&=c_2^*\frac{\mathfrak{s}_-}{x^1} + \cdots
\end{align}
where we have only indicated the singular terms at the boundary,
$\mathfrak{s}_1$, $\mathfrak{s}_2$ and $\mathfrak{s}_3$ are constant elements
of the Lie algebra, $\mathfrak{s}_{\pm} = \mathfrak{s}_1 \pm i \mathfrak{s}_2$
and $c_1, c_2\in \mathbb{C}$ are some numerical constant values. 
The boundary conditions (\ref{AvmG2}), (\ref{AvmG4}) and (\ref{AvmG5}) require that 
\begin{align}
\mathfrak{s}_3 & = - \frac{\sqrt{2} ig |c_2|^2}{c_+ c_1^*-c_+^*c_1}[\mathfrak{s}_+, \mathfrak{s}_-], \label{s_eq1}\\
\mathfrak{s}_+&=\frac{\sqrt{2} ig c_1}{c_+}[\mathfrak{s}_3,\mathfrak{s}_+], \label{s_eq2} \\
\mathfrak{s}_-&=\frac{\sqrt{2} i g c_1^*}{c_+^*}[\mathfrak{s}_3,\mathfrak{s}_-] \label{s_eq3}. 
\end{align}
By setting 
\begin{align}
c_1&=-\frac{ic_{+}}{\sqrt{2} g},& 
c_2&=\frac{c_+}{g}, 
\end{align}
the equations (\ref{s_eq1}), (\ref{s_eq2}) and (\ref{s_eq3}) simply implies that 
the constant elements $\mathfrak{s}_3$, $\mathfrak{s}_+$ and $\mathfrak{s}_-$ belong to the $\mathfrak{su}(1,1)$ with 
the relation
\begin{align}
\mathfrak{s}_3&= - [\mathfrak{s}_+,\mathfrak{s}_-],\\
\mathfrak{s}_+&=[\mathfrak{s}_3,\mathfrak{s}_+],\\
\mathfrak{s}_-&=-[\mathfrak{s}_3,\mathfrak{s}_-].
\end{align}
where $\mathfrak{s}_{\pm}=\mathfrak{s}_1\pm i \mathfrak{s}_2$ are raising and lowering operators 
of $\mathfrak{su}(1,1)$.

\bibliographystyle{utphys}
\bibliography{ref}

\end{document}